\documentclass{rmp-exp}

\usepackage{tipa}

\def\HMCHI{\chi}
\def\ORder{order}

\newcommand{\ALTERNATIVE}[2]{#1}

\def\podr{&&\hspace{0pt}}
\def\nnb{\nonumber \\}
\def\bnn{\\ \nonumber}
\def\nnbnn{\nonumber \\ \nonumber}
\newcommand{\beqs}{\begin{eqnarray*}}
\newcommand{\eeqs}{\end{eqnarray*}}

\newcommand{\beq}{\begin{equation}}
\newcommand{\eeq}{\end{equation}}
\newcommand{\beqa}{\begin{eqnarray}}
\newcommand{\eeqa}{\end{eqnarray}}
\newcommand{\nn}{\nonumber \\}
\def \podr {&& \hspace{0pt}}

\def \setcntrs {\setcounter{equation}{0}\setcounter{theorem}{0}\setcounter{definition}{0}\setcounter{remark}{0}\setcounter{example}{0}}

\newcommand{\medskp}{

  \medskip

  }

\newenvironment{Proof}[1][\it Proof]{\noindent\textit{#1.}${}$\hspace{7pt}${}$}{\nolinebreak$\quad$\nolinebreak$\Box$}

\newcounter{tmpc}
\newlength{\tmplenght}
\newlength{\tmplenghta}
\newlength{\tmplenghtb}
\newlength{\tmplenghtc}

\newenvironment{LIST}[1]{%
\setlength{\tmplenghta}{#1}
\setlength{\tmplenghtb}{#1}
\setlength{\tmplenghtc}{#1}
\advance\tmplenghtb-5pt
\advance\tmplenghtc 42pt
\setcounter{tmpc}{0}
\begin{list}{{\rm (\alph{tmpc})}}{\usecounter{tmpc}
\setlength{\leftmargin}{\tmplenghta}
\setlength{\rightmargin}{0cm}
\setlength{\itemsep}{1pt}
\setlength{\topsep}{3pt}
\setlength{\labelsep}{5pt}
\setlength{\labelwidth}{\tmplenghtb}
\setlength{\listparindent}{\tmplenghta}}
}{\end{list}}

\DeclareMathAlphabet{\mathbbm}{U}{bbm}{m}{n}
\DeclareMathAlphabet{\mathpzc}{OT1}{pzc}{m}{it}

\DeclareSymbolFont{ltrs}     {OT1}{pzc}{m}{it}
\DeclareSymbolFont{ltrsa}     {OMS}{cmsy}{m}{n}
\DeclareSymbolFont{ltrsA}{U}{txmia}{m}{it}
\DeclareSymbolFont{symbolsC}{U}{txsyc}{m}{n}
\DeclareSymbolFont{ltrsB}{U}{rsfs}{m}{n}

\DeclareSymbolFontAlphabet{\mfrak}{ltrsA}
\DeclareMathAlphabet{\mathpzc}{OT1}{pzc}{m}{it}
\DeclareMathAlphabet{\mathrsfs}{U}{rsfs}{m}{n}

\def \R {{\mathbb R}}
\def \C {{\mathbb C}}
\def \Z {{\mathbb Z}}
\def \N {{\mathbb N}}

\def \Sr {{\mathbb S}}

\newcommand{\vrestr}[2]{\hspace{-2pt}\left.\raisebox{#1}{$\,$}\hspace{-2pt}\right|_{\,\raisebox{1pt}{\small \(#2\)}}}

\newcommand{\Txfrac}[2]{\frac{\raisebox{2pt}{$#1$}}{\raisebox{-5pt}{$#2$}}}

\def \z {{\mathrm{z}}}

\def \u {{\mathrm{u}}}
\def \x {{\mathrm{x}}}

\def \y {{\mathrm{y}}}
\def \p {{\mathrm{p}}}

\def \vq {\vec{\mathrm{q}}}

\def \Ss {\mathrsfs{S}}
\def \Dd {\mathrsfs{D}}

\def \spr {\cdot}

\def \di {\partial}

\def \DP {\Dd'}
\def \CI {\mathcal{C}^{\infty}}

\def\CI{\mathcal{C}^{\infty}}

\newcommand{\tlr}[2]{(\hspace{1pt} r^{#1} \hspace{1pt} (\log r)^{#2})_{+}}
\newcommand{\tlrb}[1]{(r^{#1} \bigr)_{+}}
\newcommand{\tlub}[1]{\vartheta(r) \hspace{1pt} r^{#1}}
\newcommand{\TLB}[1]{r^{#1}}
\def\RE{\mathrm{Re}}

\newcommand{\dltan}[1]{\delta^{(#1)}}

\def\VSP{\mathrsfs{D}}
\def\Oo{\mathcal{O}}

\DeclareMathAlphabet{\mathbbm}{U}{bbm}{m}{n}

\DeclareSymbolFont{ltrs}     {OT1}{pzc}{m}{it}
\DeclareSymbolFont{ltrsa}     {OMS}{cmsy}{m}{n}
\DeclareSymbolFont{ltrsA}{U}{txmia}{m}{it}
\DeclareSymbolFont{symbolsC}{U}{txsyc}{m}{n}
\DeclareSymbolFont{ltrsB}{U}{rsfs}{m}{n}

\DeclareSymbolFontAlphabet{\mfrak}{ltrsA}
\DeclareMathAlphabet{\mathpzc}{OT1}{pzc}{m}{it}
\DeclareMathAlphabet{\mathrsfs}{U}{rsfs}{m}{n}

\date{\DATE}

\def \DATE {\today}
\def \TITLE {Renormalization of massless Feynman amplitudes}
\usepackage{ulem}

\def\R{\mathbb{R}}
\def\N{\mathbb{N}}
\def\MREL{\mathfrak{R}}
\def\MIRL{\MCMP \MREL}
\def\mIRL{\mCMP \MREL}

\def\mNRL{\text{$\MREL\hspace{-5pt}\raisebox{1pt}{$\bigl|$}$}}
\def\MCMP{\raisebox{1pt}{\small \textstretchc}}
\def\mCMP{\raisebox{1pt}{\footnotesize \textstretchc}}
\def\Mrel{\text{\rm O}}

\def\MAND{\!,\ }

\def\MDIA{\Delta}
\def\MSET{X}
\def\MISE{I}

\def\MJSE{J}

\def\ECEL{\mathcal{C}}
\def\MCEL{\mathcal{C}^{\text{\tiny $\MCAU$}}}
\def\MPRT{\mathfrak{P}}
\def\MQRT{\mathfrak{Q}}
\def\MEQU{\sim}
\newcommand{\MST}[1]{\underline{#1}}
\def\xx{\vec{\mathrm{x}}}

\def\zz{\vec{\mathrm{z}}}
\def\rr{\vec{\mathrm{q}}}
\def\ee{\vec{\mathrm{e}}}
\def\uu{\mathrm{\bf u}}
\def\uu{\vec{\mathrm{u}}}
\def\uu{\vec{\mathit{u}}}
\def\Mxx{\mathbf{s}}
\def\mxx{s}
\def\mXx{\mathrm{x}}
\def\ESPA{\mathbb{E}}
\def\MSPA{{\mathbb{M}}}
\def\DD{\mathcal{D}}
\def\DP{\DD'}
\def\disjuni{\dot{\cup}}
\def\MKSE{K}
\newcommand{\NSIM}[1]{\nsim\raisebox{-3pt}{\hspace{-4pt}}_{#1}}
\newcommand{\NSIm}[1]{\nsim\raisebox{-3pt}{\hspace{-1pt}}_{#1}}
\newcommand{\SIM}[1]{\sim\raisebox{-3pt}{\hspace{-4pt}}_{#1}}
\newcommand{\SIm}[1]{\sim\raisebox{-3pt}{\hspace{-1pt}}_{#1}}

\def\MCON{\Gamma}
\def\MHYP{\Sigma}
\def\MEUL{\mathrsfs{E}}
\def\MRAD{\varrho}
\def\MUD{\mathcal{G}}
\def\MUH{\widehat{\MUD}}
\def\MUHZ{\MUH{}^0}

\def\MUDZ{\MUD^0}
\def\MDFF{\sigma}

\def\MVOL{\mathrm{Vol}}
\def\MDGR{\kappa}
\def\MLIE{\mathcal{L}}
\newcommand{\Mcnt}[3]{\langle #1 \hspace{0pt},\hspace{1pt} #2 \rangle\raisebox{-1.5pt}{\hspace{-1pt}}_{#3}}
\newcommand{\MCnt}[3]{\bigl\langle #1 \hspace{1pt},\hspace{1pt} #2 \bigr\rangle\raisebox{-4pt}{\hspace{-1.5pt}}_{#3}}
\newcommand{\MCNt}[3]{\Bigl\langle #1 \hspace{1pt},\hspace{1pt} #2 \Bigr\rangle\raisebox{-7pt}{\hspace{-2pt}}_{#3}}
\newcommand{\RMCNt}[1]{\Bigl\langle #1 \hspace{1pt},}
\newcommand{\LMCNt}[2]{#1 \Bigr\rangle\raisebox{-7pt}{\hspace{-2pt}}_{#2}}

\def\MVCT{\mathcal{X}}

\newcommand{\Mref}[1]{{\rm (\ref{#1})}}
\def\MCAU{\precsim}

\def\MDFO{\mathcal{X}}
\def\MIse{\raisebox{-2pt}{\tiny $\MISE$}}
\def\MISe{\raisebox{-1pt}{\scriptsize $\MISE$}}
\def\MJse{\raisebox{-2pt}{\tiny $\MJSE$}}
\def\MJSe{\raisebox{-1pt}{\scriptsize $\MJSE$}}

\def\RCEL{\ECEL^{\MREL}}

\def\MDGR{\alpha}
\def\MORD{n}
\def\MORd{m}
\def\MORDD{n'}

\newcounter{Intrem}

\def\RENMAP{\mathcal{R}}
\def\AMPSPA{\mathrsfs{O}}
\def\BVMAP{\text{\rm b.v.}}

\def\BLT{*}
\def\PRIMAP{\mathcal{P}}
\def\SECMAP{{\mathop{\RENMAP}\limits^{\raisebox{-1pt}{\scriptsize $\bullet$}}}}
\def\sECMAP{{\mathop{\RENMAP}\limits^{\raisebox{0pt}{\tiny $\bullet$}}}}
\def\REGSPA{\mathrsfs{A}}
\def\QU{Q}

\def\RES{\text{\rm res}}
\def\FFU{F}

\def\MODUL{\mathfrak{M}}
\def\Res{\text{\rm Res}}

\def\ZERN{0_N}

\def\PERMGR{\mathcal{S}}

\newcounter{Statement}

\def \setcntrs {\setcounter{equation}{0}\setcounter{Statement}{0}}

\renewcommand{\theStatement}{\arabic{section}.\arabic{Statement}}

\newenvironment{Statement}[1]{%
        \renewcommand{\theStatement}{\thesection.\arabic{Statement}}
        \refstepcounter{Statement}\noindent\textbf{\bf #1\hspace{3.5pt}\arabic{section}.\arabic{Statement}.}${}$\hspace{1pt}${}$\it}{}

\begin{document}

\markboth{N.M. Nikolov, R. Stora and I. Todorov}
{Renormalization of massless Feynman amplitudes in configuration space}

\catchline{}{}{}{}{}

\title{RENORMALIZATION OF MASSLESS FEYNMAN AMPLITUDES IN CONFIGURATION SPACE}

\author{NIKOLAY M. NIKOLOV,${}^{1,\,3}$ \ \ RAYMOND STORA${}^{2,\,3}$ \ \ and \ \ IVAN TODOROV${}^{1,\,3}$}

\address{${}$ \\ \begin{minipage}{250pt}\begin{LIST}{10pt}
\item[${}^1$] 
INRNE, Bulgarian Academy of Sciences, 
Tsarigradsko chaussee 72 Blvd., Sofia 1784, Bulgaria
\item[${}^2$]
Laboratoire d'Annecy-le-Vieux de Physique Th\'eorique (LAPTH),
F-74941 Annecy-le-Vieux Cedex, France
\item[${}^3$]
Theory Division, Department of Physics, CERN, CH-1211 Geneva 23,
Switzerland
\end{LIST}\end{minipage}}

\maketitle

\begin{abstract}
A systematic study of recursive renormalization of Feynman amplitudes is carried out both in 
Euclidean and in Minkowski configuration space. For a massless quantum field theory (QFT) we use the
technique of extending associate homogeneous distributions to complete the renormalization recursion.
A {\it homogeneous} (Poincar\'e covariant) amplitude is said to be convergent if it admits a (unique
covariant) extension as a {\it homogeneous distribution}. 
{\it For any amplitude without subdivergences}
--  i.e. for a Feynman distribution that is homogeneous off the full (small) diagonal -- we define a
{\it renormalization invariant residue}. Its vanishing is a necessary and sufficient condition for the
convergence of such an amplitude. 
It extends to arbitrary - not necessarily primitively divergent - Feynman amplitudes.
This notion of convergence is finer than the usual power counting
criterion and includes cancellation of divergences.
\end{abstract}

\keywords{Renormalization, Feynman amplitudes, residues, dilation anomaly}

\ccode{Mathematics Subject Classification 2010: 81T15, 81T05, 81T18, 81T50, 81Q30 }

\tableofcontents

\section{Introduction}\label{INTR}

Ultraviolet divergences were first discovered and renormalization theory was developed for momentum space integration.
E.C.G. Stueckelberg and A. Petermann \cite{SP}, followed by N.N.~Bogolubov, a mathematician who set himself to master quantum field theory (QFT), realized that (perturbative) renormalization can be formulated as a problem of extending products of distributions, originally defined for non-coinciding arguments in {\it position space} (or, {\it $\x$--space}).
Such  extensions are naturally restricted by {\it locality} or {\it causality}, a concept introduced in QFT by Stueckelberg \cite{Stu} and further developed by Bogolubov and collaborators (for a review and references see \cite{BS}).
Whereas $\x$-space renormalization was straightened out in all generality \cite{BP} \cite{Step} \cite{Hep}, it took some more time to settle the $\p$-space problem \cite{Z} \cite{Zi} \cite{L} \cite{LZ}, resulting in what is now termed the BPHZ theory.
The idea of causal renormalization in position space was taken up and implemented  systematically by H. Epstein and V. Glaser \cite{EG} \cite{EGS}(see also parallel work by O. Steinmann \cite{St}; for later contributions and surveys see 
\cite{Scha1}, \cite{Scha2}, \cite{S82/93}, \cite{S06}). 
It is conceptually clear and represents a crucial step in turning QFT renormalization into a mathematically respectable theory. 
By the late 1990's when the problem of developing perturbative QFT and operator product expansions on a curved background became the order of the day, it was realized that it is just the $\x$-space approach that offers a way to its solution 
\cite{BF,BFV,DF,HW}.

The aim of the present paper is twofold.

First, we develop the view of renormalization as a problem of causal extension of (numerical) distributions,
separating the renormalization problem from concrete QFT models.
We 
therefore use the terms ``renormalization'' and ``extension of distributions'' interchangeably.
Following Epstein and Glaser who construct time ordered products as operator valued distributions we formulate causal requirements similar to theirs directly on ($\x$--space) {\it Feynman amplitudes} attached to 
Feynman graphs.\footnote{%
In this article we are only concerned with the ultraviolet problem,
without any reference to the ``adiabatic limit'', i.e., consider all vertices of a Feynman graph as external.}
This can be viewed as an analogue of the BPHZ procedure in position space.
The extension of the $\x$--space Feynman amplitudes is performed recursively with respect to subgraphs, i.e., with respect to the number of vertices of the graph.
The main condition which leads to this recursion is the {\it causal factorization} for Feynman amplitudes.
One of the important properties of 
a (unrenormalized) Feynman amplitude
is its multiplicative structure
(which in QFT originates from the 
Wick expansion theorem):
in the case of scalar field theories it is a product of two--point functions\footnote{%
For a general graph in a scalar field theory, $G_{j,k} (\x_j,\x_k) = \bigl(\tau(\x_j-\x_k)\bigr)^{\mu_{j,k}}$, where $\mu_{j,k}=0,1,\dots$ and $\tau(\x-\y)$ is the field propagator.}
\beq\label{FAMP}
\mathop{\prod}\limits_{1 \, \leqslant \, j \, < \, k \, \leqslant \, n} G_{j,k} (\x_j,\x_k) \,.
\eeq
In a massless field theory including composite fields of any spin there is a basis of 2-point 
functions defined for non isotropic $\x=\x_j-\x_k$ by 
\beq\label{Gij}
G_{ij}(\x) = \frac{P_{ij}(\x)}{(\x^2)^{\mu_{ij}}}
\,, \quad 
\x^2 = \eta_{\mu \nu} x^\mu x^\nu
\eeq
($\eta$ being the euclidean or the (space-like, i.e., $(D-1,1)$) Minkowski metric, respectively). Here $P_{ij}$ is 
a homogeneous harmonic polynomial, $\mu_{ij} = 0, 1, 2, ...$ in an even dimensional space-time and
$ \mu_{ij} = 0, 1/2, 1, 3/2, ...$ in an odd dimensional space-time.
In a general field theory a Feynman amplitude is a finite sum of expressions of type (\ref{FAMP}) and hence, without loss of generality, we can focus on renormalization of expressions of this form.
Causal factorization for Feynman amplitudes can be drawn
schematically by the following picture \\
\begin{minipage}{\textwidth}
\raisebox{-10pt}{${}$} \\
\ALTERNATIVE{%
${}$ \hfill \includegraphics{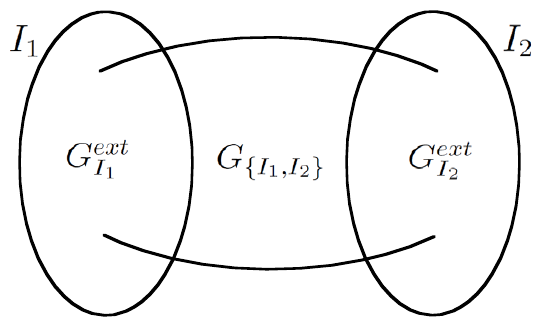}  \hfill ${}$}{%
${}$ \hfill \includegraphics{NSTfig01.eps}  \hfill ${}$}
\\
${}$ \hfill pic. 1 \hfill ${}$
\\
\end{minipage}\\
for every partition $\MISE_1 \disjuni \MISE_2$ of the set of vertices $\MISE=\{1,\dots,n\}$ of the graph in the domain where the vertices from $\MISE_1$ and $\MISE_2$ are 
{\it causally separated} and $G_{\MISE_k}^{ext}$ ($k=1,2$) is the extended (renormalized) Feynman amplitude for the subset $\MISE_k$, while $G_{\{\MISE_1,\MISE_2\}}$ contains all the remaining intermediate two--point functions (propagators) which do not need to be extended when the sub-pieces $\MISE_1$ and $\MISE_2$ are causally separated.
The concrete graphs and corresponding Feynman rules of a given QFT model are not important in this condition.
We address the problem of recursive renormalization based on causal factorization in a systematic fashion in Sect. \ref{TSn1-N}.
It is considered in both the Euclidean and (the technically more complicated) Minkowski case.%
\footnote{%
\label{FOOTNT2}%
The causal factorization condition for renormalized Euclidean Feynman amplitudes was 
sketched in \cite{S82/93} and investigated in more detail in \cite{Ni}.
For Minkowski space it was given in 
\cite{N}
where it was demonstrated that it implies the Epstein-Glaser causal factorization for time--ordered products.}
Causal factorization is combined with a glueing procedure over an open covering of the complement of the total diagonal.  
This open covering is induced by the causal separation which is different in the Euclidean and in the Minkowski space.
We have studied in Appendix \ref{ASe1} the most general combinatorics related to such coverings of the complement of the total diagonal.
The relations between the hierarchy of partitions and of the partial diagonals and causal domains is considered in  
\ref{TSn1.3-N}.
 
Secondly,
each step of the renormalization recursion is completed by an extension of a distribution defined off the full diagonal. 
Since Feynman amplitudes are translation invariant that is, they only depend on coordinate differences, the problem is reduced to the extension of distributions outside the origin of a vector space -- the space of independent coordinate differences.
On the other hand, the extension problem of distributions defined outside the origin has an explicit solution in the case of a massless QFT which gives rise to homogeneous distributions -- studied in Sect. 3.2 of H\"ormander's book \cite{H}.
As noted there the extension of a homogeneous distribution is not (in general) homogeneous.
Section \ref{SECC-3} is based on the observation that H\"ormander's theory naturally closes on the class of associate homogeneous distributions 
(a fact noted previously \cite{HW} but not exploited in a constructive way). 

This paper is organized as follows.
Section \ref{TSn1-N} is devoted to the description of the recursive reduction of the renormalization process to a sequence of extensions of distributions obtained by exploiting the causal factorization property, both in the Euclidean and in the Minkowskian cases.

Anticipating on the prominent role of scaling properties in the concrete construction of the required extensions, Sects. \ref{SECC-3} and  \ref{SectNew4} respectively review the main features of associate homogeneous distributions and of their relevant extensions.
These developments are rather necessary generalizations of H\"ormander's theory of homogeneous distributions.

Section \ref{SEC-new5} is devoted to the solution of the renormalization program formulated in Sect. \ref{TSn1-N}.

A number of technical details which would interrupt the logic of this construction are developed in four Appendices.

Appendices 
A and B
deal with the geometry of causal gluing.
Appendix 
C
goes into some details about the renormalization maps constructed in Sect. \ref{NSEC-4.2NN}.
Appendix 
D 
is devoted to the detailed treatment of nontrivial examples.

\section{Recursive renormalization of Feynman amplitudes}\label{TSn1-N}

Recall that in a general field theory an unrenormalized Feynman amplitude is a finite sum of expressions of form 
\(
\prod G_{j,k} (\x_j,\x_k)
\)
(\ref{FAMP})
and without loss of generality we focus on renormalization of functions of this form.
Such 
expressions are, in particular, defined as distributions for pairwise non-coinciding points.
More detailed properties will be specified later.

\subsection{The recursive procedure in Euclidean space}\label{TSn1.1-N}

Let $\ESPA \cong \R^D$ be a $D$--dimensional Euclidean space.
We start with a collection
\beq\label{EUCSYS}
\bigl\{G_{j,k}(\mXx_j,\mXx_k) \, = \, G_{k,j}(\mXx_k,\mXx_j) \, \bigl( \, = \, G_{j,k}(\mXx_j-\mXx_k) \bigr) \,\bigl|\,  1 \leqslant j < k \leqslant n\bigr\}
\eeq
of (translation invariant) smooth functions on the complement $(\ESPA \times \ESPA)\backslash\MDIA_{j,k}$ of the diagonal $\MDIA_{j,k} = \{\mXx_j=\mXx_k\}$.
The task is to construct a distribution $G^{ext}_{\MISE}$ on $\ESPA^{\MISE}$, where $\MISE=\{1,\dots,n\}$, such that for every nontrivial partition\footnote{%
$\MISE_1 \disjuni \MISE_2 = \MISE$ stands for $\MISE_1 \cup \MISE_2 = \MISE$ and $\MISE_1 \cap \MISE_2 = \emptyset$; ``nontrivial'' means that $\MISE_1 \neq \emptyset \neq \MISE_2$.}
$\MISE_1 \disjuni \MISE_2 = \MISE$ we have the following ``{\it causal factorization property}''  (cf. pic. 1):
\beq\label{teq19}
G^{ext}_{\MISE} 
\hspace{1pt}\Bigl|\raisebox{-7pt}{\hspace{1pt}}_{\ECEL_{\{\MISE_1,\MISE_2\}}} \, = \, G^{ext}_{\MISE_1} \, G^{ext}_{\MISE_2} 
\, G_{\{\MISE_1,\MISE_2\}}
\hspace{1pt}\Bigl|\raisebox{-7pt}{\hspace{1pt}}_{\ECEL_{\{\MISE_1,\MISE_2\}}} 
,\quad
G_{\{\MISE_1,\MISE_2\}} \, = \,
\mathop{\prod}\limits_{\mathop{}\limits^{j_1 \in \MISE_1}_{j_2 \in \MISE_2}}
G_{j_1,j_2} 
(\x_{j_1},\x_{j_2})
\,,
\eeq
where 
\beq\label{MCEL2}
\ECEL_{\{\MISE_1,\MISE_2\}} 
\, = \,
\bigl\{(\x_j)_{j \in \MISE} \in \ESPA^{\MISE} \,\bigl|\, \x_{j_1} \neq \x_{j_2} \text{ for } j_1 \in \MISE_1,\, j_2 \in \MISE_2\bigr\} 
\hspace{1pt} (\, = \, \ECEL_{\{\MISE_2,\MISE_1\}} )
\hspace{1pt}.
\eeq
We note that the right hand side of (\ref{teq19}) contains a well defined product of distributions due to the restriction on the domain (\ref{MCEL2}), where $G_{\{\MISE_1,\MISE_2\}}$ is a smooth function, and hence, is a multiplier.

We shall assume recursively that this problem is already solved for every proper subset $\emptyset \subsetneqq \MJSE \subsetneqq \MISE$ of points.
Namely, we suppose that for every such $\MJSE$  we are given a distribution $G^{ext}_{\MJSE} \in \DP (\ESPA^{\MJSE})$ with the property that for every nontrivial partition $\MJSE_1 \disjuni \MJSE_2 = \MJSE$ 
Eq. (\ref{teq19}) holds with $\MISE$, $\MISE_1$ and $\MISE_2$ replaced by $\MJSE$, $\MJSE_1$ and $\MJSE_2$, respectively.
It is convenient to set
\beq\label{CNVNT}
G^{ext}_{\MKSE} \, = \, 1 \,,
\quad \text{if} \quad |\MKSE| \leqslant 1
\eeq
($G_{j_1,j_2} = G_{j_1,j_2}(\x_{j_1},\x_{j_2})$).

\medskp

Thus, 
the starting point of the renormalization recursion is the two point case, $|\MISE|=2$.
In this case the extension can be performed either with the technique of extending associate homogeneous distributions developed in Sects. \ref{SECC-3} and \ref{SectNew4}, or directly as in Sect. \ref{NSEC-4.2NN}.

\medskp

\begin{Statement}{Theorem}\label{THM-EUCL-REC}
The above recursion hypothesis implies that there exists a unique distribution $G^{0}_{\MISE}$ on the complement $\ESPA^{\MISE} \backslash \MDIA_{\MISE}$ of the total diagonal\,\footnote{%
$\MDIA_{\MISE} = \bigl\{\mXx_{j_1}= \cdots =\mXx_{j_n}\bigr\}$ if $\MISE = \{j_1,\dots,j_n\}$} 
$\MDIA_{\MISE}$ with the property $\Mref{teq19}$, i.e., such that
\beq\label{teq19a}
G^{0}_{\MISE} 
\hspace{1pt}\Bigl|\raisebox{-7pt}{\hspace{1pt}}_{\ECEL_{\{\MISE_1,\MISE_2\}}} \, = \, G^{ext}_{\MISE_1} \, G^{ext}_{\MISE_2} 
\, G_{\{\MISE_1,\MISE_2\}}
\hspace{1pt}\Bigl|\raisebox{-7pt}{\hspace{1pt}}_{\ECEL_{\{\MISE_1,\MISE_2\}}} 
\eeq
for every nontrivial partition $\MISE_1 \disjuni \MISE_2 = \MISE$.
\end{Statement}

\medskp

We 
start the {\it proof} of Theorem \ref{THM-EUCL-REC} by considering first
the uniqueness of~$G^{0}_{\MISE}$. 
It is a consequence of Eq. (\ref{teq19a}) and of the fact that the domains 
$\ECEL_{\{\MISE_1,\MISE_2\}}$ cover $\ESPA^{\MISE} \backslash \MDIA_{\MISE}$
(a distribution is determined by its restrictions on a covering, cf. \cite[Theorem 2.2.4]{H}).
Here we used the following geometric fact known as
``Euclidean Diagonal Lemma''
(for a generalization of this lemma see 
\ref{ASe1}):

\medskp

\begin{Statement}{Lemma}\label{EucDiaLem}
The family  
$$
\bigl\{\ECEL_{\{\MISE_1,\MISE_2\}}\,\bigl|\,\MISE_1 \disjuni \MISE_2 = \MISE \text{ is a nontrivial partition}\bigr\}
$$ 
covers $\ESPA^{\MISE} \backslash \MDIA_{\MISE}$.
\end{Statement}

\medskp

\noindent
{\it Proof.}
Let $(\mXx_1,\dots,\mXx_n) \in \ESPA^{\MISE} \backslash \MDIA_{\MISE}$.
Then there are at least two different points, $\mXx_{j_1} \neq \mXx_{j_2}$.
We define $\MISE_1$ to be the set of indices $j$ in $\MISE$ such that $\mXx_j=\mXx_{j_1}$
and $\MISE_2 = \MISE\backslash \MISE_1$.
Then the partition $\MISE = \MISE_1 \dot{\cup} \MISE_2$ is proper and $(\mXx_1,\dots,\mXx_n)$ $\in$ $\ECEL_{\{\MISE_1,\MISE_2\}}$.$\quad\Box$

\medskp

Continuing with the proof of Theorem \ref{THM-EUCL-REC}, the existence is implied again by Theorem 2.2.4 of \cite{H} and Eq. (\ref{teq19a}) but combined now with the glueing property:

\medskp

\begin{Statement}{Lemma}\label{EucGlue}
For every two nontrivial partitions $\MISE_1 \disjuni \MISE_2 = \MISE$ and $\MKSE_1 \disjuni \MKSE_2 = \MISE$ we have
\beq\label{GL-eq1}
G^{ext}_{\MISE_1} \, G^{ext}_{\MISE_2} 
\, G_{\{\MISE_1,\MISE_2\}}
\hspace{1pt}\Bigl|\raisebox{-7pt}{\hspace{1pt}}_{\ECEL_{\{\MISE_1,\MISE_2\}} \cap \, \ECEL_{\{\MKSE_1,\MKSE_2\}}}
= \,
G^{ext}_{\MKSE_1} \, G^{ext}_{\MKSE_2} 
\, G_{\{\MKSE_1,\MKSE_2\}}
\hspace{1pt}\Bigl|\raisebox{-7pt}{\hspace{1pt}}_{\ECEL_{\{\MISE_1,\MISE_2\}} \cap \, \ECEL_{\{\MKSE_1,\MKSE_2\}}} .
\eeq
\end{Statement}

\medskp

\noindent
{\it Proof.} 
Let us introduce the sets $\MJSE_{a,b} := \MISE_a \cap \MKSE_b$ some of which can be empty.
They form a partition of $\MISE$ $=$ $\MJSE_{1,1}$ $\hspace{-1pt}\dot{\cup}\hspace{-1pt}$ $\MJSE_{1,2}$ $\hspace{-1pt}\dot{\cup}\hspace{-1pt}$ $\MJSE_{2,1}$ $\hspace{-1pt}\dot{\cup}\hspace{-1pt}$ $\MJSE_{2,2}$ and $\MISE_a$ $=$ $\MJSE_{a,1}$ $\hspace{-1pt}\dot{\cup}\hspace{-1pt}$ $\MJSE_{a,2}$, $\MKSE_b$ $=$ $\MJSE_{1,b}$ $\hspace{-1pt}\dot{\cup}\hspace{-1pt}$ $\MJSE_{2,b}$. 
By the recursively assumed condition (\ref{teq19}) it follows that
\beqs
G^{ext}_{\MISE_a} \, = \podr
G^{ext}_{\MJSE_{a,1}} \, G^{ext}_{\MJSE_{a,2}} 
\, G_{\{\MJSE_{a,1},\MJSE_{a,2}\}} 
\hspace{1pt}\Bigl|\raisebox{-7pt}{\hspace{1pt}}_{\ECEL_{\{\MJSE_{a,1},\MJSE_{a,2}\}}}
\,, \quad
\\
G^{ext}_{\MKSE_b} \, = \podr
G^{ext}_{\MJSE_{1,b}} \, G^{ext}_{\MJSE_{2,b}} 
\, G_{\{\MJSE_{1,b},\MJSE_{2,b}\}} 
\hspace{1pt}\Bigl|\raisebox{-7pt}{\hspace{1pt}}_{\ECEL_{\{\MJSE_{1,b},\MJSE_{2,b}\}}}
\,,
\eeqs
where in addition to the convention (\ref{CNVNT}) we set
$G_{\MJSE,\MKSE} = 1$ if $\MJSE=\emptyset$ or $\MKSE=\emptyset$.
Now Eq. (\ref{GL-eq1}) follows by the above identities together with the following ones
\beqs
\podr \hspace{-15pt}
\ECEL_{\{\MISE_1,\MISE_2\}} \cap \, \ECEL_{\{\MKSE_1,\MKSE_2\}}
\, = \,
\ECEL_{\{\MJSE_{1,1},\MJSE_{1,2}\}} \cap \, \ECEL_{\{\MJSE_{2,1},\MJSE_{2,2}\}}
\cap
\ECEL_{\{\MJSE_{1,1},\MJSE_{2,1}\}} \cap \, \ECEL_{\{\MJSE_{1,2},\MJSE_{2,2}\}}
\,,
\\ \podr
G_{\{\MJSE_{1,1},\MJSE_{1,2}\}} \, G_{\{\MJSE_{2,1},\MJSE_{2,2}\}} \,
G_{\{\MISE_1,\MISE_2\}}
\\ \podr
= \,
G_{\{\MJSE_{1,1},\MJSE_{2,1}\}} \, G_{\{\MJSE_{1,2},\MJSE_{2,2}\}} \,
G_{\{\MKSE_1,\MKSE_2\}}
\\ \podr
= G_{\{\MJSE_{1,1},\MJSE_{1,2}\}}
G_{\{\MJSE_{1,1},\MJSE_{2,1}\}}
G_{\{\MJSE_{1,1},\MJSE_{2,2}\}}
G_{\{\MJSE_{1,2},\MJSE_{2,1}\}}
G_{\{\MJSE_{1,2},\MJSE_{2,2}\}}
G_{\{\MJSE_{2,1},\MJSE_{2,2}\}}
\\ \podr
=: \,
G_{\{\MJSE_{1,1},\MJSE_{1,2},\MJSE_{2,1},\MJSE_{2,2}\}}
\qquad \text{on} \qquad \ECEL_{\{\MISE_1,\MISE_2\}} \cap \, \ECEL_{\{\MKSE_1,\MKSE_2\}}
\eeqs
since substituting all above expressions in both sides of Eq. (\ref{GL-eq1}) gives the same result
$$
G^{ext}_{\MJSE_{1,1}} \, G^{ext}_{\MJSE_{1,2}} \, G^{ext}_{\MJSE_{2,1}} \, G^{ext}_{\MJSE_{2,2}} 
\, G_{\{\MJSE_{1,1},\MJSE_{1,2},\MJSE_{2,1},\MJSE_{2,2}\}}
$$
on the domain $\ECEL_{\{\MISE_1,\MISE_2\}} \cap \, \ECEL_{\{\MKSE_1,\MKSE_2\}}$.$\quad\Box$

\medskp

The 
above lemma completes the proof of Theorem \ref{THM-EUCL-REC}.

\medskp

In addition we have a general result
which allows to preserve the symmetries throughout
the construction of $G^{0}_{\MISE}$.
To this end let us consider a vector field $\MDFO_{(\x)}$ in $\x$, i.e., a homogeneous first order partial differential operator (i.e., a derivation) 
and let $d_{j,k}=d_{k,j}$ be (real) 
numbers given for every pair $j,k \in \MISE$ of different $j \neq k$.
We set for every index set~$\MISE$
\beq\label{MDFO}
\MDFO_{\MISE} \, := \, \mathop{\sum}\limits_{j \, \in \, \MISE} \MDFO_{(\x_j)}
\,,\qquad
d_{\MISE} \, := \, \frac{1}{2}\mathop{\sum}\limits_{\mathop{}\limits^{j,k \, \in \, \MISE}_{j \, \neq \, k}} d_{j,k} \, = \, \mathop{\sum}\limits_{\mathop{}\limits^{j,k \, \in \, \MISE}_{j \, < \, k}} d_{j,k}  \,.
\eeq

\medskp

\begin{Statement}{Lemma}\label{EucSym}
Under the assumptions of Theorem $\ref{THM-EUCL-REC}$ let the system $(\ref{EUCSYS})$ satisfies the condition
\beq\label{MDFOCND1}
(\MDFO_{(\mXx_j)}+\MDFO_{(\mXx_k)} - d_{j,k}) \, G_{j,k} (\mXx_j,\mXx_k) \vrestr{12pt}{\ESPA^{\times 2} \backslash \MDIA_{j,k}} \, = \, 0
\eeq
$(j<k)$.
Furthermore, let
for every $\emptyset \subsetneqq\MJSE \subsetneqq \MISE$ the distribution $G^{ext}_{\MJSE}$ satisfies the equation
\beq\label{MDFOCND2}
\bigl(\MDFO_{\MJSE} - d_{\MJSE}\bigr)^{\mu_{\MJse}} G^{ext}_{\MJSE} \, = \, 0
\eeq
for some positive integer $\mu_{\MJSe}$. Then the distribution $G^{0}_{\MISE}$ satisfies the equation
\beq\label{MDFOCND3}
\bigl(\MDFO_{\MISE} - d_{\MISE}\bigr)^{\mu_{\MIse}} G^{0}_{\MISE} \, = \, 0
\eeq
on $\ESPA^{\MISE}\backslash\MDIA_{\MISE}$ with
\beq\label{MUEST}
\mu_{\MISe} \, = \, \mathop{\max}\limits_{\mathop{}\limits^{\MISE \, = \, \MISE_1 \, \dot{\cup} \, \MISE_2}_{ \MISE_1,\, \MISE_2 \, \neq \, \emptyset}} (\mu_{\MISe_1} + \mu_{\MISe_2}-1) \,.
\eeq
\end{Statement}

\medskp

\noindent
{\it Proof.}
Let us apply to both sides of Eq.~(\ref{teq19a}) the operator $\bigl(\MDFO_{\MISE} - d_{\MISE}\bigr)^{\mu_{\MIse}}$ and use the combined Leibniz rule and Newton binomial formula
$$
\bigl(\MDFO + a+b\bigr)^n \bigl(FG\bigr) \, = \,
\mathop{\sum}\limits_{k \, = \, 0}^n
{n \choose k}
\Bigl(\bigl(\MDFO + a \bigr)^{n-k}F\Bigr) 
\Bigl(\bigl(\MDFO + b \bigr)^k G\Bigr)  \,
$$
valid for any derivation $\MDFO$.
The result is
\beqa\label{PrfCalc1}
\podr
\bigl(\MDFO_{\MISE} - d_{\MISE}\bigr)^{\mu_{\MIse}}
G^{0}_{\MISE} 
\hspace{1pt}\Bigl|\raisebox{-7pt}{\hspace{1pt}}_{\ECEL_{\{\MISE_1,\MISE_2\}}} 
\, 
\nnb \podr = \, 
\mathop{\sum}\limits_{\gamma \, = \, 0}^{\mu_{\MIse}}
{\mu_{\MISE} \choose \gamma}
\Bigl(\bigl(\MDFO_{\MISE} - d_{\MISE_1} -d_{\MISE_2}\bigr)^{\mu_{\MIse}-\gamma}G^{ext}_{\MISE_1}G^{ext}_{\MISE_2}\Bigr) 
\nnb \podr \times \, 
\Bigl(\bigl(\MDFO_{\MISE} -d_{\{\MISE_1,\MISE_2\}}\bigr)^{\gamma}G_{\{\MISE_1,\MISE_2\}}\Bigr)
\hspace{1pt}\Bigl|\raisebox{-7pt}{\hspace{1pt}}_{\ECEL_{\{\MISE_1,\MISE_2\}}}
\nnb \podr = \, 
\mathop{\sum}\limits_{\alpha+\beta+\gamma = \mu_{\MIse}}
\frac{\mu_{\MISe}!}{\alpha!\beta!\gamma!} \,
\Bigl(\bigl(\MDFO_{\MISE_1} - d_{\MISE_1}\bigr)^{\alpha}G^{ext}_{\MISE_1}\Bigr) 
\Bigl(\bigl(\MDFO_{\MISE_2} - d_{\MISE_2}\bigr)^{\beta} G^{ext}_{\MISE_2} \Bigr)
\, 
\nnb \podr \times \, 
\Bigl(\bigl(\MDFO_{\MISE} - d_{\{\MISE_1,\MISE_2\}}\bigr)^{\gamma}G_{\{\MISE_1,\MISE_2\}}\Bigr)
\hspace{1pt}\Bigl|\raisebox{-7pt}{\hspace{1pt}}_{\ECEL_{\{\MISE_1,\MISE_2\}}}
\,,
\eeqa
where
$$
d_{\{\MISE_1,\MISE_2\}} \,
 := \,
\mathop{\sum}\limits_{\mathop{}\limits^{j_1 \in \MISE_1}_{j_2 \in \MISE_2}}
d_{j_1,j_2} \, = \, d_{\MISE_1}+d_{\MISE_2}-d_{\MISE} 
\,.
$$
By Eq. (\ref{MDFOCND1}) it follows that if $\gamma > 0$ then the right hand side of (\ref{PrfCalc1}) is zero and similarly, it is zero if either $\alpha \geqslant \mu_{\MISe_1}$ or $\beta \geqslant \mu_{\MISe_2}$ by (\ref{MDFOCND2}).
Thus, Eq.~(\ref{MUEST}) ensures that 
one of these conditions will necessarily hold
and hence, 
$\bigl(\MDFO_{\MISE} - d_{\MISE}\bigr)^{\mu_{\MIse}} G^{0}_{\MISE} = 0$ on every
$\ECEL_{\{\MISE_1,\MISE_2\}}$. Thus the lemma follows by Lemma~\ref{EucDiaLem}.$\quad\Box$

\medskp

\begin{Statement}{Corollary}\label{CrXX1}
In the case when $\MDFO_{(\x)} = \di_{x^{\mu}}$ and all $d_{j,k}=0$, $\mu_{\MJSe}=1$ we conclude that if all $G^{ext}_{\MJSE}$ are translation invariant distributions then so is $G^{0}_{\MISE}$.
\end{Statement}


\begin{remark}\label{RnXXX1}
$(a)$
The case of the Euler vector field $\MDFO_{(\x)}$ $=$
$\MEUL_{(\mXx)}$ $:=$ $\mXx \cdot \frac{\partial}{\partial \mXx} \equiv
\mathop{\sum}\limits_{\mu \, = \, 1}^D x^{\mu} \frac{\partial}{\partial x^{\mu}}$
corresponds to associate homogeneous distributions and will be of central interest in the next section.

$(b)$
One can extend Lemma \ref{EucSym} to the case when the differential operators  (\ref{MDFO}) act on tensor valued (multicomponent) distributions $G^{ext}_{\MJSE}$ and the numbers $d_{j,k}$ are replaced by matrices coming from Lie algebraic representation while $\mu_{\MJSe} = 1$ for all $\MJSE$.
This is the case of rotation symmetry; we deduce as a consequence that if the distributions $G^{ext}_{\MJSE}$ are Euclidean invariant then so is~$G^{0}_{\MISE}$.
\end{remark}

\subsection{The recursive procedure in Minkowski space}\label{TSn1.2-N}

Consider now Minkowski space $\MSPA \cong \R^{D-1,1}$.
We start with a system of 
(translation invariant) distributions $W_{j,k} (\x_1,\x_2)$ $=$ $(W_{j,k}(\x_j-\x_k))$ on $\MSPA \times \MSPA$ defined for every pair of different indices $j,k =1, \dots, n$.
We now  have a weaker symmetry
\beq\label{MinSymCnd}
W_{j,k} (\x_j,\x_k) \, = \, 
W_{k,j} (\x_k,\x_j) \quad \text{for space--like separated} \ (\x_j,\,\x_k) 
\,
.
\eeq
Furthermore, 
for every $j \neq k$ we assume that
the distribution
$W_{j,k} (\x_j,\x_k)$ 
is the
boundary value
of an
analytic function on 
the backward tube
\beq\label{BACKTUBE}
\bigl\{(\x_j+i\y_j,\x_k+i\y_k) \, \bigl| \, \y_k-\y_j \in -V_+\bigr\} 
\eeq
($V_+$ being the open future light cone in $\MSPA$).
From now on we shall denote for short 
$$
W_{j,k} \, := \, W_{j,k} (\x_j,\x_k) \,.
$$


\begin{remark}\label{NREM-1n}
The symmetry condition (\ref{MinSymCnd}) (as well as its Euclidean version in (\ref{EUCSYS})) should not be mixed with the {\it Bose--Fermi statistics symmetry}.
In our notation this will be an additional symmetry that is expressed by flipping the arguments but not the indices:
\beq\label{BosFer}
W_{j,k} (\x_j,\x_k) \, = \, 
(-1)^{\epsilon_{j,k}} \,
W_{j,k} (\x_k,\x_j) \quad \text{for space--like separated} \ (\x_j,\,\x_k)
\,, 
\eeq
where the sign factor $(-1)^{\epsilon_{j,k}}$ reflects the Bose--Fermi statistics.
\end{remark}


\begin{example}\label{EXM-1nn}
In massless QFT models we generally have
$$
W_{j,k} (\x_j,\x_k) 
\, = \, 
\frac{P_{j,k}(\x_j,\x_k)}{\bigl((\x_j-\x_k)^2+i0(x_j^0-x_k^0)\bigr)^{\nu_{j,k}}} \,,
$$
where $P_{j,k}(\x_j,\x_k) = P_{j,k}(\x_j-\x_k) = P_{k,j}(\x_k,\x_j)$ are homogeneous polynomials and $\nu_{j,k}=\nu_{k,j}$ are nonnegative indices (for $j \neq k$).%
\end{example}


Then the factorization property (\ref{teq19}) is modified as follows:
\beq\label{teq19M}
G^{ext}_{\MISE} 
\hspace{1pt}\Bigl|\raisebox{-7pt}{\hspace{1pt}}_{\MCEL_{(\MISE_1,\MISE_2)}} \, = \, G^{ext}_{\MISE_1} \, G^{ext}_{\MISE_2} 
\, W_{(\MISE_1,\MISE_2)}
\hspace{1pt}\Bigl|\raisebox{-7pt}{\hspace{1pt}}_{\MCEL_{(\MISE_1,\MISE_2)}} 
,\quad
W_{(\MISE_1,\MISE_2)} \, = \,
\mathop{\prod}\limits_{\mathop{}\limits^{j_1 \in \MISE_1}_{j_2 \in \MISE_2}}
W_{j_1,j_2} 
\,,
\eeq
for every nontrivial partition $\MISE_1 \disjuni \MISE_2 = \MISE$, where the {\it Minkowski causal domains} are defined by 
\beq\label{MCEL2M}
\MCEL_{(\MISE_1,\MISE_2)} 
\, = \,
\bigl\{(\x_j)_{j \in \MISE} \in \MSPA^{\MISE} \,\bigl|\, \x_{j_1} \MCAU \x_{j_2} \text{ for } j_a \in \MISE_a \ (a=1,2)\bigr\} 
\hspace{1pt} (\, \neq \, \MCEL_{(\MISE_2,\MISE_1)} )
\hspace{1pt},
\eeq
$\x\MCAU\y$ standing for $\x \notin \y+\overline{V}_+$ (the closed future cone with a tip~$\y$).
Note that the domains $\MCEL_{(\MISE_1,\MISE_2)}$ depend on an additional order in the partition; that is why we use an ``ordered pair'' notation.


\begin{remark}\label{NotRem}
Sometimes 
the following notations are convenient. Let $\xx_{\MJSE}$ be a collection of points $(\x_j)_{j \,\in\, \MJSE}$ indexed in the set $\MJSE$. 
and write $\xx_{\MJSE} \MCAU \xx_{\MKSE}$ iff $\x_j \MCAU \x_k$ for all $j \in \MJSE$ and all $k \in \MKSE$.
Then (\ref{MCEL2M}) reads
\beq\label{MCEL2M2}
\MCEL_{(\MISE_1,\MISE_2)} 
\, = \,
\bigl\{\xx_{\MISE} \,\bigl|\, \xx_{\MISE_1} \MCAU \xx_{\MISE_2}\bigr\}
\hspace{1pt}.
\eeq
More generally,
for a partition $\MISE=\MISE_1 \dot{\cup} \cdots \dot{\cup} \MISE_n$ we introduce domains
\beq\label{MCEL2Mn}
\MCEL_{(\MISE_1,\dots,\MISE_n)} 
\, = \,
\bigl\{\xx_{\MISE} \,\bigl|\, \xx_{\MISE_a} \MCAU \xx_{\MISE_b} \text{ for all } 1 \leqslant a < b \leqslant  n\bigr\}
\hspace{1pt}.
\eeq
The condition in the right hand side of Eq.~(\ref{MCEL2Mn}) may be also written as $\xx_{\MISE_1} \MCAU \cdots \MCAU \xx_{\MISE_n}$ but one should keep in mind that $\MCAU$ is not a transitive relation.
\end{remark}


The problem of existence of the product of distributions in (\ref{teq19M}) is more involved than in Euclidean space.
In contrast to (\ref{teq19}) the product (\ref{teq19M}) now globally exists as it is shown in 
\cite{N}.
For the sake of completeness we repeat the proof here.

\medskp

\begin{Statement}{Lemma}\label{ExProd0}
Let $\MISE = \{1,\dots,n\} = \MISE_1 \dot{\cup} \cdots \dot{\cup} \MISE_k$ be an arbitrary non trivial partition $($with nonempty $\MISE_j$$)$ and let $G_{\MISE_j}$ be an arbitrary {\rm translation invariant} distribution on $\MSPA^{\MISE_j}$ for $j=1,\dots,k$. 
Let $W_{\MISE}$ be a distribution on $\MSPA^{\times n}$ that is a boundary value of an analytic function in the backward tube\footnote{%
The backward tube in the space of $n$ complex variables $\z_j = \x_j+i\y_j$ ($j=1,\dots,n$) is defined by the condition $\y_j-\y_{j+1} \in -V_+$ ($j=1,\dots,n-1$).
More generally, in what follows we shall deal also with backward tubes in $(\MSPA+i\MSPA)^{\MISE}$ for any index set $\MISE$ endowed with an arbitrary linear order
$\sigma:\MISE \cong \{1,\dots,n\}$. 
In this case the backward tube is defined just by a relabeling the arguments according to $\sigma$.} 
in $(\MSPA+i\MSPA)^{\times n}$.
Then the product of distributions 
$$
\Biggl(\mathop{\prod}\limits_{j \, = \, 1}^k G_{\MISE_j} \Biggr) \, W_{\MISE}
$$
exists on $\MSPA^{\times n}$.
\end{Statement}

\medskp

\noindent
{\it Proof.}
Clearly, the product $\mathop{\prod}\limits_{j \, = \, 1}^k G_{\MISE_j}$ exists on $\mathop{\prod}\limits_{j \, = \, 1}^k \MSPA^{\MISE_j} = \MSPA^{\times n}$ as a tensor product of distributions.
To establish the existence of the product of $W_{\MISE}$ with $\mathop{\prod}\limits_{j \, = \, 1}^k G_{\MISE_j}$ we shall use the wave front 
criterion 
\cite[Theorem 8.2.10]{H}.

Without loss of generality we may assume that the subsets $\MISE_k$ are ordered with respect to the order of their minimal elements:
$$
\min \, \MISE_1 \, < \, \cdots \, < \, \min \, \MISE_k \,.
$$
Let us introduce new coordinates 
$$
(\y_1,\zz_1,\dots,\y_k,\zz_k)
$$
on $\MSPA^{\times n}$ such that $\y_j = \x_{\min \, \MISE_j}$ and $\zz_j$ is a basis of (vector) differences in $\MSPA^{\MISE_j}$ (i.e., $\zz_j$ are relative coordinates in $\MSPA^{\MISE_j}$).
In particular, $(\y_j,\zz_j)$ are coordinates on $\MSPA^{\MISE_j}$ for $j=1,\dots,k$.

The wave front sets of distributions on $\MSPA^{\times n}$ are subbundles of the \textit{co-tangent bundle} $T^* \bigl(\MSPA^{\times n}\bigr)$.
The latter is isomorphic to $\MSPA^{\times n} \times  \MSPA^{\times n}$ and the coordinates $(\y_1,\zz_1,\dots,\y_k,\zz_k)$ in the base $\MSPA^{\times n}$ induce coordinates in the fibers $\cong \MSPA^{\times n}$, which we shall denote by $(\p_1,\vq_1,\dots,\p_k,\vq_k)$  (i.e., $\p_j$ is the momentum corresponding to $\y_j$ and $\vq_j$ is the collection of momenta corresponding to the collection of relative positions $\zz_j$).

In the above new coordinates the wave front set of the product $\mathop{\prod}\limits_{j \, = \, 1}^k G_{\MISE_j}$ is contained in the sub-bundle 
$$
\MSPA^{\times n} \times
\bigl\{(\p_1,\vq_1,\dots,\p_k,\vq_k) \, \bigl| \,\p_1 \, = \, \cdots \, = \, \p_k = 0\bigr\}
\, \subset \, \MSPA^{\times n} \times \MSPA^{\times n} \, \cong \, T^* \bigl(\MSPA^{\times n}\bigr)
$$ 
of the co-tangent bundle over $\MSPA^{\times n}$
since each of the distributions $G_{\MISE_j}$ is translation invariant
and hence, its wave front set over $\MSPA^{\MISE_j}$ is contained in the sub-bundle $\MSPA^{\MISE_j} \times \{(\p_j,\vq_j) \,|\, \p_j=0\}$.
On the other hand, the wave front set of $W_{\MISE}$ is contained in the tube
$$
\MSPA^{\times n} \times \bigl\{(\p_1,\vq_1,\dots,\p_k,\vq_k) \, \bigl| \,\p_2 - \p_1 ,\, \dots,\, \p_k - \p_{k-1} \, \in \, V_+ \bigr\} \,
$$
since it is a boundary value distribution. 
Hence, the sum of the wave front sets of the distributions $\mathop{\prod}\limits_{j \, = \, 1}^k G_{\MISE_j}$ and $W_{\MISE}$ cannot contain zero tangent vectors and thus, according to 
Theorem 8.2.10 of \cite{H}
their product exists.$\quad\Box$

\medskp

Since Lemma \ref{ExProd0} is valid for any permutation of the forward tube $(\MSPA+i\MSPA)^{\times n}$ we obtain as a consequence that

\medskp

\begin{Statement}{Corollary}\label{ExProd}
The product of distributions $\Mref{teq19M}$ exists.
\end{Statement}

\medskp

\begin{Statement}{Theorem}\label{THM-MINK-REC}
Assume that there is a distribution $G^{ext}_{\MJSE}$ for every proper subset $\emptyset \subsetneqq \MJSE \subsetneqq \MISE$ such that for every nontrivial partition $\MJSE_1 \disjuni \MJSE_2 = \MJSE$ the causal factorization property holds: 
\beq\label{teq18M}
G^{ext}_{\MJSE} 
\hspace{1pt}\Bigl|\raisebox{-7pt}{\hspace{1pt}}_{\MCEL_{(\MJSE_1,\MJSE_2)}} \, = \, G^{ext}_{\MJSE_1} \, G^{ext}_{\MJSE_2} 
\, W_{(\MJSE_1,\MJSE_2)} 
\hspace{1pt}\Bigl|\raisebox{-7pt}{\hspace{1pt}}_{\MCEL_{(\MJSE_1,\MJSE_2)}}
,\quad
W_{(\MJSE_1,\MJSE_2)} \,:= \,
\mathop{\prod}\limits_{\mathop{}\limits^{j_1 \in \MJSE_1}_{j_2 \in \MJSE_2}}
W_{j_1,j_2}
\,
\eeq
$($setting again $G^{ext}_{\MJSE} = 1$ if $|\MJSE|=1$$)$.
Then there exists a unique distribution $G^{0}_{\MISE}$ on $\MSPA^{\MISE} \backslash \MDIA_{\MISE}$ with the property \Mref{teq19M}, i.e., such that
\beq\label{teq19Ma}
G^{0}_{\MISE} 
\hspace{1pt}\Bigl|\raisebox{-7pt}{\hspace{1pt}}_{\MCEL_{(\MISE_1,\MISE_2)}} \, = \, G^{ext}_{\MISE_1} \, G^{ext}_{\MISE_2} 
\, W_{(\MISE_1,\MISE_2)}
\hspace{1pt}\Bigl|\raisebox{-7pt}{\hspace{1pt}}_{\MCEL_{(\MISE_1,\MISE_2)}} 
\eeq
for every nontrivial partition $\MISE_1 \disjuni \MISE_2 = \MISE$.
\end{Statement}

\medskp

The {\it proof} of Theorem \ref{THM-MINK-REC}
is analogous to the proof of Theorem \ref{THM-EUCL-REC}.
The uniqueness of $G^{0}_{\MISE}$ according to \cite[Theorem 2.2.4]{H} is a consequence of Eq. (\ref{teq19Ma}) and the  ``Minkowski Diagonal Lemma''
(in 
\ref{ASe1} 
we give a ``maximal'' generalization of this statement):

\medskp

\begin{Statement}{Lemma}\label{MinDiaLem}
The family $$\bigl\{\MCEL_{(\MISE_1,\MISE_2)}\,\bigl|\,\MISE_1 \disjuni \MISE_2 = \MISE \text{ is a nontrivial partition}\bigr\}$$ covers $\MSPA^{\MISE} \backslash \MDIA_{\MISE}$.
\end{Statement}

\medskp

\noindent
{\it Proof.}
Let $(\x_1,\dots,\x_n) \in \MSPA^{\MISE} \backslash \MDIA_{\MISE}$.
Then there are at least two different points, $\x_{j_1} \neq \x_{j_2}$.
Given such a pair, there exists a Lorentz frame in which their time coordinates are different, say $x_{j_1}^0 < x_{j_2}^0$.
Let $\MISE_1$ be the set of indices $j \in \MISE$ such that $x_j^0 \leqslant x_{j_1}^0$ and let $\MISE_2 = \MISE\backslash \MISE_1$.
Then $\MISE_1,\MISE_2 \neq \emptyset$ and $(\x_1,\dots,\x_n)\in \MCEL_{(\MISE_1,\MISE_2)}$.$\quad\Box$

\medskp
 
To complete the proof of Theorem \ref{THM-MINK-REC} it remains to check the glueing property
\beq\label{GL-eq1M}
G^{ext}_{\MISE_1} \, G^{ext}_{\MISE_2} 
\, W_{(\MISE_1,\MISE_2)}
\hspace{1pt}\Bigl|\raisebox{-7pt}{\hspace{1pt}}_{\MCEL_{(\MISE_1,\MISE_2)} \cap \, \MCEL_{(\MKSE_1,\MKSE_2)}}
= \,
G^{ext}_{\MKSE_1} \, G^{ext}_{\MKSE_2} 
\, W_{(\MKSE_1,\MKSE_2)}
\hspace{1pt}\Bigl|\raisebox{-7pt}{\hspace{1pt}}_{\MCEL_{(\MISE_1,\MISE_2)} \cap \, \MCEL_{(\MKSE_1,\MKSE_2)}} 
\eeq
for every two nontrivial partitions $\MISE_1 \disjuni \MISE_2 = \MISE$ and $\MKSE_1 \disjuni \MKSE_2 = \MISE$.
To this end we introduce, as in the proof of Lemma \ref{EucGlue},
the sets $\MJSE_{a,b} := \MISE_a \cap \MKSE_b$ (some of which can be empty) such that $\MISE$ $=$ $\MJSE_{1,1}$ $\hspace{-1pt}\dot{\cup}\hspace{-1pt}$ $\MJSE_{1,2}$ $\hspace{-1pt}\dot{\cup}\hspace{-1pt}$ $\MJSE_{2,1}$ $\hspace{-1pt}\dot{\cup}\hspace{-1pt}$ $\MJSE_{2,2}$ and $\MISE_a$ $=$ $\MJSE_{a,1}$ $\hspace{-1pt}\dot{\cup}\hspace{-1pt}$ $\MJSE_{a,2}$, $\MKSE_b$ $=$ $\MJSE_{1,b}$ $\hspace{-1pt}\dot{\cup}\hspace{-1pt}$ $\MJSE_{2,b}$. 
A key point is that we have similarly to the Euclidean case the property:
$$
\MCEL_{(\MISE_1,\MISE_2)} \cap \, \MCEL_{(\MKSE_1,\MKSE_2)}
\, = \,
\MCEL_{(\MJSE_{1,1},\MJSE_{1,2})} \cap \, \MCEL_{(\MJSE_{2,1},\MJSE_{2,2})}
\cap
\MCEL_{(\MJSE_{1,1},\MJSE_{2,1})} \cap \, \MCEL_{(\MJSE_{1,2},\MJSE_{2,2})}
\,,
$$
Then by the recursively assumed condition (\ref{teq19M}) it follows that
\beqs
G^{ext}_{\MISE_a} \, = \podr
G^{ext}_{\MJSE_{a,1}} \, G^{ext}_{\MJSE_{a,2}} 
\, W_{(\MJSE_{a,1},\MJSE_{a,2})} 
\hspace{1pt}\Bigl|\raisebox{-7pt}{\hspace{1pt}}_{\MCEL_{(\MJSE_{a,1},\MJSE_{a,2})}}
\,, \quad
\\
G^{ext}_{\MKSE_b} \, = \podr
G^{ext}_{\MJSE_{1,b}} \, G^{ext}_{\MJSE_{2,b}} 
\, W_{(\MJSE_{1,b},\MJSE_{2,b})} 
\hspace{1pt}\Bigl|\raisebox{-7pt}{\hspace{1pt}}_{\MCEL_{(\MJSE_{1,b},\MJSE_{2,b})}}
\,,
\eeqs
($G_{\MJSE,\MKSE} := 1$ if $\MJSE=\emptyset$ or $\MKSE=\emptyset$).
Thus, the left and the right hand sides of Eq. (\ref{GL-eq1M}) read, respectively:
\beqa\label{LHS-TEMP}
\podr
G^{ext}_{\MISE_1} \, G^{ext}_{\MISE_2} 
\, W_{(\MISE_1,\MISE_2)}
\hspace{1pt}\Bigl|\raisebox{-7pt}{\hspace{1pt}}_{\MCEL_{(\MISE_1,\MISE_2)} \cap \, \MCEL_{(\MKSE_1,\MKSE_2)}}
\nnb
\podr
= \,
G^{ext}_{\MJSE_{1,1}} \, G^{ext}_{\MJSE_{1,2}} 
\, W_{(\MJSE_{1,1},\MJSE_{1,2})} \, 
G^{ext}_{\MJSE_{2,1}} \, G^{ext}_{\MJSE_{2,2}} 
\, W_{(\MJSE_{2,1},\MJSE_{2,2})} 
\, W_{(\MISE_1,\MISE_2)} \,,
\\ \label{RHS-TEMP}
\podr
G^{ext}_{\MKSE_1} \, G^{ext}_{\MKSE_2} 
\, W_{(\MKSE_1,\MKSE_2)}
\hspace{1pt}\Bigl|\raisebox{-7pt}{\hspace{1pt}}_{\MCEL_{(\MISE_1,\MISE_2)} \cap \, \MCEL_{(\MKSE_1,\MKSE_2)}}
\nnb
\podr
= \,
G^{ext}_{\MJSE_{1,1}} \, G^{ext}_{\MJSE_{2,1}} 
\, W_{(\MJSE_{1,1},\MJSE_{2,1})} \,
G^{ext}_{\MJSE_{1,2}} \, G^{ext}_{\MJSE_{2,2}} 
\, W_{(\MJSE_{1,2},\MJSE_{2,2})}
\, W_{(\MKSE_1,\MKSE_2)} \,.
\eeqa
Now Eq. (\ref{GL-eq1M}) follows since
\beqa
\podr
W_{(\MJSE_{1,1},\MJSE_{1,2})} \, 
W_{(\MJSE_{2,1},\MJSE_{2,2})} \, 
W_{(\MISE_1,\MISE_2)}
\nnb
\podr
= \, 
W_{(\MJSE_{1,1},\MJSE_{1,2})} \,
W_{(\MJSE_{2,1},\MJSE_{2,2})} \,
W_{(\MJSE_{1,1},\MJSE_{2,1})} \,
W_{(\MJSE_{1,2},\MJSE_{2,2})} \,
W_{(\MJSE_{1,1},\MJSE_{2,2})} \,
W_{(\MJSE_{1,2},\MJSE_{2,1})}
\nnb
\podr
= \, 
W_{(\MJSE_{1,1},\MJSE_{2,1})} \,
W_{(\MJSE_{1,2},\MJSE_{2,2})} \,
W_{(\MJSE_{1,1},\MJSE_{1,2})} \,
W_{(\MJSE_{2,1},\MJSE_{2,2})} \,
W_{(\MJSE_{1,1},\MJSE_{2,2})} \,
W_{(\MJSE_{2,1},\MJSE_{1,2})}
\nnbnn
\podr
= \, 
W_{(\MJSE_{1,1},\MJSE_{2,1})} \, 
W_{(\MJSE_{1,2},\MJSE_{2,2})} \, 
W_{(\MKSE_1,\MKSE_2)} \,,
\eeqa
where we have used that
$$
W_{(\MJSE_{1,2},\MJSE_{2,1})}
\, \equiv \,
W_{(\MISE_{1} \cap \MKSE_{2},\MISE_{2}\cap\MKSE_{1})}
\, = \,
W_{(\MISE_{2}\cap\MKSE_{1},\MISE_{1}\cap\MKSE_{2})}
\, \equiv \,
W_{(\MJSE_{2,1},\MJSE_{1,2})}
$$
on $\MCEL_{(\MISE_1,\MISE_2)} \cap \, \MCEL_{(\MKSE_1,\MKSE_2)}$ 
as a consequence of (\ref{MinSymCnd}).

\label{END-cit}
This completes the proof of Theorem \ref{THM-MINK-REC}.

\medskp

We shall 
propose now a construction that transforms the Minkowski renormalization recursion to Euclidean--like procedure.
``Euclidean-like'' means that we shall pass to the Euclidean type causal domains $\ECEL_{\{\MISE_1,\MISE_2\}}$ but considered on Minkowski space, i.e., we shall pass to domains
$$
 \ECEL_{\{\MISE_1,\MISE_2\}} 
\, = \,
\bigl\{(\x_j)_{j \in \MISE} \in \MSPA^{\MISE} \,\bigl|\, \x_{j_1} \neq \x_{j_2} \text{ for } j_a \in \MISE_a \ (a=1,2)\bigr\} 
\hspace{1pt} (\, = \, \ECEL_{\{\MISE_2,\MISE_1\}} ) \,
$$
instead of $\MCEL_{(\MISE_1,\MISE_2)}$~(\ref{MCEL2M}).

To begin with observe that we have
\beq\label{ecelmcel}
\ECEL_{j,k} \, = \, \MCEL_{j,k} \cup \MCEL_{k,j} 
\, = \, \MSPA^{\times 2} \backslash \MDIA_{j,k} \,,
\eeq
where we denoted $\ECEL_{j,k}:=\ECEL_{\{\{j\},\{k\}\}}$ and $\MCEL_{j,k}:=\MCEL_{(\{j\},\{k\})}$ for short.
Furthermore,
the distributions $W_{j,k}(\x_j,\x_k)\vrestr{12pt}{\MCEL_{j,k}}$ and
$W_{k,j}(\x_k,\x_j)\vrestr{12pt}{\MCEL_{k,j}}$ 
can be glued on (\ref{ecelmcel}) to a distribution on
$\ECEL_{j,k}$
\beq\label{MinkG}
G_{j,k} (\x_j,\x_k) \, := \, 
\left\{\begin{array}{ll}
W_{j,k}(\x_j,\x_k) & \text{on }\ \MCEL_{j,k}\raisebox{-8pt}{} \\
W_{k,j}(\x_k,\x_j) & \text{on }\ \MCEL_{k,j} \,
\end{array}\right.
\eeq
(using the same sign factor as in condition (\ref{MinSymCnd})).
It is now 
symmetric
$$
G_{j,k}(\x_j,x_k) \, = \, 
G_{k,j}(\x_k,x_j) 
\qquad
(\x_j \, \neq \, \x_k)
\,.
$$


\begin{example}\label{Gelfand}
For $W_{j,k}$ given in Example \ref{EXM-1nn} the distributions $G_{j,k} (\x_j,\x_k)$ defined by (\ref{MinkG}) can be written as
\beqs
G_{j,k} (\x_j-\x_k) \, = \, \frac{P_{j,k}(\x)}{\bigl(\x^2+i0\bigr)^{\nu_{j,k}}} \qquad (\x = \x_j-\x_k \neq 0)
\,.
\eeqs
It coincides with the distribution (with the same notation) defined in Sect. IV.2.4 of
\cite{GS}.
It follows from the analysis of 
\cite{GS}
that $G(\x)$ is a well defined (and homogeneous) distribution for all $\nu$ in $\R\backslash \{0\}$.
Moreover, for $\nu < \frac{D}{2} - \text{the degree of } P$, it has a unique homogeneous continuation to $\R^D$.%
\end{example}


\begin{Statement}{Proposition}\label{LmXXz}
$(a)$
For every nontrivial partition $\MISE = \{1,\dots,n\}=\MISE_1\dot{\cup}\MISE_2$ the product 
\beq\label{MinkGIJ}
G_{\{\MISE_1,\MISE_2\}} \, = \,
\mathop{\prod}\limits_{\mathop{}\limits^{j_1 \in \MISE_1}_{j_2 \in \MISE_2}}
G_{j_1,j_2} 
(\x_{j_1},\x_{j_2}) \qquad
\text{exists on} \qquad \ECEL_{\{\MISE_1,\MISE_2\}} \,.
\eeq
Furthermore,
\beq\label{Consistence1}
G_{\{\MISE_1,\MISE_2\}} \vrestr{12pt}{\MCEL_{(\MISE_1,\MISE_2)}}
\, = \,
W_{(\MISE_1,\MISE_2)}
\eeq

$(b)$
For every nontrivial partition $\MISE=\MISE_1\dot{\cup}\MISE_2$ the product 
$$
G^{ext}_{\MISE_1} \, G^{ext}_{\MISE_2} 
\, G_{\{\MISE_1,\MISE_2\}}
\hspace{1pt}\Bigl|\raisebox{-7pt}{\hspace{1pt}}_{\ECEL_{\{\MISE_1,\MISE_2\}}}
$$
exists. Furthermore,
$$
G^{ext}_{\MISE_1} \, G^{ext}_{\MISE_2} 
\, G_{\{\MISE_1,\MISE_2\}}
\hspace{1pt}\Bigl|\raisebox{-7pt}{\hspace{1pt}}_{\MCEL_{(\MISE_1,\MISE_2)}}
\, = \,
G^{ext}_{\MISE_1} \, G^{ext}_{\MISE_2} 
\, W_{(\MISE_1,\MISE_2)}
\hspace{1pt}\Bigl|\raisebox{-7pt}{\hspace{1pt}}_{\MCEL_{(\MISE_1,\MISE_2)}} \,.
$$
\end{Statement}

\medskp

\noindent
{\it Proof.}
We split the proof of part $(a)$ into two steps.

1.) 
{\it For every point $\xx_{\MISE} \in \ECEL_{\{\MISE_1,\MISE_2\}}$ there exists a splitting
\beq\label{tmp-split}
\MISE_1 \, = \, \MISE^{(1)} \dot{\cup} \MISE^{(3)} \dot{\cup} \cdots \dot{\cup} \MISE^{(2\ell-1)}
\quad \text{and} \quad
\MISE_2 \, = \, \MISE^{(0)} \dot{\cup} \MISE^{(2)} \dot{\cup} \cdots \dot{\cup} \MISE^{(2\ell)}
\eeq
$($with $2\ell \leqslant n$, where some $\MISE^{(m)}$ may be empty$)$ such that
\beq\label{XIcond}
\xx_{\MISE} \, \in \, \MCEL_{(\MISE^{(0)},\dots,\MISE^{(2\ell)})}  
\eeq
$($in the notations of Eq. $(\ref{MCEL2Mn})$$)$.} 
Indeed, we first split $\MISE$ $=$ $\MJSE_0$ $\dot{\cup}$ $\cdots$ $\dot{\cup}$ $\MJSE_{\ell}$
in such a way that: (i) the points $\x_j$ indexed in each subset $\MJSE_m$ have equal time coordinates, i.e.,
$x_j^0 = t_m$ for $j \in \MJSE_m$ ($m=0,\dots,\ell$); (ii) the time coordinates increase passing from $\MJSE_m$ to $\MJSE_{m+1}$, i.e., $t_0 < \cdots < t_{\ell}$.
Then we set $\MISE^{(2m+1)}$ $:=$ $\MJSE_m \cap \MISE_1$ and 
$\MISE^{(2m)}$ $:=$ $\MJSE_m \cap \MISE_2$ for $m=0,\dots,\ell$.
Thus,
$\xx_{\MISE} \, \in \, \MCEL_{(\MJSE_0,\dots,\MJSE_{\ell})}$.
Condition (\ref{XIcond}) follows since for every $m=0,\dots,\ell$ 
and every $j \in \MISE^{(2m+1)}$ and $k \in \MISE^{(2m)}$ the points $\x_j$ and $\x_k$ are space--like
(they are different due to $\xx_{\MISE} \in \ECEL_{\{\MISE_1,\MISE_2\}}$ and have equal time coordinates).

2.) 
Recall that by definition (Eq. (\ref{MinkG})) $G_{j_1,j_2}$ are distribution defined for $\x_{j_1} \neq \x_{j_2}$ (i.e., on $\ECEL_{j_1,j_2}$).
Then in a neighborhood of every $\xx_{\MISE} \in \ECEL_{\{\MISE_1,\MISE_2\}}$ each factor $G_{j_1,j_2}$ in (\ref{MinkGIJ}) equals to a Wightman function
$W_{s(j_1,j_2),t(j_1,j_2)}$, where $(j_1,j_2)$ $\mapsto$ $(s(j_1,j_2),t(j_1,j_2))$ is a transposition that is identity iff $j_1 \in \MISE^{(k_1)}$ and $j_2 \in \MISE^{(k_2)}$.
Furthermore, the product of all these $W_{s(j_1,j_2),t(j_1,j_2)}$ is well defined as a product of boundary value distributions with respect to a common (backward) tube
(in other words, it satisfies the analytic wave--front set condition for existence of a product).
Indeed, the latter tube corresponds to any linear order in the set $\MISE$ that respects the order of the subsets $\MISE^{(0)},\dots,\MISE^{(2\ell)}$
(note that there are no factors $G_{j_1,j_2}$ in (\ref{MinkGIJ}) with $j_1$ and $j_2$ belonging to a same peace $\MISE^{(k)}$).
Hence, we see that the product of distributions $G_{j_1,j_2}$ in (\ref{MinkGIJ}) exists in a neighborhood of each point $\xx_{\MISE} \in \ECEL_{\{\MISE_1,\MISE_2\}}$ according to the wave--front set criterion.
Since the wave-front set criterion is local, the product (\ref{MinkGIJ}) is uniquely determined.

3.) 
Thus, we have proved the existence of the product (\ref{MinkGIJ}).
Equation (\ref{Consistence1}) then follows from the definition of 
$W_{(\MISE_1,\MISE_2)}$ in (\ref{teq18M}).

\medskp

We continue with the proof of part $(b)$ of Proposition \ref{LmXXz}.
The product of $G_{\{\MISE_1,\MISE_2\}}$ with $G^{ext}_{\MISE_1} G^{ext}_{\MISE_2}$ exists on $\ECEL_{\{\MISE_1,\MISE_2\}}$ since in a neighborhood of each element of $\ECEL_{\{\MISE_1,\MISE_2\}}$, $G_{\{\MISE_1,\MISE_2\}}$ coincides with a boundary value distribution and Lemma \ref{ExProd0} can be applied.
The second part follows from Eq. (\ref{Consistence1}).$\quad\Box$


\begin{remark}\label{REM-JUNE13} 
Using the same arguments as above and the considerations 
of \ref{TSn1.3-N} one can extend the statement of Eq. (\ref{MinkGIJ}) to finer partitions (cf. Proposition \ref{UNPrpp}).
In particular, for the case of the finest partition we recover the old result that
{\it 
the Feynman amplitudes are always well defined as distribution outside the large diagonal $($i.e., for pairwise distinct arguments$)$, i.e., they do not need to be renormalized in this region}.
In fact, there is a simple direct argument for the latter statement:
if all the points of an amplitude are distinct then we can choose a Lorentz reference frame in which these points have different time components in a neighborhood;
it follows that the product 
\(
\prod_{\mathop{}\limits^{j_1 \, \leqslant \, j_2}_{j_1, j_2 \in \MISE}}
G_{j_1,j_2} (\x_{j_1},\x_{j_2})
\)
coincides in this neighborhood with a product of Wightman functions due to Eq. (\ref{MinkG})
and hence, it exists.
\end{remark}

Thus, while the proof of the causal factorization theorem in Minkowski space uses at each step Epstein--Glaser domains labeled by ordered partitions the final result can be formulated in terms of Euclidean domains (corresponding to unordered partitions).

\medskp

Symmetries in the Minkowski space are treated essentially in the same way as in the Euclidean. In particular, Lemma \ref{EucSym} and Corollary \ref{CrXX1} remain true without any change.


\section{Associate homogeneous distributions}\label{SECC-3}
\setcntrs

The step which completes the renormalization in both the Euclidean and the Minkowski case consists in an extension of distributions to the full diagonal.
As the Feynman amplitudes are translation invariant (even
after the renormalization recursion, cf. Corollary~\ref{CrXX1}) the problem reduces to extending distributions defined outside the origin of $\R^N$ for some $N \in \N$.
(The full diagonal modulo translations reduces to a point.)
For {\it associate homogeneous distributions} \cite{GS} (encountered in massless QFT) this extension can be 
performed
quite explicitly.
An associate homogeneous distribution $\MUD(\xx)$ of degree $\MDGR$ is a distribution which under the dilations 
$\MUD \mapsto \MUD_{\lambda} := \lambda^{-\MDGR} \, \MUD (\lambda\x)$ spans a finite dimensional space.
We shall initially define the associate homogeneous distributions as generalized eigenvectors of the Euler operator and later we shall prove this to be equivalent to the above definition that uses global dilations.

\subsection{Infinitesimal definition}\label{SECT-3.1-newww}

We shall denote the pairing between distributions $\MUD(\xx) \in \DP(\mathcal{U})$ on an open domain $\mathcal{U} \subseteq \R^N$  and test functions $f(\xx) \in \DD(\mathcal{U})$ by
$$
\MCnt{\MUD \, \MVOL}{f}{} \, \equiv \,
\MCnt{\MUD (\xx) \, \MVOL}{f(\xx)}{\xx} \,,
$$
thus making explicit 
its 
dependence 
on the volume form
$$
\MVOL \, \equiv \, d x^1 \wedge d x^2 \wedge \ldots \wedge d x^N
$$
on $\R^N$
and 
it will be also useful to indicate as a subscript in $\MCnt{\cdot}{\cdot}{\xx}$ the variable $\xx$ with respect to which the pairing is considered.
This pairing extends integration of products of smooth functions:
\(
\int_{\mathcal{U}} 
\MUD (\xx) f(\xx) \MVOL
\).
In particular, the derivatives $\di_{x^{a}} \MUD(\xx)$ ($\di_{x^{a}} := \frac{\di}{\di x^{a}}$) are defined by integration by parts
$$
\MCnt{\bigl(\di_{x^{a}} \MUD\bigr) \, \MVOL}{f(\xx)}{\xx} \, := \, 
- \, \MCnt{\MUD \, \MVOL}{\bigl(\di_{x^{a}} f(\xx)\bigr)}{\xx}
\,,
$$
which implies the formula for 
the action of an arbitrary vector field $\MVCT_{(\xx)}$
($=$ $\MVCT^{a}(\xx)$ $\di_{x^{a}}$)
\beq\label{MVCT}
\MCnt{\bigl(\MVCT_{(\xx)} \MUD\bigr) \, \MVOL}{f(\xx)}{\xx} \, := \, 
- \, \MCnt{\MUD \, \MVOL}{\MVCT_{(\xx)} f+f\,\text{\rm Div} (\MVCT)}{\xx} \,,
\eeq
where the divergence $\text{\rm Div} \MVCT$ 
of the vector field $\MVCT$
is defined with respect to the volume form $\MVOL$ by its Lie derivative 
$$
\MLIE_{\MVCT} (\MVOL) \, =: \, \bigl(\text{\rm Div} \MVCT\bigr) \MVOL
\,
$$
(then 
$\text{\rm Div} \MVCT$ $=$ $\di_{x^{a}} \MVCT^{a}(\xx)$).
Whenever the action of the vector field $\MVCT$ can be integrated to a globally defined one parameter group of diffeomorphisms $\MDFF_t$, its action on $\MUD$ is given according to (\ref{MVCT}) by the inverse action on the test 
functions\footnote{%
For the readers used to the differential geometric notions,
$\MDFF_t(\MVOL) \equiv \MDFF_t^{-1*}(\MVOL)$ is the pullback of $\MVOL$ with respect to
the flow $\MDFF_t^{-1}$.}
\beqa\label{CHCOR}
\podr
\MCnt{\MDFF_t (\MUD) \, \MDFF_t(\MVOL)}{f}{} \, = \, \MCnt{\MUD \, \MVOL}{\MDFF_t^{-1}(f)}{} \, \equiv \, \MCnt{\MUD \, \MVOL}{f \bigl(\MDFF_t(\xx)\bigr)}{\xx} \,,
\bnn 
\podr
\hspace{-18pt} (\Rightarrow \quad
\MCnt{\bigl(\MVCT \MUD\bigr) \, \MVOL}{f}{}
+\MCnt{\MUD \, \MVOL}{f\,\text{\rm Div} (\MVCT)}{}
\, = \, 
\frac{d}{dt} \,
\MCnt{\MDFF_t (\MUD) \, \MDFF_t(\MVOL)}{f}{}
\vrestr{12pt}{t \, = \, 0}
).
\eeqa

Let $\MCON$ be an open cone in $\R^N$ and let $\MUD (\xx)$ be a distribution on $\MCON$. Then $\MUD (\xx)$ is called {\bf associate homogeneous distribution of 
scaling degree
$\MDGR \in \R$ and 
scaling {\ORder}}\footnote{%
From now on we will refer to these briefly as degree and order, the latter not to be confused with the order in the sense of distributions.}
$\MORD$ $(=0,1,2,\dots)$ iff it satisfies the equation:
\beq\label{MEUL}
\bigl(\MEUL_{(\xx)}-\MDGR\bigr)^{\MORD+1} \MUD (\xx) \, = \, 0 \,
\eeq
but 
$\bigl(\MEUL_{(\xx)}-\MDGR\bigr)^{\MORD} \MUD (\xx) \neq 0$,
where $\MEUL_{(\xx)}$ is the Euler vector field
\beq\label{MEUL-vf}
\MEUL_{(\xx)} \, := \, \xx \cdot \frac{\partial}{\partial \xx} \, \equiv \,
\mathop{\sum}\limits_{a \, = \, 1}^N x^{a} \frac{\partial}{\partial x^{a}} \,.
\eeq
In particular, 
$\MUD (\xx) \neq 0$ when we speak about an associate homogeneous distribution of a given non-negative 
{\ORder}, 
and formally, we can define $\MUD (\xx) = 0$ to be an associate homogeneous distribution of 
{\ORder}
$-1$ and any degree.
Note 
also that the (scaling) degree and the order depend on the domain $\MCON$ on which we consider the associate homogeneous distribution and in particular, if we diminish the domain the order may decrease.
Using Eq. (\ref{MVCT}) we can rewrite condition (\ref{MEUL}) in terms of test functions
\beq\label{MEUL-1}
0 =
\MCnt{\bigl(\MEUL_{(\xx)}-\MDGR\bigr)^{\MORD+1} \MUD (\xx) \, \MVOL}{f(\xx)}{\xx} 
\equiv
\MCnt{\MUD (\xx) \, \MVOL}{\bigl(-\MEUL_{(\xx)}-N-\MDGR\bigr)^{\MORD+1} f(\xx)}{\xx} 
\,,
\eeq

\subsection{Generalized Euler's decomposition}\label{SeGenEu}

The above definition of associate homogeneity is clearly local and so, $\MUD (\xx)$ does not need to be defined on a cone domain.
Since the Euler vector field generates the flow of the dilation group $\xx \mapsto \lambda \xx$ 
($\lambda = e^t$) then condition (\ref{MEUL}) has an equivalent global form on a dilation invariant domain $\MCON$, i.e. on a cone.
To state this result let us call a {\bf cone section} of $\MCON$ every smooth hyper-surface $\MHYP \subseteq \MCON$ such that it intersects {\it transversely} every ray $\{\lambda \xx\}_{\lambda > 0} \subseteq \MCON$
(for $\xx \neq 0$) at a 
unique point.
Then there is an induced smooth function 
$\MRAD_{\MHYP} (\xx) \in \R^+ \equiv (0, \infty)$
on $\MCON \backslash \{0\}$ 
defined by the condition
\beq\label{UNNUMBERED}
\MRAD_{\MHYP} (\xx)^{-1} \, \xx \, \in \, \MHYP 
\qquad (\xx \in \MCON \backslash \{0\})
\,.
\eeq
Hence, the correspondence,
\beqa\label{MDIF}
\begin{array}{rccl}
\MCON\backslash\{0\} \, \ni 
&
\xx 
&
\mathop{\longmapsto}\limits^{\MDFF} 
&
(\uu,r) \, := \, (\MRAD_{\MHYP} (\xx)^{-1} \, \xx,\, \MRAD_{\MHYP} (\xx)) 
\, 
\in \, \MHYP \times \R^+ 
\\
\MHYP \times \R^+  \, \ni 
&
(\uu,r) 
& 
\mathop{\longmapsto}\limits^{\MDFF^{-1}} 
& 
r\uu 
\,
\in \, \MCON\backslash\{0\} \,,
\end{array}
\eeqa
is a diffeomorphism. 

\medskp

\begin{Statement}{Theorem}\label{THM-EXT}
Every associate homogeneous distribution $\MUD (\xx)$ 
on an open cone 
$\MCON \backslash \{0\}$
is uniquely decomposable 
for every cone section $\MHYP$ into the sum
\beqa\label{MDEC}
\podr
\MUD \vrestr{12pt}{\xx \mapsto (\uu,r)} 
\, = \,
\mathop{\sum}\limits_{\MORd \, = \, 0}^{\MORD} \, 
\frac{1}{\MORd!} \
\MUH_{\MHYP,\MORd} (\uu) \otimes
r^{\MDGR} \bigl(\ln r\bigr)^{\MORd}
\qquad
\bnn \podr
\bigl( \,
\MUD \vrestr{12pt}{\xx \mapsto (\uu,r)} 
\, := \, \MDFF (\MUD) \bigl(\uu,r\bigr) 
\, \bigr)
\eeqa
under the diffeomorphism $(\ref{MDIF})$, where $\MUH_{\MHYP,\MORd} (\uu)$ are distributions on $\MHYP$.
The numbers $\MORD$ in Eqs.~$(\ref{MEUL})$ and $(\ref{MDEC})$ coincide. 

In particular, the distribution $\MUD (\x)$ admits a restriction on $\MHYP$ and 
fur\-ther\-mo\-re,
\beq\label{restriction-on-mhyp}
\Bigl((\MEUL_{(\xx)}-\MDGR)^{\MORd} \,
\MUD\Bigr) \vrestr{12pt}{\MHYP} 
\, = \,
\MUH_{\MHYP,\MORd} \,,
\eeq
for $\MORd=0,1,\dots,\MORD$ $((\MEUL_{(\xx)}-\MDGR)^{0} \equiv 1)$.
\end{Statement}

\medskp

\noindent
{\it Proof.}
We prove the statement by induction in $\MORD$. For $\MORD = 0$ the statements is a generalization of Euler's theorem to homogeneous distributions. 
It follows from the fact that if we set
$$
\MUH_{\MHYP} (\uu,r) \, := \, r^{-\MDGR} \cdot \MUD \vrestr{12pt}{\xx \mapsto (\uu,r)}
$$
then $\frac{\di}{\di r}\MUH_{\MHYP} (\uu,r) = 0$ and hence, $\MUH_{\MHYP} (\uu,r)$ is independent of $r$.

Let then by induction assume that the theorem is proven for all associate homogeneous distributions of 
{\ORder}
$\MORDD < \MORD$. Let $\MUD$ be of 
{\ORder}
$\MORD$ (i.e., satisfy (\ref{MEUL})). Then we can apply the inductive hypothesis to $(\MEUL_{(\xx)}-\MDGR) \, \MUD$  and conclude that we have a unique decomposition
\beq\label{HELP-eq}
\MDFF \Bigl(
(\MEUL_{(\xx)}-\MDGR) \, \MUD \Bigr)
\, = \,
\mathop{\sum}\limits_{\MORd \, = \, 0}^{\MORD-1} \,
\frac{1}{\MORd!} \
\MUH_{\MHYP,\MORd+1} (\uu) \otimes
r^{\MDGR} \bigl(\ln r\bigr)^{\MORd} \,,
\eeq
which can be rewritten as
$$
(\MEUL_{(\xx)}-\MDGR) \, 
\Biggl(
\MUD
-
\MDFF^{-1}
\Biggl(
\mathop{\sum}\limits_{\MORd \, = \, 1}^{\MORD} \,
\frac{1}{\MORd!} \
\MUH_{\MHYP,\MORd} (\uu) \otimes
r^{\MDGR} \bigl(\ln r\bigr)^{\MORd}
\Biggr)\Biggr) \, = \, 0 \,.
$$
Thus, applying again the base of induction we obtain Eq.~(\ref{MDEC}) as a unique decomposition of $\MUD$ and the theorem is proven.$\quad\Box$

\medskp

We shall now write the decomposition (\ref{MDEC}) when both sides are applied to a test function belonging to $\Dd\bigl(\MCON \backslash \{0\}\bigr)$.

\medskp

Let us define first the restricted volume form on $\MHYP$: to this end we introduce 
\beqa\label{Kron-f}
\MVOL_{\MHYP} \podr := \, \MRAD_{\MHYP}(\xx)^{-N} \iota_{\MEUL_{(\xx)}} \MVOL 
\nnb
\podr = \,
\MRAD_{\MHYP}(\xx)^{-N} \,
\mathop{\sum}\limits_{a \, = \, 1}^{N} (-1)^{a-1} \, x^{a} \, d x^1 \wedge \cdots \wedge \widehat{dx^{a}} \wedge \cdots \wedge dx^N
\,,
\qquad
\eeqa
where $\iota_{\MEUL_{(\xx)}} \MVOL$ stands for the contraction of $\MEUL_{(\xx)}$ with $\MVOL$
(the hat, $\widehat{\cdot}\,$, over an argument means, as usual, that this argument is omitted).
So defined $\MVOL_{\MHYP}$ is a differential $(N-1)$--form on $\R^N \backslash \{0\}$, which is closed and $\MEUL$--invariant
$$
d \MVOL_{\MHYP} \, = \, 0
\,,\quad
\MLIE_{\MEUL_{(\xx)}} \MVOL_{\MHYP} \, = \, 0
$$
($\MLIE_{\MEUL_{(\xx)}}$ standing for the Lie derivative along $\MEUL_{(\xx)}$).
Under diffeomorphism (\ref{MDIF}) $\MVOL_{\MHYP}$ is transformed to an $N-1$ form on $\MHYP \times \R^+$, which does not depend on $r \in \R^+$.\footnote{%
This means that the transformed $\MVOL_{\MHYP}$ on $\MHYP \times \R^+$ is a pullback of a form on $\MHYP$ under the projection $\MHYP \times \R^+\to\MHYP$.}
This induces a volume form on $\MHYP$, which we shall denote again by $\MVOL_{\MHYP}$ with a slight abuse of the notations. Since we have\footnote{%
\(
\MRAD_{\MHYP}^{N-1} \, d\MRAD_{\MHYP} \wedge \MVOL_{\MHYP}
= 
\MRAD_{\MHYP}^{N-1} \, d\MRAD_{\MHYP} \wedge 
\MRAD_{\MHYP}(\xx)^{-N} \iota_{\MEUL_{(\xx)}} \MVOL
=
\MRAD_{\MHYP}^{-1} \, d\MRAD_{\MHYP} \wedge \iota_{\MEUL_{(\xx)}} \MVOL
=
\MRAD_{\MHYP}^{-1} \, \bigl(\iota_{\MEUL_{(\xx)}} d\MRAD_{\MHYP} \bigr)  \, \MVOL
=
\MVOL
\)
}
\(
\MVOL = 
\MRAD_{\MHYP}^{N-1} \, d\MRAD_{\MHYP} \wedge \MVOL_{\MHYP}
\)
it follows that
$$
\MDFF (\MVOL) \, = \, r^{N-1} \, dr \wedge \MVOL_{\MHYP} \,.
$$

Now according to Eq.~(\ref{MDEC}) for every test-function $f(\xx)$ supported at $\MCON\backslash\{0\}$ we have:
\beq\label{MDEC1}
\MCnt{\MUD (\xx) \, \MVOL}{f(\xx)}{\xx} \, = \,
\mathop{\sum}\limits_{\MORd \, = \, 0}^{\MORD} \, 
\frac{1}{\MORd!} \,
\MCNt{r^{N+\MDGR-1} \, \bigl(\ln r\bigr)^{\MORd}
\, dr
}{
\MCnt{\MUH_{\MHYP,\MORd} (\uu)\,\MVOL_{\MHYP}}{f(r\uu)}{\uu}
}{r}
\,.
\eeq
Note that $f(r\uu)$ is a test-function on $\DD(\MHYP)$ smoothly depending on $r$, which means that $\Mcnt{\MUH_{\MHYP,\MORd} (\uu)\MVOL_{\MHYP}}{f(r\uu)}{\uu}$ is a test function belonging to $\DD (0,\infty)$.

\subsection{The global dilation law for associate homogeneous distributions}

\begin{Statement}{Proposition}\label{CrZZZ1}
Let $\MUD (\xx)$ be a distribution on an open cone $\MCON$, which is associate homogeneous of degree $\MDGR$ and 
{\ORder}
$\MORD$.
Then there exists a unique sequence $\MUD$ $=$ $\MUD_0(\xx),$ $\dots,$ $\MUD_{\MORD} (\xx)$ of distributions on $\MCON$ such that for all $\lambda \in \R^+:$
\beq\label{DLTLW}
\left(\hspace{-2pt}
\begin{array}{c}
\lambda^{-\MDGR}\,\MUD_0 (\lambda\xx)
\raisebox{14pt}{}\raisebox{-10pt}{}
\\
\lambda^{-\MDGR}\,\MUD_1 (\lambda\xx)
\raisebox{17pt}{}\raisebox{-7pt}{}
\\
\vdots
\\
\raisebox{16pt}{}\raisebox{-6pt}{}
\lambda^{-\MDGR}\,\MUD_{\MORD} (\lambda\xx)
\end{array}
\hspace{-2pt}\right)
\, = \,
\left(\hspace{-2pt}
\begin{array}{cccc}
1 & \ln \lambda & \cdots & \Txfrac{(\ln \lambda)^n}{n!_{}} 
\\
0 & 1 & \cdots & \Txfrac{(\ln \lambda)^{n-1}}{(n-1)!}
\raisebox{18pt}{}\raisebox{-15pt}{}
\\
\cdots & \cdots & \cdots & \cdots
\\
0 & 0 & \cdots & 1
\end{array}
\hspace{-2pt}\right)
\left(\hspace{-2pt}
\begin{array}{c}
\MUD_0 (\xx)
\raisebox{14pt}{}\raisebox{-10pt}{}
\\
\MUD_1 (\xx)
\raisebox{17pt}{}\raisebox{-7pt}{}
\\
\vdots
\\
\raisebox{16pt}{}\raisebox{-6pt}{}
\MUD_{\MORD} (\xx)
\end{array}
\hspace{-2pt}\right)
\eeq
or, 
\(
\left(\hspace{0pt}
\begin{array}{c}
\lambda^{-\MDGR}\,\MUD_0 (\lambda\xx)
\\
\vdots
\\
\lambda^{-\MDGR}\,\MUD_{\MORD} (\lambda\xx)
\end{array}
\hspace{0pt}\right)
= 
e^{A \, \ln \lambda} \,
\left(\hspace{0pt}
\begin{array}{c}
\MUD_0 (\xx)
\\
\vdots
\\
\MUD_{\MORD} (\xx)
\end{array}
\hspace{0pt}\right)
\)
with
\(
A = 
\left(\hspace{0pt}
\raisebox{12pt}{\(\begin{array}{cc}
0 & 1 \\
0 & 0 
\end{array}\)}
\raisebox{0pt}{\(
\raisebox{5pt}{\hspace{4pt}$\ddots$\hspace{-4pt}}
\hspace{-10pt}\raisebox{-7pt}{\hspace{-13pt} $\ddots$}\hspace{14pt}
\raisebox{-7pt}{\hspace{-19pt} $\ddots$}
\)}
\raisebox{-10pt}{\(\begin{array}{c}
1 \\
\hspace{-10pt}0\hspace{8pt}0
\end{array}\)}
\hspace{0pt}\right)\).

In particular,
each distribution $\MUD_{m} (\xx)$ is associate homogeneous of degree $\MDGR$ and 
{\ORder}
$\MORD-m$.
\end{Statement}

\medskp

\noindent
{\it Proof.}
We start with the case where $0 \notin \MCON$, which we obtain as a consequence of Theorem \ref{THM-EXT}. 
The sequence $\MUD_m (\xx)$ that satisfies the matrix law (\ref{DLTLW}) is constructed by using Eq. (\ref{MDEC}):
\beq\label{MDECm}
\MUD_m \vrestr{12pt}{\xx \mapsto (\uu,r)} 
\, := \,
\mathop{\sum}\limits_{\MORd' \, = \, 0}^{\MORD-\MORd} \, 
\frac{1}{\MORd'!} \
\MUH_{\MHYP,\MORd+\MORd'} (\uu) \otimes
r^{\MDGR} \bigl(\ln r\bigr)^{\MORd'}
\eeq
($m=0,1,\dots,n$).
Then (\ref{DLTLW}) is a consequence of the Newton binomial formula.

It 
remains to prove Eq. (\ref{DLTLW}) for the case when $\MCON=\R^N$.
Let $\MUD(\xx) \in \DP(\R^N)$ and let us denote its restriction on $\R^N \backslash \{0\}$ by $\MUDZ(\xx)$ ($=\MUD(\xx)\vrestr{10pt}{\R^N \backslash \{0\}}$).
Then $\MUDZ(\xx)$ satisfies Eq. (\ref{DLTLW}) with a sequence
$\MUDZ$ $=$ $\MUDZ_0(\xx),$ $\dots,$ $\MUDZ_{\MORD'} (\xx)$
with some (possibly different) order $\MORD'$.
In fact,\footnote{%
\label{FTNTXX21}%
Note that the distribution valued function 
$\lambda \mapsto \MUD_{\lambda} := \lambda^{-\MDGR} \, \MUD (\lambda\x)$ 
is infinitely differentiable}
$$
\MUDZ_{\ell}(\xx) =
\Bigl(\frac{d}{d \ln \lambda}\Bigr)^{\ell}
\bigl(\lambda^{-\MDGR}\,\MUDZ (\lambda\xx)\bigr)\vrestr{10pt}{\lambda = 1}
\equiv
\bigl(\MEUL_{(\xx)}-\MDGR\bigr)^{\ell} \MUDZ(\xx)
$$
for $\ell = 0,\dots,\MORD'$. 
Let us set
$\MUD_{\ell}(\xx)$ $:=$ $\bigl(\MEUL_{(\xx)}-\MDGR\bigr)^{\ell} \MUD(\xx)$.
Then $\MUD_{\MORD'+1}(\xx)\vrestr{10pt}{\R^N \backslash \{0\}}$
$=$ $\MUDZ_{\MORD'+1}(\xx)$ $=$ $0$ and hence, it must be proportional to a homogeneous derivative of the delta function (in particular, it may vanish).
Hence, $\MUD_{\MORD'+2}(\xx)$ $=$ $0$ and we obtain the scaling law (\ref{DLTLW}) with an 
{\ORder}
$\MORD$ $=$ $\MORD'+1$.$\quad\Box$


\begin{remark}\label{NewRem-3.1}
We 
see in the second part of the above proof an important feature of associate homogeneous distributions: when one makes a restriction from $\R^N$ to $\R^N\backslash \{0\}$ the order $\MORD$ of $\MUD$ may decrease by one and if this happens then the $\MORD$'th successor $\MUD_n$ of $\MUD$ is necessarily a homogeneous distribution supported at zero.
In the next section we shall reverse the point of view and consider the problem of extending associate homogeneous distributions from $\R^N\backslash\{0\}$ to $\R^N$.
In particular, we shall see that the $\MORD$'th successor $\MUD_n$ does not depend on the extension.
\end{remark}


\begin{Statement}{Corollary}\label{CrZZZ2}
Let $\MUD (\xx)$ be a distribution on an open cone $\MCON$.
Then it is an associate homogeneous distribution of degree $\MDGR$ and 
{\ORder}
$\MORD$ iff under the transformation law 
$\MUD \mapsto \MUD_{\lambda} := \lambda^{-\MDGR} \, \MUD (\lambda\x)$
it spans a finite dimensional vector space of dimension
$\leqslant$ $\MORD+1$.
\end{Statement}

\medskp

\noindent
{\it Proof.}
The necessity follows from Proposition \ref{CrZZZ1}.
To prove the sufficiency we introduce the sequence
\(
\MUD_m (\x) = 
\bigl(\lambda \frac{d}{d\lambda}\bigr)^m \, \MUD_{\lambda} \vrestr{12pt}{\text{\scriptsize $\lambda = 1$}}
\)
for $m=0,1,\dots$
(cf. footnote \ref{FTNTXX21}).
There exists $n \in \{0,1,\dots\}$ such that
$\MUD_0 (\x),$ $\dots,$ $\MUD_n (\x)$ form a basis within the infinite sequence.
Since this basis spans a representation of the multiplicative group $\R^+$ with a generator that is a Jordan block with zero diagonal we get the transformation law (\ref{DLTLW}).
But then it follows that 
\(\bigl(\lambda \frac{d}{d\lambda}\bigr)^{n+1} \, \MUD_{\lambda} = 0\),
which in turn is equivalent to (\ref{MEUL}).$\quad\Box$


\section{Extension of distributions and homogeneity}\label{SectNew4}

The subsequent discussion is a generalization of H\"ormander's treatment of homogeneous distributions \cite[Sect. 3.2]{H} to the case of associate homogeneous ones.

\subsection{The extension problem for associate homogeneous distributions: ambiguity and reduction to one-dimension}\label{SSE3.4NN}

Let us first describe the extension problem.

Assume that we have a distribution $\MUDZ(\xx)$ defined on $\R^N \backslash \{0\}$, which is associate homogeneous of degree $\MDGR$. 
We look for a distribution $\MUD(\xx)$ on $\R^N$ which is again  associate homogeneous of degree $\MDGR$ and extends $\MUDZ(\xx)$  in the sense that
\beq\label{EXTPROB}
\MUD\vrestr{12pt}{\R^N\backslash \{0\}} \, = \, \MUDZ \,.
\eeq

Let us consider first the uniqueness.  By (\ref{EXTPROB}) the difference $\MUD'-\MUD$ for two different extensions of $\MUDZ$ is a distribution supported at zero, i.e., a linear combinations of derivatives of 
the
delta-functions. Since the 
{\ORder}
$k$ derivatives of the delta-function are associate homogeneous  of degree $-N-k$ we obtain the following result

\medskp

\begin{Statement}{Proposition}\label{PeEXT}
If $\MUDZ(\xx)$ is associate homogeneous of degree $\MDGR \neq -N,-N-1,-N-2\dots$ then $\MUDZ(\xx)$ has {\rm exactly} one associate homogeneous extension $\MUD(\xx)$ of the same degree $\MDGR$. If $\MDGR=-N-k$ then any two extensions $\MUD$ and $\MUD'$ of $\MUDZ$ of the same degree of associate homogeneity differ by a linear combination of 
{\ORder}
$k$ derivatives of the delta-function. 
\end{Statement}

\medskp

We will now prove the asserted existence of extensions. 
We shall demonstrate that it reduces to a one dimensional problem.

Let us try to use decomposition (\ref{MDEC1}) in order to construct an extension of $\MUDZ(\xx)$:
$$
\MCnt{\MUDZ (\xx) \, \MVOL}{f(\xx)}{\xx} =
\mathop{\sum}\limits_{\MORd \, = \, 0}^{\MORD} \,
\frac{1}{\MORd!} \, 
\MCNt{r^{N+\MDGR-1} \, \bigl(\ln r\bigr)^{\MORd}
\, dr
}{
\MCnt{\MUHZ_{\MHYP,\MORd} (\uu)\,\MVOL_{\MHYP}}{f(r\uu)}{\uu}
}{r}
\,,
$$
where $f(\xx)$ is a test-function on $\R^N \backslash \{0\}$.
Now if $f(\xx)$ is a test-function on $\R^N$ 
we first note that $\Mcnt{\MUH_{\MHYP,\MORd} (\uu)\,\MVOL_{\MHYP}}{f(r\uu)}{\uu}$ is a compactly supported smooth function in $r \in [0,\infty)$, which can be continued (arbitrarily) to a test-function over $\R$ (possibly, with a larger support).
With a slight abuse of notations we shall treat $\Mcnt{\MUH_{\MHYP,\MORd} (\uu)\,\MVOL_{\MHYP}}{f(r\uu)}{\uu}$ as a test function on $\R$, which will not cause a problem since the results will not depend on the continuation of $\Mcnt{\MUH_{\MHYP,\MORd} (\uu)\,\MVOL_{\MHYP}}{f(r\uu)}{\uu}$ outside $[0,\infty)$.
Hence, the extension of $\MUDZ(\xx)$ to a globally defined distribution $\MUD(\xx) \in \DP(\R^N)$ is solved by defining the following actions of distributions
$$
\MCNt{r^{N+\MDGR-1} \, \bigl(\ln r\bigr)^{\MORd} \vartheta(r) \, dr}{
\MCnt{\MUHZ_{\MHYP,\MORd} (\uu)\,\MVOL_{\MHYP}}{f(r\uu)}{\uu}
}{r}
$$
where $\vartheta (r)$ stands for the characteristic function of $[0,\infty)$.
In other words, we need to define the distributions
$r^{\MDGR} \bigl(\ln r\bigr)^{\MORD} \vartheta(r)$ for a general $\MDGR \in \R$ and integer $\MORD$.
We shall denote them by
\beq\label{rlogeq}
\tlr{\MDGR}{n} \, := \, \text{extension to $\R$ of } \ \Bigl(\vartheta(r) \, r^{\MDGR} (\ln r)^n\Bigr)\hspace{-1pt}\vrestr{12pt}{\R\backslash \{0\}} \,.
\eeq

Thus, as soon as we have defined the distributions $(r^{\MDGR} \bigl(\ln r\bigr)^{\MORD})_+$ ($\MDGR \in \R$, $\MORD =0,1,\dots$) we will obtain a formula for an extension $\MUD(\xx)$ of $\MUDZ (\xx)$:
\beqa\label{MDEC2}
\podr
\MCnt{\MUD (\xx) \, \MVOL}{f(\xx)}{\xx} 
\nnb \podr
= \,
\mathop{\sum}\limits_{\MORd \, = \, 0}^{\MORD} \, 
\frac{1}{\MORd!} \,
\MCNt{(r^{N+\MDGR-1} \, \bigl(\ln r\bigr)^{\MORd})_+\, dr}{
\MCnt{\MUHZ_{\MHYP,\MORd} (\uu)\, \MVOL_{\MHYP}}{f(r\uu)}{\uu}
}{r}
\,.
\eeqa
Let us summarize:

\medskp

\begin{Statement}{Corollary}\label{CRR1}
If we construct the extensions $(\ref{rlogeq})$ for all $\MDGR \in \C$ and $n =0,1,\dots$ then Eq. $(\ref{MDEC2})$ provides an extension of $\MUDZ (\xx)$ to $\R^N$.
If the distributions $(\ref{rlogeq})$ are associate homogeneous on $\R$ then $\MUD(\xx)$ are associate homogeneous on $\R^N$.
\end{Statement}

\medskp

\noindent
{\it Proof.}
Concerning the associate homogeneity of the extension we use Eq. (\ref{MEUL-1}):
\beqs
\podr
\MCnt{\bigl(\MEUL_{(\xx)}-\MDGR\bigr)^{n'+1} \MUD (\xx) \, \MVOL}{f(\xx)}{\xx} 
\\ \podr
= 
\mathop{\sum}\limits_{\MORd \, = \, 0}^{\MORD}
\frac{1}{\MORd!} \,
\Bigl\langle
(r^{N+\MDGR-1} \, \bigl(\ln r\bigr)^{\MORd})_+\, dr
\,,
\\ \podr
\phantom{
= 
\mathop{\sum}\limits_{\MORd \, = \, 0}^{\MORD}
\frac{1}{\MORd!} \,
\Bigl\langle
}
\Bigl(-r\,\frac{d}{dr}-N-\MDGR\Bigr)^{n'+1}
\MCnt{\MUHZ_{\MHYP,\MORd} (\uu)\, \MVOL_{\MHYP}}{f(r\uu)}{\uu}
\Bigr\rangle_r
\\ \podr
=
\mathop{\sum}\limits_{\MORd \, = \, 0}^{\MORD}
\frac{1}{\MORd!} \,
\Bigl\langle
\Bigl(r\,\frac{d}{dr}-N-\MDGR+1\Bigr)^{n'+1}
(r^{N+\MDGR-1} \, \bigl(\ln r\bigr)^{\MORd})_+\, dr
\,,
\\ \podr
\phantom{
= 
\mathop{\sum}\limits_{\MORd \, = \, 0}^{\MORD}
\frac{1}{\MORd!} \,
\Bigl\langle
}
\MCnt{\MUHZ_{\MHYP,\MORd} (\uu)\, \MVOL_{\MHYP}}{f(r\uu)}{\uu}
\Bigr\rangle_r
\,.
\eeqs

\subsection{Extension of associate homogeneous distributions on $\R^+$ via analytic renormalization}\label{An-Sect}

In this section we give a construction for the
extended (or, renormalized) distributions (\ref{rlogeq}).
We shall follow the idea of the so called analytic renormalization \cite{Sp}, and 
we recall first some results about analytic vector valued functions.

\medskp

\begin{Statement}{Theorem}\label{Th2.1-1}
$(\cite[Chapt. I.3]{Rud})$
Let $\VSP$ be a locally convex, complete topological vector space
and let $\{v_z\}_{z \, \in \, \Oo}$ be collection of continuous linear functionals
$v_z$ on $\VSP$ labeled by $z$ belonging to an open subset $\Oo \subset \C$.
The following conditions are equivalent:
\begin{LIST}{26pt}
\item[$(a)$]
The function $\Oo \to \VSP'$ $:$ $z \mapsto v_z$ is differentiable in the weak topology
of the dual space  $\VSP'$ -- that is, the weak limit
$\displaystyle \mathop{\text{\rm w-\hspace{-1pt}}\lim}\limits_{z' \, \to \, z}\frac{v_{z'} - v_z}{z'-z}$
exists for every $z \in \Oo$.
\item[$(b)$]
For every $z \in \C$ there exists a weakly convergent expansion
$$
v_z \, = \, \mathop{\sum}\limits_{n \, = \, 0}^{\infty} \frac{1}{n!} \, u_n \, (z'-z)^n
\,\qquad
(u_n=u_{n,z'})
$$
for $u_n \in \VSP'$,
if the disc $\bigl\{z'' \in \C : |z''-z| \leqslant |z'-z|\bigr\}$ is contained~in~$\Oo$.
\item[$(c)$]
For every $f \in \VSP$ the complex valued function
$\Oo \to \C$ $:$ $z \mapsto v_z [f]$ is analytic in $\Oo$.
\end{LIST}
In each of these cases we say that $z \mapsto v_z$ is an {\rm\bf analytic} distribution valued function.
Then it also follows that
$$
v_z[f] \, = \, \mathop{\sum}\limits_{n \, = \, 0}^{\infty} \frac{1}{n!} \, u_n[f] \, (z'-z)^n
$$
for every $f \in \VSP$,
the series being absolutely convergent
if the disc $\bigl\{z'' \in \C : |z''-z| \leqslant |z'-z|\bigr\}$ is contained in $\Oo$.
\end{Statement}


\begin{definition}\label{Df2.1-1}
The function $\Oo_0 \to \VSP'$ $:$ $z \mapsto v_z$ defined in a dense subset $\Oo_0$
of an open set $\Oo$ $\subseteq \C$ is called \textbf{meromorphic} on $\Oo$
if for every $z \in \Oo$ ($\subseteq \C$) there exists $N_z \in \{0,1,\dots\}$
and an analytic function
$w_{z,z'}$ in $z'$ belonging to a neighborhood $U_z$ of $z$ such that
$w_{z,z'}$ $=$ $(z'-z)^{N_z} \bigl(v_{z'}-v_{z}\bigr)$ for $z' \in U_z \backslash \{z\}$
and $w_{z,z} \neq 0$.
We say that $v_z$ has a simple pole at $z \in \C$ if $N_z=1$.
\end{definition}


We shall mainly consider distribution valued analytic functions $v_z (r)$ ($z \in \Oo$),
where $v_z (r)$ is a distribution for every fixed $z$ (and thus, $v_z [f]$
is an analytic function in $z \in \Oo$ for every test function $f \in \Dd (\R)$).
By the above theorem the derivative $\di_r v_z(r)$ is also a distribution
which analytically depends on $z \in \Oo$.

It is shown in \cite[Chapt. 3.2]{H} that the family of distributions
\beq\label{war}
\HMCHI_{\MDGR} (r) \, := \,
\frac{\tlrb{\MDGR}}{\Gamma(\MDGR+1)} \quad \MDGR \neq -1,-2,\dots
\eeq
is uniquely extendable to a distribution valued entire analytic function
$\MDGR \mapsto \HMCHI_{\MDGR}$.
{\it Indeed}, $\tlrb{\MDGR}$ is a locally integrable function for $\RE \, \MDGR > -1$ and
hence the function $\MDGR$ $\mapsto$ $\HMCHI_{\MDGR} [f]$ is analytic for every $f \in \Dd(\R)$
when $\RE \, \MDGR > -1$.
Using further the property $\Gamma(\MDGR+1)$ $=$ $\MDGR \Gamma(\MDGR)$ we see that
$$
\di_r \HMCHI_{\MDGR}  \, = \, \HMCHI_{\MDGR-1} \,,
$$
and hence, $\HMCHI_{\MDGR}$ possesses an extension to an entire analytic function in $\MDGR \in \C$.
Since $\HMCHI_0 (r)$ $=$ $\vartheta (r)$ we obtain that for $k=0,1,\dots$
\beq\label{wn-1}
\HMCHI_{-k-1} (r) \, = \, \dltan{k} (r) \,.
\eeq
Thus, we conclude that $\tlrb{\MDGR}$ $\equiv$ $\tlub{\MDGR}$ is a meromorphic function in $\MDGR \in \C$.
We already know that for a noninteger $\MDGR$
the distribution $\tlrb{\MDGR}$ is uniquely determined as a homogeneous extension of $\TLB{\MDGR}$ on $[0,\infty)$.
The distribution valued function $\tlub{\MDGR}$ has simple poles at $\MDGR=-k-1$, $k=0,1,\dots$ and furthermore,
it possesses a Laurent expansion around these $\MDGR$'s of the form:
\beq\label{Eq2n.0}
\tlub{-k-1+\varepsilon}
\, = \,
\frac{(-1)^k}{k!} \, \dltan{k} (r) \, \frac{1}{\varepsilon}
\, + \,
\mathop{\sum}\limits_{\MORD \, = \, 0}^{\infty} \, L_{-k-1,\MORD} (r) \, \varepsilon^{\MORD}
\,.
\eeq
Indeed, the polar part in Eq.~(\ref{Eq2n.0}) 
is obtained by Eqs.~(\ref{war}), (\ref{wn-1}) and
$$
\displaystyle
\Gamma(-k+\varepsilon) =
\frac{\pi}{\Gamma(k+1-\varepsilon) \, \sin (\pi (\varepsilon-k))}
\, = \,
\frac{(-1)^k}{k!} \, \frac{1}{\varepsilon} + \Phi(\varepsilon)
$$
where 
$\Phi(\varepsilon)$ is analytic at $\varepsilon=0$.
Thus, we set
\beqa\label{Lkn}
\frac{1}{\MORD !} \, \tlr{-k-1}{\MORD} \, := \podr L_{-k-1,\MORD} (r) \,,
\nn
\frac{1}{\MORD !} \,
\tlr{k}{\MORD} \, :=  \podr L_{k,\MORD} (r) \, \equiv \, \frac{1}{\MORD !} \, \vartheta(r) \, r^{k} (\ln r)^{\MORD}
\eeqa
for $k,\MORD=0,1,\dots$.
The distributions $L_{-k-1,\MORD}(r)$ for $k=0,1,\dots$ can be explicitly expressed by 
\beqa\label{Eq2.12qq}
L_{-k-1,\MORD}(r) 
\, = \podr 
\lim_{\varepsilon \to 0} \,
\frac{1}{\MORD !}
\frac{d^{\MORD}}{d\varepsilon^{\MORD}} \Bigl( \vartheta(r)\, r^{\varepsilon - k - 1} - \frac{1}{\varepsilon} \, \frac{\delta^{(k)} (-r)}{k!} \Bigr) 
\nn
\, = \podr
\frac{(-1)^{k}}{k!} \left( \frac{d}{dr} \right)^{k + 1} \ \sum_{\MORd=0}^{\MORD+1} \sigma_{k\MORd} \, L_{0\MORd} (r) \,,
\eeqa
where $L_{0\MORd} (r)=\frac{1}{\MORd !}\vartheta(r)(\ln r)^{\MORd}$ are (integrable) powers of $\ln$ and the constants $\sigma_{k\MORd}$ are, in fact, determined by:
$$
\sigma_{k 0} = 1  \, , \qquad 
\sigma_{0\MORd} = 0 \quad (\MORd = 1,\ldots , \MORD+1) \,, 
$$
$$
\sigma_{k\MORd} =
\sigma_{k-1,\MORd} + \frac{\sigma_{k,\MORd-1}}{k} = 
\sum_{1 \leqslant j_1 \leqslant \ldots \leqslant j_{\MORd} \leqslant k} \frac{1}{j_1 \ldots j_{\MORd}} 
$$
for $k = 1,2,\ldots$, $\MORd=1,2,\dots,\MORD+1$.

\medskp

\begin{Statement}{Proposition}\label{ProN1}
The distributions $L_{\MDGR,\MORD} (r)$ defined for $\MDGR \in \Z$, $\MORD=0,1,\dots$ by Eqs.~$(\ref{Eq2n.0})$ and $(\ref{Lkn})$ have the properties
\begin{LIST}{26pt}
\item[$(a)$]
$L_{\MDGR,\MORD}(r)= \frac{1}{\MORD !} \, \vartheta(r) \, r^{\MDGR} (\ln r)^{\MORD}$ for $r \neq 0$ and all $\MDGR \in \Z$, $\MORD=0,1,\dots;$
\item[$(b)$]
$(\MEUL_{(r)} - \MDGR)\, L_{\MDGR,\MORD} (r) = L_{\MDGR,\MORD-1} (r)$ for $\MORD=1,2,\dots$ and $(\MEUL_{(r)} - \MDGR)\, L_{\MDGR,0} (r)=0$ for $\MDGR \neq -1,-2,\dots$, 
while 
$(\MEUL_{(r)} +k+1)\, L_{-1-k,0} (r)=\frac{(-1)^k}{k!} \, \delta^{(k)}(r)$ for $k=0,1,\dots$
$(\MEUL_{(r)}=r\frac{d}{dr}$ and in particular, $L_{\MDGR,\MORD}(r)$ are associate ho\-mo\-ge\-ne\-o\-us$);$
\item[$(c)$]
$r L_{\MDGR,\MORD}(r)=L_{\MDGR+1,\MORD}$ for all $\MDGR \in \Z$, $\MORD=0,1,\dots;$
\item[$(d)$]
$\frac{d}{dr} L_{\MDGR,\MORD} (r) = \MDGR L_{\MDGR-1,\MORD}+L_{\MDGR-1,\MORD-1}$ for all $\MDGR \in \Z$, $\MORD=1,2,\dots$.
\end{LIST}

Conversely, the first two properties $(a)$--$(b)$ uniquely determine the system of distributions $L_{\MDGR,\MORD}(r)$.
\end{Statement}

\medskp

\noindent
{\it Proof.}
Property ({\it a}) is satisfied by (\ref{Eq2n.0}).
Properties ({\it b})-({\it d}) follow in a straightforward way for $\MDGR > -1$
Property ({\it c}) for $\MDGR = -k-1 \leqslant -1$ 
($k=0,1,\dots$) 
follows from
$$
r \, \tlub{-k-1+\varepsilon} \, = \, \tlub{-k+\varepsilon}
$$
for the distribution valued meromorphic functions $\tlub{-k-1+\varepsilon}$ and $\tlub{-k+\varepsilon}$ and their Laurent expansions according to (\ref{Eq2n.0}).
Similarly, ({\it b}) and ({\it d}) follow for integral $\MDGR = -k-1 \leqslant -1$ from
\(
r \, \frac{d}{dr} \Bigl( \tlub{-k-1+\varepsilon} \Bigr) 
=
-(k+1-\varepsilon) \, \tlub{-k-1+\varepsilon}
\) and \(
\frac{d}{dr} \Bigl( \tlub{-k-1+\varepsilon} \Bigr) 
=
-(k+1-\varepsilon) \, \tlub{-k-2+\varepsilon},
\)
respectively,
and the comparison of the Laurent expansions of both sides in $\varepsilon$
according to Eq. (\ref{Eq2n.0}).

Regarding the second part of the proposition, let us assume that there is a second such system of distributions $L_{\MDGR,n}(r)'$. The difference $L_{\MDGR,\MORD}(r)-L_{\MDGR,\MORD}'(r)$ is an associate homogeneous distribution of degree $\MDGR$ supported at zero, according to ({\it b}) and ({\it a}).
Hence, $L_{\MDGR,\MORD}(r)-L_{\MDGR,\MORD}'(r)=0$ for $\MDGR >-1$ 
and $L_{-k-1,\MORD}(r)-L_{-k-1,\MORD}'(r)=C_{k,n}\delta^{(k)}(r)$ for $k=0,1,\dots$.
From ({\it b}) it follows that
$0=(\MEUL_{(r)} -k-1) C_{k,\MORD+1}\delta^{(k)}(r) = C_{k,\MORD}\delta^{(k)}(r)$ and so, $C_{k,\MORD}=0$ for all $n \geqslant 0$.$\quad\Box$

\medskp

Let us now consider the distribution $\tlub{-k-1+\varepsilon}\bigl(\ln r \bigr)^n$ on the semiaxis $r>0$ for $k \in \Z$. 
For a non-integer $-k-1+\varepsilon$ it is uniquely extendable to a distribution on $\R$ that we again denote with $\tlub{-k-1+\varepsilon}\bigl(\ln r \bigr)^n$.
The uniqueness follows by the same arguments used in the proof of Proposition \ref{PeEXT}.
The existence of $\tlub{-k-1+\varepsilon}\bigl(\ln r \bigr)^m$ on $\R$ is a consequence of Eqs. (\ref{Eq2n.0}) and (\ref{Lkn}) as it 
can be expanded as
\beqa\label{Eq2n.0-1}
\podr
\tlub{-k-1+\varepsilon}
(\ln r)^m
\nnb \podr 
= \,
\frac{(-1)^{k+m}m!}{k!} \, \dltan{k} (r) \, \frac{1}{\varepsilon^{m+1}}
\, + \,
\mathop{\sum}\limits_{\MORD \, = \, 0}^{\infty} \, \frac{1}{\MORD !} \, \tlr{-k-1}{m+\MORD} \, \varepsilon^{\MORD}
\,.
\eeqa
In particular, it follows that the distribution $\tlub{-k-1+\varepsilon}\bigl(\ln r \bigr)^m$ depends meromorphically on $\varepsilon$.

\subsection{Constructing extensions of associate homogeneous distributions. Re\-si\-du\-es}\label{Se3.6GGG}

Let us combine the technique of analytic regularization and renormalization of the previous subsection with the extension formula (\ref{MDEC2}).
To this end let us assume that $\MUDZ(\xx)$ is an associate homogeneous distribution on $\R^N \backslash \{0\}$ of degree $-N-k$ for an integer $k \geqslant 0$.
Then by Eqs. (\ref{MDEC2}) and (\ref{Eq2n.0-1}) we have:
\beqa\label{PRIMAP-constr}
\podr
\MCnt{\MUD (\xx) \, \MVOL}{f(\xx)}{\xx} 
\nnb \podr
= \,
\mathop{\sum}\limits_{\MORd \, = \, 0}^{\MORD} \, 
\frac{1}{\MORd!} \,
\MCNt{(r^{-k-1} \, \bigl(\ln r\bigr)^{\MORd})_+\, dr}{
\MCnt{\MUHZ_{\MHYP,\MORd} (\uu)\, \MVOL_{\MHYP}}{f(r\uu)}{\uu}
}{r}
\bnn \podr
= \,
\mathop{\sum}\limits_{\MORd \, = \, 0}^{\MORD} \, 
\frac{1}{\MORd!} \,
\RMCNt{
\mathop{\lim}\limits_{\varepsilon \, \to \, 0}
\Bigl(
\vartheta(r)
r^{-k-1+\varepsilon} \, \bigl(\ln r\bigr)^{\MORd}
\bnn \podr
\phantom{= \,
\mathop{\sum}\limits_{\MORd \, = \, 0}^{\MORD} 
\frac{1}{\MORd!} \,
\ }
-
\frac{(-1)^{k+m}\,m!}{k!\,\varepsilon^{m+1}} \,
\dltan{k} (r)
\Bigr)
\, dr}
\LMCNt{
\MCnt{\MUHZ_{\MHYP,\MORd} (\uu)\, \MVOL_{\MHYP}}{f(r\uu)}{\uu}
}{r}
\,.
\eeqa
Now, we observe that for noninteger $\varepsilon$ the expression \\
\(
\mathop{\sum}\limits_{\MORd \, = \, 0}^{\MORD} \, 
\MCNt{\vartheta(r)r^{-k-1+\varepsilon} \, \bigl(\ln r\bigr)^{\MORd}\, dr}{
\MCnt{\MUHZ_{\MHYP,\MORd} (\uu)\, \MVOL_{\MHYP}}{f(r\uu)}{\uu}
}{r}
\)
gives the unique extension on $\R^N$ 
of the associate homogeneous distribution
$\MRAD_{\MHYP} (\xx)^{\varepsilon}\,\MUDZ (\xx)$
(cf. Eqs. (\ref{MDEC1}), (\ref{UNNUMBERED})).
On the other hand, the second term in the above formula gives 
\beq\label{TEMP-Eq-4.3}
\Bigl(-\frac{d}{dr}\Bigr)^k f(r\uu) \vrestr{14pt}{r \, = \, 0}
\, = \,
(-1)^{k}
\mathop{\sum}\limits_{|\hspace{1pt}\rr\,| \, = \, k} \, 
\frac{k!}{\rr\hspace{1pt}!} \,
f^{(\rr)} (0) \, \uu^{\,\rr}
\,,
\eeq
where 
$f^{(\rr)} (\xx)$ $:=$ 
$\bigl(\frac{\di}{\di \xx}\bigr)^{\rr} f (\xx)$ $:=$ 
$\Bigl(\mathop{\prod}\limits_{a \, = \, 1}^N \bigl(\frac{\di}{\di x^a}\bigr)^{q_a}\Bigr) f(\xx)$,
$\uu^{\,\rr}$ $:=$
$\mathop{\prod}\limits_{a \, = \, 1}^N \bigl(u^a\bigr)^{q_a}$
and
$|\rr|$ $:=$ $q_1+\cdots+q_N$
are multi-index notations for $\rr = (q_1,\dots,q_N)$ $\in$ 
$\{0,1,\dots\}^{\times N}$.
Thus, for the resulting extension $\MUD(\xx)$ of $\MUDZ(\xx)$ we get the following result:

\medskp

\begin{Statement}{Theorem}\label{Th-extension-ann}
If 
$\MUDZ(\xx)$ is an associate homogeneous distribution on $\R^N \backslash \{0\}$ of degree $-N-k$ for an integer $k \geqslant 0$ then for $\varepsilon \in \C \backslash \Z$ the distribution
\(
\Bigl[\MRAD_{\MHYP} (\xx)^{\varepsilon}\,\MUDZ (\xx) \Bigr]
\raisebox{12pt}{\hspace{0pt}}^{\text{extended on } \R^N}
\)
is analytic and subtracting its singular part around $\varepsilon = 0$ we obtain an extension $\MUD(\xx)$ of $\MUDZ(\xx)$ on $\R^N$$:$\footnote{%
Similar formulas are used in \cite{K}, \cite{K10} under the name of ``dimensional regularization in position space''.} 
\beqa\label{AnSubtr}
\MUD (\xx) \podr = \, 
\mathop{\lim}\limits_{\varepsilon \, \to \, 0}
\,\Bigl(
\Bigl[\MRAD_{\MHYP} (\xx)^{\varepsilon}\,\MUDZ (\xx) \Bigr]
\raisebox{12pt}{\hspace{0pt}}^{\text{extended on } \R^N}
\nn \podr
- \,
\mathop{\sum}\limits_{\MORd \, = \, 0}^{\MORD} \, 
\frac{(-1)^{m}}{\varepsilon^{m+1}} \,
\mathop{\sum}\limits_{|\hspace{1pt}\rr\,| \, = \, k} \, 
\frac{1}{\rr\hspace{1pt}!} \,
\MCnt{\MUHZ_{\MHYP,\MORd} (\uu)\, \MVOL_{\MHYP}}{(-\uu)^{\,\rr}\,}{\uu}
\
\delta^{(\rr)} (\xx)
\Bigr)
\,
\qquad
\eeqa
$($note the appearance of $(-\uu)^{\hspace{1pt}\rr}=(-1)^k \, \uu^{\hspace{1pt}\rr}$ due to the similar sign factor in Eq. $(\ref{TEMP-Eq-4.3})$$)$.
\end{Statement}

\medskp

Next, 
let us study in more details the structure of the singular part in $\varepsilon$ in Eq. (\ref{AnSubtr}) of the above theorem.
For this purpose, we denote the distribution coefficient to $1/\varepsilon$ in (\ref{AnSubtr}) by $\Res_{\MHYP} (\MUDZ) (\xx)$
and we call it the {\bf analytic residue} of $\MUDZ$ with respect to the hyper-surface $\MHYP$.
Thus, it can be written as 
\beq\label{Res-def1}
\Res_{\MHYP} \bigl(\MUDZ\bigr) (\xx) \, : = \,
\mathop{\sum}\limits_{\rr \, \in \, \{0,1,\dots\}^{\times N}} \, 
\RES_{\MHYP} 
\Bigl(
\frac{(-\xx)^{\hspace{1pt}\rr}}{\rr\hspace{1pt}!} \, \MUDZ (\xx) 
\Bigr)
\,
\delta^{(\rr)} (\xx) 
\eeq
by means of a linear functional $\RES_{\MHYP}$ on $\DP \bigl(\R^N \backslash \{0\}\bigr)$ defined by
\beq\label{RES-def1}
\RES_{\MHYP}
(
\MUDZ
)
\, := \,
\left\{
\begin{array}{rl}
\MCnt{\MUDZ\vrestr{10pt}{\MHYP}\hspace{-1pt} (\uu)\, \MVOL_{\MHYP}}{1\,}{\uu}
\quad & \text{if $\MUDZ$ is associate homogeneous}
\\
\quad & \text{of degree $-N$ and 
{\ORder}
zero},
\\[9pt]
0 \quad & \text{otherwise.}
\end{array}
\right.
\,
\eeq
Indeed,
one obtains first that $\Res_{\MHYP} (\MUDZ)$ $=$
\(
\mathop{\sum}\limits_{|\hspace{1pt}\rr\,| \, = \, k} \, 
\frac{1}{\rr\hspace{1pt}!} \,
\MCnt{(-\uu)^{\,\rr} \, \MUHZ_{\MHYP,0} (\uu)\, \MVOL_{\MHYP}}{1\,}{\uu}
\) \(
\delta^{(\rr)} (\xx)
\)
provided that the degree of associate homogeneity of $\MUDZ$  is $-N-k$
and since 
\(
\bigr(\xx^{\,\rr} \, \MUDZ \bigl) \vrestr{10.5pt}{\hspace{-1pt}\MHYP} = \uu^{\,\rr} \, \MUHZ_{\MHYP,0} (\uu)
\)
(cf. Eq. (\ref{restriction-on-mhyp}))
then one observes that the expression for $\Res_{\MHYP}(\MUDZ)$ uses the {\it graded} linear functional
\beq\label{RES-def1a}
\RES_{\MHYP} : 
\DP_{\BLT} \bigl(\R^N \backslash \{0\}\bigr) \, \longrightarrow \,
\R \,,
\eeq
defined by Eq. (\ref{RES-def1}).
The grading on $\DP \bigl(\R^N \backslash \{0\}\bigr)$, which we shall introduce below, is particularly important as it makes finite the sum in (\ref{Res-def1}) for a particular distribution $\MUDZ$:
the possible nonzero terms are only for $|\rr|=k$ if the degree of associate homogeneity of $\MUDZ$ is $-N-k$ ($k=0,1,\dots$).

\medskp

We 
now endow the vector spaces of associate homogeneous distributions with the grading provided by the degree of associate homogeneity in the important nontrivial case of integral degrees
(cf. Remark~\ref{Rm-4.4-1} below).
These are $\Z$--graded vector spaces: 
\beqa
\label{PRIMAP-def2} &
\DP_{\BLT} (\R^N)
\, := \, \mathop{\text{\large $\textstyle \bigoplus$}}\limits_{k \, \in \, \Z}
\DP_k (\R^N)
\,,\quad
\DP_{\BLT} \bigl(\R^N \backslash \{0\}\bigr)
\, := \, \mathop{\text{\large $\textstyle \bigoplus$}}\limits_{k \, \in \, \Z}
\DP_k \bigl(\R^N \backslash \{0\}\bigr)
\,,
& \qquad
\eeqa
where the graded peaces
$\DP_k(\R^N)$
and
$\DP_k \bigl(\R^N \backslash \{0\}\bigr)$
consist of all associate homogeneous distributions of degree $k$ belonging to
$\DP (\R^N)$
and
$\DP \bigl(\R^N \backslash \{0\}\bigr)$,
respectively.
Thus, the linear functional $\RES_{\MHYP}$ (\ref{RES-def1})--(\ref{RES-def1a}) has a grading degree $-N$, while the linear map $\Res_{\MHYP}$ (\ref{Res-def1}) is a grading preserving map
\beqa\label{Res-def0}
\podr 
\Res_{\MHYP} :
\DP_{\BLT} \bigl(\R^N \backslash \{0\}\bigr) \, \longrightarrow \,
\DP_{\BLT} [\ZERN] \,,
\\[5pt] \nonumber
\podr
\DP_{\BLT} [\ZERN]
\, := \, \mathop{\text{\large $\textstyle \bigoplus$}}\limits_{k \, \in \, \Z}
\DP_k [\ZERN]
\,,\quad
\DP_k [\ZERN] \, := \,
\bigl\{\MUD \in \DP_k \bigl(\R^N\bigr) \, \bigl| \, 
\text{supp} \ \MUD \, \subseteq \, \{0\}\bigr\}
\,,
\eeqa
where $\DP_{\BLT} [\ZERN]$ stands the vector space of distribution supported at the origin 
$0 \equiv \ZERN$ 
in $\R^N$, which is graded again by the degree of homogeneity.
Note that $\DP_{\BLT}(\R^N \backslash \{0\})$ is actually a bi-graded vector space if we consider in addition to the degree of associate homogeneity also the 
{\ORder}
of the associate homogeneous distributions (as defined in Sect.~\ref{SECT-3.1-newww}).
The latter is used in the definition of $\RES_{\MHYP}$ (\ref{RES-def1}) as a bi--graded linear functional.

\medskp

Using 
the above definitions
and again Eq. (\ref{restriction-on-mhyp}) we can rewrite Eq. (\ref{AnSubtr}) as
\beqa\label{AnSubtr-RES}
\MUD (\xx) \podr = \, 
\mathop{\lim}\limits_{\varepsilon \, \to \, 0}
\,\Bigl(
\Bigl[\MRAD_{\MHYP} (\xx)^{\varepsilon}\,\MUDZ (\xx) \Bigr]
\raisebox{12pt}{\hspace{0pt}}^{\text{extended on } \R^N}
\nn \podr
- \,
\mathop{\sum}\limits_{\MORd \, = \, 0}^{\MORD} \, 
\frac{(-1)^{m}}{\varepsilon^{m+1}} \,
\Res_{\MHYP}\bigl((\MEUL-\MDGR)^{\MORd} \, \MUDZ\bigr) (\xx)
\Bigr)
\,,
\qquad
\eeqa
where 
$\MUDZ \in \DP_{\MDGR} (\R^N\backslash\{0\})$ and
$\MEUL := \MEUL_{(\xx)} \equiv \xx \cdot \frac{\partial}{\partial \xx}$ 
is the Euler vector field.

\medskp

Now consider 
the important special case of a homogeneous distribution $\MUDZ (\xx)$
(i.e., $\MUHZ_{\MHYP,\MORd} (\uu)$ $=$ $0$ for $\MORd > 0$),
which is a locally integrable function on $\R^N \backslash \{0\}$.
In massless QFT we have such a situation for the so called {\it primitively divergent} Feynman amplitudes and we shall consider such an example in Appendix~\ref{Spokes}.
Thus, the restriction $\MUHZ_{\MHYP,0} (\uu)$ $\equiv$ $\MUHZ (\xx)\vrestr{10pt}{\MHYP}$ 
($\xx$ $\equiv$ $\uu$ $\in$ $\MHYP$)
on the hyper-surface $\MHYP$ is also a locally integrable function.
Since $\MUDZ (\xx)$ is homogeneous, i.e., its (scaling) order $\MORD$ is zero,
then Eq. (\ref{AnSubtr-RES}) 
reads
\beqa\label{AnSubtr-1}
\podr 
\MUD (\xx) 
\,
= \, 
\mathop{\lim}\limits_{\varepsilon \, \to \, 0}
\,\Biggl\{
\Bigl[\MRAD_{\MHYP} (\xx)^{\varepsilon}\,\MUDZ (\xx) \Bigr]
\raisebox{12pt}{\hspace{0pt}}^{\text{extended on } \R^N}
- \,
\frac{1}{\varepsilon} \,
\Res_{\MHYP}\bigl(\MUDZ\bigr)(\xx)
\biggr\}
\qquad
\\
\podr \text{with} \quad 
\Res_{\MHYP}\bigl(\MUDZ\bigr)(\xx) \, = \,
\mathop{\sum}\limits_{|\hspace{1pt}\rr\,| \, = \, k} \, 
\biggl(
\mathop{\int}\limits_{\hspace{-5pt}\MHYP}
\frac{(-\xx)^{\hspace{1pt}\rr}}{\rr\hspace{1pt}!} \,
\MUDZ (\xx) 
\, 
\nnb \podr
\times \,
\Bigl(
\mathop{\sum}
\limits_{j \, = \, 1}^{N} (-1)^{j-1} \, x^j \, d x^1 \wedge \cdots \wedge \widehat{dx^j} \wedge \cdots \wedge dx^N
\Bigr)
\Biggr)
\delta^{(\rr)} (\xx)
\,\quad
\eeqa
(cf. Eq. (\ref{Kron-f})).
The integral that appears in (\ref{AnSubtr-1}) expresses in this case the linear functional $\RES_{\MHYP}$,
\beqa\label{W-RES}
\podr
\RES_{\MHYP} 
(\FFU)
\bnn
\podr = \,
\left\{
\begin{array}{l}
0 \hspace{60pt}\qquad\qquad \text{if $F$ is not homogeneous of degree $-N$}
\\[9pt]
\mathop{\text{\Large $\textstyle \int$}}\limits_{\hspace{-5pt}\MHYP}
\FFU (\xx)
\, \Bigl(
\mathop{\sum}
\limits_{j \, = \, 1}^{N} (-1)^{j-1} \, x^j \, d x^1 \wedge \cdots \wedge \widehat{dx^j} \wedge \cdots \wedge dx^N
\Bigr)
\\
\phantom{0} \hspace{60pt}\qquad\qquad \text{if $F$ is homogeneous of degree $-N$},
\end{array}
\right.
\eeqa
for a locally integrable function $\FFU (\xx)$ on $\R^{N} \backslash \{0\}$.
It is a special case of the {\it Wodzicki residue}
(see e.g. \cite{Ka,Sch} 
and references therein).
Furthermore, the {\rm residue} functional defined by Eq. $(\ref{W-RES})$ does not depend on the choice of the hyper-surface $\MHYP$.
Indeed,
the integrand in (\ref{W-RES}) is a {\it closed} $N-1$ form in the case when 
$\FFU (\xx)$ is a homogeneous (locally integrable) function on $\R^N\backslash \{0\}$.
This is the reason why the integral in (\ref{W-RES}) does not depend on the used hyper-surface~$\MHYP$.


\begin{remark}\label{Rm-Horm-4}
We 
note that Eq. (\ref{AnSubtr-RES}) takes the form (\ref{AnSubtr-1}) for every homogeneous distribution $\MUDZ$ on $\R^N\backslash\{0\}$ (i.e., associate homogeneous of order zero).
We shall prove later in Corollary \ref{COR-new-4X} that in this case $\RES_{\MHYP} (\MUDZ)$ (and hence, $\Res_{\MHYP} (\MUDZ) (\xx)$) is again independent of $\MHYP$.
However, for associate homogeneous distributions of higher order $\MORD$ the term in the singular part of Eq. (\ref{AnSubtr-RES}) that is independent of $\MHYP$ is the most singular nonzero term
(cf. Corollary~\ref{COR-new-4X}).
\end{remark}


Let us stress the special correspondence between the linear map $\Res_{\MHYP}$ (\ref{Res-def1}) and the linear functional $\RES_{\MHYP}$ (\ref{RES-def1}).
It is due to more general decompositions described below (\cite{N}).
Let us introduce first the situation in which these decompositions appear.
Let $\MODUL$ be a $\Z$--graded vector space endowed with an action of the polynomial algebra $\R[\xx]$. 
We say also that $\MODUL$ is an $\R[\xx]$--module.
In the application to the map $\Res$ we will have $\MODUL=\DP_{\BLT} (\R^N \backslash \{0\})$
and the action of a polynomial $p \in \R[\xx]$ on $\MODUL$ is via a multiplication by $p$.
Let us assume that the action of a polynomial $p \in \R[\xx]$ on $\MODUL$
has a degree equal to the degree of homogeneity of $p$.
We say also that $\MODUL$ is a $\Z$--graded $\R[\xx]$--module.
Of course, the latter condition is satisfied in the example
$\MODUL=\DP_{\BLT} (\R^N \backslash \{0\})$.

\medskp

\begin{Statement}{Theorem}\label{Th-4.6-IMP}
$\cite{N}$
Let us have, as above, a $\Z$--graded $\R[\xx]$--module $\MODUL$ and~let
$$
\gamma : \MODUL \, \rightarrow \, \DP_{\BLT} [\ZERN]
$$ 
be a linear map of degree $k$ $\in$ $\Z$.
Then $\gamma$ has
the property
\beq\label{p-MULT}
\gamma (p \, m) \, = \, p \, \gamma(m) \quad
(\forall p \in \R[\xx] \quad\text{and}\quad \forall m \in \MODUL)
\eeq
$($in other words, $\gamma$ is a $\R[\xx]$--module map of degree $k$, i.e., 
$\gamma \in \text{\rm Hom}_{\R[\xx]}^k \bigl(\MODUL,$ $\DP_{\BLT}[\ZERN]\bigr)$$)$
iff $\gamma$ has the following $($unique$)$ decomposition$:$
\beq\label{DECOM-1}
\gamma (m) (\xx) \, = \,
\mathop{\sum}\limits_{\rr \, \in \, \{0,1,\dots\}^{\times N}} \, 
\Gamma 
\Bigl(
\frac{(-\xx)^{\hspace{1pt}\rr}}{\rr\hspace{1pt}!} \, m 
\Bigr)
\,
\delta^{(\rr)} (\xx)
\,
\eeq
for a linear functional $\Gamma : \MODUL \to \R$ of degree $k$
$($considering $\R$ as a $\Z$--graded vector space with a non-zero piece at degree zero$)$.
\end{Statement}

\medskp

\noindent
{\it Proof.}
In general, we have a decomposition
\beq\label{DECOM-2}
\gamma (m) (\xx) \, = \,
\mathop{\sum}\limits_{\rr \, \in \, \{0,1,\dots\}^{\times N}} \, 
\frac{(-1)^{|\rr|}}{\rr\hspace{1pt}!} \
\Gamma_{\rr} 
(m)
\,
\delta^{(\rr)} (\xx)
\,,
\eeq
where $\Gamma_{\rr} : \MODUL \to \R$ are linear functionals of degree $k-|\rr|$ and the pre-factors $\frac{(-1)^{|\rr|}}{\rr\hspace{1pt}!}$ are introduced for convenience.
Then, the condition (\ref{p-MULT}) is equivalent to the condition
\beq\label{p-MULT1}
\gamma (x^a \, m) \, = \, x^a \, \gamma(m) \quad
(\forall a = 1,\dots,N \quad\text{and}\quad \forall m \in \MODUL) \,,
\eeq
which in view of (\ref{DECOM-2}) reads
$$
\Gamma_{\rr \, + \, \ee_a} (m) \, = \, \Gamma_{\rr} (x^a \, m) \, 
$$
($\ee_a$ being the $a$th basic vector in $\R^N$).
We thus conclude that the property (\ref{p-MULT}) holds iff in the decomposition (\ref{DECOM-2}) one has 
$\Gamma_{\rr} (m) \, = \, \Gamma_{0} (\xx^{\hspace{1pt}\rr} \, m)$.
But the latter is exactly (\ref{DECOM-1}) with $\Gamma = \Gamma_0$.$\quad\Box$

\subsection{Primary renormalization maps and the dilation anomaly}\label{Se4.4NEWWW}

Equation 
(\ref{AnSubtr-RES}) (or (\ref{PRIMAP-constr})) defines a {\it grading preserving} linear map 
\beq\label{PRIMAP-MHYP}
\PRIMAP^{\MHYP} : \,
\MUDZ \, \longmapsto \, \MUD : \,
\DP_{\BLT} \bigl(\R^N \backslash \{0\}\bigr)
\, \longrightarrow \,
\DP_{\BLT} (\R^N)\,,
\eeq
for every hyper-surface $\MHYP \subset \R^N \backslash \{0\}$ that encircles the origin.
An equivalent expression for $\PRIMAP^{\MHYP}$ in terms of Eq.~(\ref{PRIMAP-constr})~is:
\beq\label{SECDEF}
\MCnt{\PRIMAP^{\MHYP} (\MUDZ ) \, \MVOL}{f}{} 
= \hspace{-3pt} \mathop{\sum}\limits_{\MORd \, = \, 0}^{\infty} \,
\frac{1}{\MORd!} \,
\MCNt{(r^{-k-1} \, \bigl(\ln r\bigr)^{\MORd})_+ dr}{
\MCnt{\MUHZ_{\MHYP,\MORd} (\uu) \MVOL_{\MHYP}}{f(r\uu)}{\uu}
}{r}
\hspace{1pt}
\eeq
(provided that the degree of the associate homogeneous distribution $\MUDZ$ is $-N-k$;
we remind also that the sequence $(\MUHZ_{\MHYP,\MORd})_{m \, = \, 0}^{\infty}$ is defined by Theorem \ref{THM-EXT} applied to $\MUDZ$).
We assume that for $k < 0$ the whole sums in the second term of the right hand side of Eq. (\ref{SECDEF}) vanish 
(to this end one may set $\delta^{(\rr)}(\xx)=0$ iff $\rr \notin \{0,1,\dots\}^{\times N}$). 

\medskp

\begin{Statement}{Theorem}\label{Lemma-C.4-NN}
Every linear map 
$\PRIMAP := \PRIMAP^{\MHYP}$
$(\ref{PRIMAP-MHYP})$ satisfies the following properties:
\begin{LIST}{26pt}
\item[$(p_0)$] {\it Extension.} \\
For every $\MUDZ \in \DP_{\BLT} \bigl(\R^N \backslash \{0\}\bigr)$ we have
$\PRIMAP \bigl(\MUDZ\bigr) \vrestr{11pt}{\R^N \backslash \{0\}}$
$=$
$\MUDZ$.
\item[$(p_1)$] {\it Scaling.} \\
$\PRIMAP$ are grading preserving.
\item[$(p_2)$] {\it $\text{\rm GL}(N)$-equivariance.} \\
For every linear transformation $\Lambda \in \text{\rm GL}(N)$ and every $\MUDZ \in \DP_{\BLT} \bigl(\R^N \backslash \{0\}\bigr)$ we have \
$\PRIMAP (\Lambda^* \hspace{1pt} \MUDZ)$ $=$ $\Lambda^* \hspace{1pt} \PRIMAP' (\MUDZ)$,
where $\PRIMAP' := \PRIMAP^{\MHYP'}$, with
$\MHYP' = \Lambda^{-1} \MHYP$.
\item[$(p_3)$] {\it Commutativity with multiplication by polynomials.} \\
$\PRIMAP (p \hspace{1pt}\MUDZ)$ $=$ $p \hspace{1pt}\PRIMAP (\MUDZ)$ 
for every $($complex$)$ polynomial $p$ on $\R^N$ and every $\MUDZ \in \DP_{\BLT} \bigl(\R^N \backslash \{0\}\bigr)$.
\end{LIST}
\end{Statement}

\medskp

\begin{Proof}
The verifications of conditions $(p_0)$, $(p_1)$ and $(p_2)$ is straightforward from the construction.
To prove $(p_3)$ it is sufficient to verify it for the multiplication by coordinates,
$\PRIMAP^{\MHYP} (x^a \hspace{1pt}\MUDZ)$ $=$ $x^a \hspace{1pt}\PRIMAP^{\MHYP} (\MUDZ)$, 
which follows from
Eq.~(\ref{SECDEF})
applied to~$x^a \MUDZ$
(of a scaling degree $-N-k+1$):
\beqs
\podr
\MCnt{\PRIMAP^{\MHYP} (x^a\MUDZ ) \, \MVOL}{f}{} 
\\ \podr
= \hspace{-3pt} \mathop{\textstyle \sum}\limits_{\MORd \, = \, 0}^{\infty} 
\ \frac{1}{\MORd!} \,
\MCNt{(r^{-k} \, \bigl(\ln r\bigr)^{\MORd})_+ dr}{
\MCnt{\widehat{\bigl(x^a\MUDZ\bigr)}_{\MHYP,\MORd} (\uu) \MVOL_{\MHYP}}{f(r\uu)}{\uu}
}{r}
\\ \podr
= \hspace{-3pt} \mathop{\textstyle \sum}\limits_{\MORd \, = \, 0}^{\infty} 
\ \frac{1}{\MORd!} \,
\MCNt{(r^{-k} \, \bigl(\ln r\bigr)^{\MORd})_+ dr}{
\MCnt{\MUHZ_{\MHYP,\MORd} (\uu) \MVOL_{\MHYP}}{u^a f(r\uu)}{\uu}
}{r} \,,
\\ \podr
\MCnt{\PRIMAP^{\MHYP} (\MUDZ ) \, \MVOL}{x^a f}{} 
\\ \podr
= \hspace{-3pt} \mathop{\textstyle \sum}\limits_{\MORd \, = \, 0}^{\infty} 
\ \frac{1}{\MORd!} \,
\MCNt{(r^{-k-1} \, \bigl(\ln r\bigr)^{\MORd})_+ dr}{
\MCnt{\MUHZ_{\MHYP,\MORd} (\uu) \MVOL_{\MHYP}}{ru^a f(r\uu)}{\uu}
}{r}
\eeqs
(since $x^a\vrestr{10pt}{\MHYP} = u^a$ by definition).
So, the equality $\PRIMAP^{\MHYP} (x^a \hspace{1pt}\MUDZ)$ $=$ $x^a \hspace{1pt}\PRIMAP^{\MHYP} (\MUDZ)$ will follow from the identity
$(r^{-k} \, \bigl(\ln r\bigr)^{\MORd})_+$
$=$
$r(r^{-k-1} \, \bigl(\ln r\bigr)^{\MORd})_+$.
But the latter is proven in Proposition~\ref{ProN1} $(c)$.
\end{Proof}

\medskp

Thus, 
we are led to consider extensions, as linear procedures, i.e., as linear maps starting from certain vector spaces of ``unrenormalized functions'' to spaces of globally defined distribution (\cite{Ni}).
Here we first introduce this notion in the case of extensions of associate homogeneous distributions defined outside the origin and later in Sect. \ref{NSEC-4.2NN} we shall extend these ideas to the case of Feynman amplitudes.  


\begin{definition}\label{Df-4.4-1}
A {\bf primary renormalization map} over $\R^N$
is a linear map
\(
\PRIMAP : 
\DP_{\BLT} \bigl(\R^N \backslash \{0\}\bigr)
\to
\DP_{\BLT} (\R^N)
\)
that satisfies properties $(p_0)$, $(p_1)$ and $(p_3)$ of Theorem \ref{Lemma-C.4-NN}.
The primary renormalization map $\PRIMAP$ is called {\bf $O(N)$--equivariant} iff in addition it satisfies 
$\PRIMAP (\Lambda^* \hspace{1pt} \MUDZ)$ $=$ $\Lambda^* \hspace{1pt} \PRIMAP' (\MUDZ)$ for all $\Lambda \in O(N)$.
Thus, by Theorem \ref{Lemma-C.4-NN}
the map $\PRIMAP^{\MHYP}$ is an $O(N)$--equivariant primary renormalization map if $\MHYP \subset \R^N$ is a sphere centered at the origin. 
\end{definition}


\begin{remark}\label{RM-p35-NEW}
Due to $(p_0)$ the operators $\PRIMAP$ define extension maps.
The linear maps $\PRIMAP$ will be basic ingredients in the construction
of the renormalization maps $\RENMAP_n$ in Sect.~\ref{NSEC-4.2NN}. 
We shall see that the conditions of Definition \ref{Df-4.4-1} for an $O(N)$ covariant primary renormalization
map correspond directly to the properties of $\RENMAP_n$ spelled out in Theorem \ref{Theorem4.2}. 
In particular, condition $(p_2)$, which may look at first sight artificial, 
will combine $O(D)$ invariance of euclidean Feynman amplitudes
with permutation symmetry (see 
Appendix~\ref{Unif-sect}). 
Condition $(p_3)$ will be necessary in order to obtain linearity
for the renormalization maps $\RENMAP_n$ (cf. Lemmas \ref{SECMAP-constr} and \ref{LEM-C2}). 
As demonstrated by Theorem 
\ref{Th-4.6-IMP} 
above these conditions strongly restrict the ambiguity in the extension maps.
\end{remark}


\begin{remark}\label{Rm-4.4-1}
One can consider primary renormalization maps $\PRIMAP$ on the $\R$--graded vector space spanned by associate homogeneous distributions of arbitrary real (or even complex) degree of homogeneity.
However, due to Proposition \ref{PeEXT} the maps $\PRIMAP$ will be uniquely defined on
associate homogeneous distributions of non-integer degree of homogeneity and hence,
the 
non-trivial and ambiguous parts of $\PRIMAP$ remain 
those considered above.
\end{remark}


The dilation anomaly characterizes the increase of the 
{\ORder}
of the associate homogeneous distributions under the extension process.
There are certain obstructions for the preservation of the 
{\ORder}
of an associate homogeneous distribution, which are independent of the ambiguity of the extension.
We shall consider these obstructions in terms of the renormalization maps~$\PRIMAP$.
It is natural to characterize the dilation anomaly by the commutator
\beq\label{gamma-N}
\bigl[\MEUL,\PRIMAP\bigr]
:
\DP_{\BLT} \bigl(\R^N \backslash \{0\}\bigr)
\, \longrightarrow \,
\DP_{\BLT} [0] \,,
\eeq
($\MEUL = \xx \cdot \frac{\partial}{\partial \xx}$ 
being the Euler vector field). 
Note that 
since $\PRIMAP$ is a linear map 
$\DP_{\BLT} (\R^N \backslash \{0\}) \to \DP_{\BLT} (\R^N)$
it follows that the commutator 
$[\MEUL,\PRIMAP]$ $\equiv$
$\MEUL\circ \PRIMAP$ $-$
$\PRIMAP\circ \MEUL$
is initially defined again as a linear map
$\DP_{\BLT} (\R^N \backslash \{0\}) \to \DP_{\BLT} (\R^N)$.
However, due to the extension property $(p_0)$ and the fact that $\MEUL$ is a local operator 
(i.e. $\MEUL$, as every vector field, commutes with taking restrictions on open domains)
it follows that the values of 
$[\MEUL,\PRIMAP]$
are distributions supported at the origin,
as it is stated in (\ref{gamma-N}).
To see explicitly how the commutator (\ref{gamma-N}) is related to the dilation anomaly
let us interpret it in terms of the dilation laws.
If we have a distribution 
$\MUDZ \in \DP_{\MDGR} (\R^N \backslash \{0\}) \to \DP_{\MDGR} (\R^N)$
and $\MUD := \PRIMAP (\MUDZ)$
(hence, $\MUDZ$ and $\MUD$ have one and the same degree $\MDGR$ of associate homogeneity)
the dilation laws described by Proposition~\ref{CrZZZ1} for $\MUDZ$ and $\MUD$
depend on the sequences of distributions
$\MUDZ_m (\xx)$ $:=$
$\bigl(\MEUL-\MDGR\bigr)^m \MUDZ(\xx)$
and
$\MUD_m (\xx)$ $:=$
$\bigl(\MEUL-\MDGR\bigr)^m \MUD(\xx)$, respectively.
On the other hand, we see that
\beqs
\MUD_{m+1} \podr = \, \PRIMAP \bigl(\MUDZ_{m+1}\bigr) + 
\bigl[\MEUL,\PRIMAP\bigr]
(\MUDZ_m) 
\quad (0 \leqslant m < n)
\,,
\quad
\\
\MUD_{\MORD+1} \podr = \, 
\bigl[\MEUL,\PRIMAP\bigr]
(\MUDZ_{\MORD}) \,,
\eeqs
where $\MORD$ is the 
{\ORder}
of the unextended distribution $\MUDZ$.
It follows that the 
{\ORder}
of $\MUDZ$ increases after the extension
$\MUDZ \mapsto \MUD$ 
iff 
\beq\label{INCR-ORD}
[\MEUL,\PRIMAP] (\MUDZ_{\MORD}) \, \equiv \,  
[\MEUL,\PRIMAP] \bigl((\MEUL-\MDGR)^{\MORD} \, \MUDZ\bigr)
\, \neq \, 0
\eeq
and the latter distribution is 
the coefficient of the highest
power of the logarithm appearing in the global dilation law for $\MUD$
(cf. Remark \ref{NewRem-3.1}).

Let us show now that the distribution (\ref{INCR-ORD}) is a renormalization invariant,
i.e., for a given $\MUDZ$ it does not depend on the used primary renormalization map.
To this end let us consider two primary renormalization maps
$\PRIMAP$ 
and
$\PRIMAP'$.
According to property $(p_1)$
the difference $\PRIMAP'-\PRIMAP$ is a linear map
$$
\PRIMAP'-\PRIMAP : 
\DP_{\BLT} \bigl(\R^N \backslash \{0\}\bigr)
\, \longrightarrow \,
\DP_{\BLT} [\ZERN]
\,,
$$
which is grading preserving.
However, 
$\MEUL$ is not in general the {\it grading operator} for 
$\DP_{\BLT} \bigl(\R^N \backslash \{0\}\bigr)$, i.e.,
the graded pieces $\DP_{k} \bigl(\R^N \backslash \{0\}\bigr)$ 
are not eigenspaces for $\MEUL$. 
This is only true in the case of homogeneous distributions belonging to 
$\DP \bigl(\R^N \backslash \{0\}\bigr)$, which happens for $\MUDZ_n$ according to the definition of the (scaling) order of an associate homogeneous distribution,
$(\MEUL-\MDGR) \, \MUDZ_n = (\MEUL-\MDGR)^{\MORD+1} \, \MUDZ=0$.
Thus,
\beq\label{ren-invar}
\bigl[\MEUL,\PRIMAP\bigr] (\MUDZ_n)
\, = \,
\bigl[\MEUL,\PRIMAP'\bigr] (\MUDZ_n) \,.
\eeq

\medskp

Let 
us stress that in general the commutator $[\MEUL,\PRIMAP]$ is not a renormalization invariant
(i.e., independent of $\PRIMAP$).
On the other hand, this commutator has a very important relationship to the previously introduced analytic residue $\RES_{\MHYP}$ (\ref{Res-def1}) according to the following:

\medskp

\begin{Statement}{Theorem}\label{Th4.4-2}
The commutator $(\ref{gamma-N})$ that characterizes the dilation anomaly coincides in the case of the primary renormalization map $\PRIMAP^{\MHYP}$ $(\ref{PRIMAP-MHYP})$ with the opposite of the analytic residue map~$(\ref{Res-def0})$$:$ 
\beq\label{COINCIDENCE}
- \Res_{\MHYP} \, = \, \bigl[\MEUL,\PRIMAP^{\MHYP}\bigr] \,.
\eeq
\end{Statement}

\medskp

\noindent
{\it Proof.}
It suffices to show that
$\Res_{\MHYP} (\MUDZ)$ $=$
$[\MEUL-\MDGR,\PRIMAP^{\MHYP}] (\MUDZ)$ for $\MUDZ \in \DP_{\MDGR}(\R^N$ $\backslash \{0\})$
and $\MDGR=-N-k$ with $k \in \{0,1,\dots\}$.
We first compute by Eq. (\ref{AnSubtr-RES}): 
\beqs
\podr
\PRIMAP^{\MHYP}\bigl((\MEUL-\MDGR) \, \MUDZ\bigr) \, = \, 
\mathop{\lim}\limits_{\varepsilon \, \to \, 0}
\,\Bigl\{
\Bigl[\MRAD_{\MHYP} (\xx)^{\varepsilon}\,(\MEUL-\MDGR)(\MUDZ) \Bigr]
\raisebox{12pt}{\hspace{0pt}}^{\text{extended on } \R^N}
\nnb
\podr
- \,
\mathop{\sum}\limits_{\MORd \, = \, 0}^{\infty} \, 
\frac{(-1)^{m}}{\varepsilon^{m+1}} \,
\Res_{\MHYP}\bigl((\MEUL-\MDGR)^{\MORd+1} \, \MUDZ\bigr) (\xx)
\Bigr\}
\,
\hspace{1pt}
\qquad
\\
\podr
= \, 
\mathop{\lim}\limits_{\varepsilon \, \to \, 0}
\,
(\MEUL-\MDGR-\varepsilon) \Bigl\{
\Bigl[\MRAD_{\MHYP} (\xx)^{\varepsilon}\,\MUDZ \Bigr]
\raisebox{12pt}{\hspace{0pt}}^{\text{extended on } \R^N}
\nnb
\podr
- \,
\mathop{\sum}\limits_{\MORd \, = \, 0}^{\infty} \, 
\frac{(-1)^{m}}{\varepsilon^{m+1}} \,
\Res_{\MHYP}\bigl((\MEUL-\MDGR)^{\MORd} \, \MUDZ\bigr) (\xx)
\Bigr\}
\, - \,
\Res_{\MHYP}\bigl(\MUDZ\bigr) (\xx)
\nnb
\podr
= \, 
(\MEUL-\MDGR) \, \bigl(\PRIMAP^{\MHYP}\bigl(\MUDZ\bigr)\bigr)
\, - \,
\Res_{\MHYP}\bigl(\MUDZ\bigr) (\xx)
\,
\hspace{1pt},
\qquad
\eeqs
where we used that: 
$(a)$
$\MEUL$ commutes with the operation of the unique extension for associate homogeneous distributions of non-integer degree of homogeneity;
$(b)$
all the distributions
$\Res_{\MHYP}\bigl((\MEUL_{(\xx)}-\MDGR)^{\MORd} \, \MUDZ\bigr) (\xx)$
are homogeneous of degree $\MDGR$ and thus,
$(\MEUL-\MDGR) \, \Res_{\MHYP}\bigl((\MEUL_{(\xx)}-\MDGR)^{\MORd} \, \MUDZ\bigr) (\xx) = 0$.
Hence, we obtain Eq. (\ref{COINCIDENCE}) which completes the proof of Theorem \ref{Th4.4-2}.$\quad\Box$

\medskp

Combining this theorem with the previous analysis of the invariance under the change in the primary renormalization maps we obtain:

\medskp

\begin{Statement}{Corollary}\label{COR-new-4X}
The 
highest nonzero pole subtraction in Eq. $(\ref{AnSubtr-RES})$ $($or, $(\ref{AnSubtr}))$ does not depend on the hyper-surface $\MHYP$.
For an associate homogeneous distribution $\MUDZ$ on $\R^N\backslash\{0\}$ of order $\MORD$ 
the coefficient to this pole $($that is of order $\MORD+1$$)$ is the distribution $\Res_{\MHYP} \bigl((\MEUL_{(\xx)}-\MDGR)^{\MORD} \, \MUDZ\bigr)$, which is supported at the origin and coincides with the coefficient to the highest logarithmic power $($that is of order $\MORD+1$$)$ in the global dilation law of the extensions of $\MUDZ$ on $\R^N$.
Another expression for the latter coefficient is given by Eq. $(\ref{INCR-ORD})$ and its non-vanishing is the necessary and sufficient condition for the increasing the order of the associate homogeneous distribution $\MUDZ$ under the extension on $\R^N$.
\end{Statement}

\medskp

To summarize, 
we have introduced two types of residues.
The first one, the {\it analytic residue} $\Res_{\MHYP}$ (\ref{Res-def1}) is related to the highest poles' subtraction in the analytic extension (renormalization) from $\R^N \backslash \{0\}$ to $\R^N$.
The second type of residue is the commutator $[\MEUL,\PRIMAP]$,
which can be called a {\it dilation residue}
as it is related to the change in the dilation law after the extension.
In particular, the vanishing of $[\MEUL,\PRIMAP]$ on the highest successor $\MUDZ_{\MORD}$ of an associate homogeneous distribution  $\MUDZ$ is the necessary and sufficient condition for the non-increasing of the 
{\ORder}
$\MORD$
of $\MUDZ$ after an extension from $\R^N \backslash \{0\}$ to $\R^N$
(that preserves the degree of associate homogeneity).
Thus, one can interpret Theorem \ref{Th4.4-2} as a coincidence of the analytic and dilation residues (up to a sign factor).
Note that we have not proven that all primary renormalization maps $\PRIMAP$ (in the sense of Definition \ref{Df-4.4-1}) are of the form $\PRIMAP^{\MHYP}$ for some $\MHYP$. Thus, the dilation residue is more general in the sense that it is defined for a (possibly) larger class of primary renormalization maps.
The reader may see in \ref{Spokes-0} an illustration of these results.


\section{A refined concept of divergence. Renormalization maps}\label{SEC-new5}
\setcntrs

\subsection{Renormalization of Feynman amplitudes as a linear procedure and its covariance}\label{NSEC-4.2NN}

Recall that renormalization of Feynman amplitudes is a recursive procedure that we explained in details in Sect. \ref{TSn1-N}.
However, this procedure was considered as a procedure on particular amplitudes.
For some purposes, in particular, preservation of symmetries, it is useful to introduce the renormalization as a system of linear maps $\{\RENMAP_n\}_{n \, = \, 2}^{\infty}$ from some vector spaces $\AMPSPA_n$ of unrenormalized Feynman amplitudes (to be defined below) to the spaces $\DP (\R^{Dn})$ of globally defined distributions of $n$ points ($n=2,3,\dots$).
These maps were introduced in 
\cite{Ni} and called {\bf renormalization maps}.
In fact, in Sect. \ref{Se4.4NEWWW} we have already introduced this point of view for the primary renormalization, that is the extension of distributions to the origin, which completes the renormalization recursion.

Since translation invariance can be preserved under renormalization (Corollary \ref{CrXX1}) we can set the target spaces of the renormalization maps to be
the spaces of translation invariant distributions on $\R^{Dn}$, i.e.,
\beq\label{DPn}
\DP_n \, := \, \DP_{\BLT}\biggl(\frac{\R^{Dn}}{\Delta_n}\biggr) \, \cong \, \DP_{\BLT}\bigl(\R^N\bigr) \,,\quad N \, := \, D (n-1) \,,
\eeq
where $\Delta_n$ $\cong$ $\R^D$
is the total diagonal in $\R^{Dn} \equiv (\R^D)^{\times \, n}$ 
and $(\cdot)_{\BLT}$ indicates the $\Z$--grading provided by the degree of associate homogeniety (cf. Eq. (\ref{PRIMAP-def2})).
The choice of the identification
$$
\frac{\textstyle \R^{Dn}}{\textstyle \Delta_n} \, \cong \, \R^N
$$
is 
important since it allows to incorporate the $\PERMGR_n$ permutation symmetry within the Euclidean $O(D)$ symmetry (cf. Eq. (\ref{IDENTIF}) below and Appendix~\ref{Unif-sect}).
Thus, we will define the renormalization maps as linear maps of the form
\beq\label{RENMAP}
\RENMAP_n : \AMPSPA_n \, \longrightarrow \, \DP_n \,.
\eeq
The space $\R^D$ above can serve for both, the $D$--dimensional Euclidean and Minkowski spaces.

\medskp

Let us now define the vector space $\AMPSPA_n$ of $n$--point unrenormalized Feynman amplitudes ($n=2,3,\dots$).
Before renormalization the Feynman amplitudes are just algebraic (or, analytic) expressions that can be added or, multiplied by scalars.
For the sake of definiteness we shall identify the vector space spanned by the $n$--point unrenormalized Feynman amplitudes with a subspace in some particular ambient space of functions.
In the Euclidean case the ambient space of functions 
can be chosen simply to be the space of all functions that are smooth outside the large diagonal~$\widetilde{\Delta}_n$:
\beq\label{EAMBSPA}
\CI \bigl(\R^{Dn}\bigr\backslash \widetilde{\Delta}_n\bigr)
\quad
(\widetilde{\Delta}_n \, = \, \{(\x_1,\dots,\x_n)| \exists j,k : j \neq k \text{ and } \x_j = \x_k\})
\,.
\eeq
The spaces $\R^{Dn}\bigr\backslash \widetilde{\Delta}_n$ are the spaces of pairwise distinct configurations of points and are called in mathematics ``configuration spaces''.
The choice of ambient space is motivated by the present situation: 
the unrenormalized  Feynman amplitude (\ref{FAMP}) is by inspection real analytic for pairwise distinct points in the Euclidean case.
In the Minkowski region it uniquely defines (after a Wick rotation) a distribution outside coinciding points as a consequence of an iterated application of 
Proposition \ref{LmXXz} $(a)$
(cf. Remark \ref{REM-JUNE13}).

A significant restriction on the vector spaces $\AMPSPA_n$ of the unrenormalized Feynman amplitudes comes from the fact that they are linearly spanned by the factorizable amplitudes 
\(
\mathop{\prod}\limits_{1 \, \leqslant \, j \, < \, k \, \leqslant \, n} G_{j,k} (\x_j-\x_k)
\).
This means that the vector space $\AMPSPA_2$ (consisting of the two--point amplitudes) determines the higher point spaces $\AMPSPA_n$.
Note that in both cases, Euclid and Minkowski, the ambient spaces of functions are algebras and hence, the products 
\(
\prod G_{j,k}
\)
are well defined.
In the case of massless QFT on even $D$ dimensional space--time this considerably simplifies the description of $\AMPSPA_n$ and makes it algebraic. 
The resulting vector spaces are the spaces of all rational (translation invariant) functions with light cone singularities:
\beq\label{RATFEYAMP}
\AMPSPA_n \, := \,
\Biggl\{
\frac{P(\x_1-\x_2,\dots,\x_{n-1}-\x_n)}{
\mathop{\prod}\limits_{1 \, \leqslant \, j \, < \, k \, \leqslant \, n}
\bigr((\x_j -\x_k)^2\bigl)^{\mu_{j,k}}
}
\, \biggl|
\hspace{1pt}
\begin{array}{l}
P \text{ is a polynomial} \\
\mu_{j,k} \, \in \, \Z
\ \
(\forall j,k)
\end{array}
\hspace{0pt}
\Biggr\} \,.
\eeq
Note that the spaces (\ref{RATFEYAMP}) include all possible Feynman amplitudes that can arise in any perturbative QFT with massless fields (for even $D$).
Another important feature of these spaces is that they 
coincide with
the so called {\it coordinate rings} of complex\footnote{%
Complex, because the polynomials $P$ in (\ref{RATFEYAMP}) are be complex in general, and in addition, from the point of view of algebraic geometry it is natural to pass to an algebraically closed field of numbers.} 
{\it affine} varieties\footnote{%
These varieties are the so called ``quadratic configuration spaces'' introduced in \cite{Ni}.}.
In this way we can interpret properties of the renormalization maps in terms of geometric properties of the latter affine varieties.

\medskp

Having introduced the vector spaces $\AMPSPA_n$ of the unrenormalized Feynman amplitudes we pass to the characterization of the renormalization maps $\RENMAP_n$ (\ref{RENMAP}).
The main condition on $\RENMAP_n$ is the renormalization recursion that reflects the causal factorization conditions (\ref{teq19}) and (\ref{teq19M}), for the Euclidean and Minkowski cases, respectively.
In terms of renormalization maps these conditions read:
\beqa\label{RENMAPCAU-E}
\RENMAP_{\MISE} \bigl(G_{\MISE}\bigr)
\vrestr{12pt}{\ECEL_{\{\MISE_1,\MISE_2\}}} 
\podr = \,
\RENMAP_{\MISE_1} \bigl(G_{\MISE_1}\bigr) \,
\RENMAP_{\MISE_2} \bigl(G_{\MISE_2}\bigr) \,
G_{\{\MISE_1,\MISE_2\}}
\qquad (\text{Euclid})\, ,
\\ \label{RENMAPCAU-M}
\RENMAP_{\MISE} \bigl(G_{\MISE}\bigr)
\vrestr{12pt}{\MCEL_{(\MISE_1,\MISE_2)}} 
\podr = \,
\RENMAP_{\MISE_1} \bigl(G_{\MISE_1}\bigr) \,
\RENMAP_{\MISE_2} \bigl(G_{\MISE_2}\bigr) \,
\BVMAP_{\hspace{1pt}\mathcal{T}(\MISE_1 \prec \, \MISE_2)} \, G_{\{\MISE_1,\MISE_2\}}
\quad \ \ (\text{Minkowski})
\,. \nonumber \\[-7pt]
\eeqa
Let us explain the notations and conventions in (\ref{RENMAPCAU-E}) and (\ref{RENMAPCAU-M}).
\begin{LIST}{26pt}
\item[$\bullet$]
$\MISE$ is a finite index set and we assume for the sake of definiteness that it is a set of positive integers;
$\MISE$ $=$ $\MISE_1 \disjuni \MISE_2$ is a nontrivial partition.
\item[$\bullet$]
For an arbitrary index set $\MISE$ of positive integers consisting of $n$ elements there is a unique monotonic isomorphism $\{1,\dots,n\}$ $\cong$ $\MISE$  and under this isomorphism we identify
$\DP_n$ $\cong$ $\DP_{\MISE}$ and $\AMPSPA_n$ $\cong$ $\AMPSPA_{\MISE}$; then, under these identifications we lift the map $\RENMAP_n$ to a linear map
$$
\RENMAP_{\MISE} : \AMPSPA_{\MISE} \, \longrightarrow \, \DP_{\MISE} \,.
$$
\item[$\bullet$]
$G_{\MISE} \in \AMPSPA_{\MISE}$ is a factorizable amplitude:
\beqa\label{GISEdef}
& G_{\MJSE} \, := \,
\mathop{\text{\large $\textstyle \prod$}}\limits_{\mathop{}\limits^{j_1,j_2 \, \in \, \MJSE}_{j_1 \, < \, j_2}}
G_{j_1,j_2} 
(\x_{j_1}-\x_{j_2})
\, \quad (\MJSE=\MISE,\MISE_1,\MISE_2)\,, & \nnb &
G_{\{\MISE_1,\MISE_2\}} \, := \,
\mathop{\text{\large $\textstyle \prod$}}\limits_{\mathop{}\limits^{j_1 \in \MISE_1}_{j_2 \in \MISE_2}}
G_{j_1,j_2} 
(\x_{j_1},\x_{j_2}) \,. &
\eeqa
\item[$\bullet$]
The map $\BVMAP_{\hspace{1pt}\mathcal{T}(\MISE_1 \prec \, \MISE_2)}$ is the boundary value map
that transforms
$G_{\{\MISE_1,\MISE_2\}}$ to the distributions $W_{(\MISE_1,\MISE_2)}$ that we used in (\ref{teq19M});
more precisely
the tube with respect to which the boundary value is taken is
\beqa\label{BACKTUBE-1}
\podr\hspace{-26pt}
\mathcal{T}(\MISE_1 \prec \, \MISE_2) \, :=
\bnn
\podr\hspace{-26pt}
\left\{
(\x_j+i\y_j)_{j \, = \, 1}^n \, \Biggl| \, \y_{j_1}-\y_{j_2} \in -V_+
\text{ if }
\Biggl\{
\begin{array}{l}
j_1 \, < \, j_2 \text{ for } j_1,j_2 \in \MISE_1 \text{ or } j_1,j_2 \in \MISE_2
\\ \text{or} \\
j_1 \, \in \, \MISE_1 \text{ and } j_2 \, \in \, \MISE_2
\end{array} 
\right\} 
.
\eeqa
\item[$\bullet$]
Finally, conditions (\ref{RENMAPCAU-E}) and (\ref{RENMAPCAU-M}) are applicable also in the case when some of the subsets $\MISE_1$ or $\MISE_2$ has one element.
This is done under the following additional convention:
\beq\label{1ptconven}
\AMPSPA_1 \, := \, \C \, =: \, \DP_1
\,,\quad \RENMAP_1 \, := \, 1 \,.
\eeq
\end{LIST}

\medskp

\begin{Statement}{Theorem}\label{Theorem4.2}
There exists a system of renormalization maps
$\{\RENMAP_n\}_{n \, = \, 2}^{\infty}$
$($$\RENMAP_1=1$$)$
satisfying the Minkowski $($resp., Euclidean$)$  causal factorization condition
$(\ref{RENMAPCAU-M})$
$($resp., $(\ref{RENMAPCAU-E})$$)$.
In addition, these maps can be chosen to satisfy the properties$:$
\begin{LIST}{26pt}
\item[$(r_1)$] 
{\rm Scaling.} \\
If $G \in \AMPSPA_n$ is a homogeneous rational function of a $($total$)$ degree $k \in \Z$
then $\RENMAP_n (G)$ is an associate homogeneous distribution of the same degree~$k$.
\item[$(r_2)$] 
{\rm Symmetries.}
\begin{LIST}{30pt}
\item[$(r_{2,1})$]
{\rm Permutation symmetry.} \\
$\RENMAP_n (\sigma^* \hspace{1pt} G)$ $=$ $\sigma^* \hspace{1pt} \RENMAP_n (G)$,
for every permutation $\sigma=(\sigma_1,\dots,\sigma_n)$,
where
$(\sigma^* F)(\x_1,\dots,\x_n)$ $:=$
$F\bigl(\x_{\sigma_1},\dots,\x_{\sigma_n}\bigr)$
for a function or a distribution~$F$.
\item[$(r_{2,2})$] 
{\rm Covariance.} \\
$\RENMAP_n (\Lambda^* \hspace{1pt} G)$ $=$ $\Lambda^* \hspace{1pt} \RENMAP_n (G)$,
for every Lorentz transformation $($resp., Euclidean rotation$)$ $\Lambda \in O(D-1,1)$
$($resp., $\Lambda \in O(D)$$)$,
where
$(\Lambda^* F)(\x_1,$ $\dots,$ $\x_n)$ $:=$
$F\bigl(\Lambda \hspace{1pt} \x_1,\dots,\Lambda \hspace{1pt} \x_n\bigr)$
for a function or a distribution~$F$.
\end{LIST}
\item[$(r_3)$] 
{\rm Commutativity with multiplication by polynomials.} \\
$\RENMAP_n (p \hspace{2pt} G)$ $=$ $p  \hspace{2pt} \RENMAP_n (G)$
for every polynomial $p \in \AMPSPA_n$
and every $G \in \AMPSPA_n$.%
\end{LIST}
\end{Statement}


\begin{remark}\label{RMR4.3}
$(a)$
Recursive conditions (\ref{RENMAPCAU-E}) and (\ref{RENMAPCAU-M}) completely fix the values of the renormalization maps $\RENMAP_n$ outside the large diagonal $\widetilde{\Delta}_n$.

$(b)$
Note that $\RENMAP_n (1) = 1$ for all $n=1,2,\dots$.
This follows by induction in the number of points $n$ as at each step one has to find an associate homogeneous extension of the constant function $1$ through the total diagonal.

$(c)$ 
In the case $n=2$ the map $\RENMAP_2$ is already constructed by Proposition~\ref{Prop-N3.1}:
$(\RENMAP_2 G)(\x_1-\x_2)$ $\equiv$
$\bigl(G(\x_1-\x_2)\bigr)^{\text{extended on } \R^D}$.
\end{remark}


The construction of renormalization maps including the proof of
Theorem \ref{Theorem4.2} 
was performed
in \cite{Ni} in the Euclidean case and sketched out 
for the Minkowski case in \cite{N}.
We will now repeat this construction using the techniques of associate homogeneous distributions developed in the present paper.
Completion of the proof of Theorem \ref{Theorem4.2} will be carried out in 
\ref{Theorem4.2-pr}.

\medskp

We present the construction of renormalization maps for the case of the spaces $\AMPSPA_n$ (\ref{RATFEYAMP}).
The renormalization recursion, which in terms of renormalization maps is based on conditions
(\ref{RENMAPCAU-E}) and (\ref{RENMAPCAU-M}), reduces 
the sequence of maps $\RENMAP_n$ to a system of {\it graded} (i.e., grading preserving) linear maps
that we have considered in Sect. \ref{Se4.4NEWWW}.
The construction of $\RENMAP_n$ is done by setting
\beq\label{RENDECOM}
\begin{array}{c}
\RENMAP_n \, = \, \PRIMAP_n \, \circ \, \SECMAP_n
\quad \ \text{with},
\\[5pt]
\AMPSPA_n
\, \mathop{\longrightarrow}\limits^{\sECMAP_n} \,
\DP_{\BLT} \biggl(\frac{\textstyle \R^{Dn} \backslash \Delta_n}{\textstyle\Delta_n} \biggr)
\, \mathop{\longrightarrow}\limits^{\PRIMAP_n} \,
\DP_n
\,,
\raisebox{23pt}{}
\end{array}
\eeq
where: 
\begin{LIST}{26pt}
\item[$\bullet$]
we remind that
$\DP_{\BLT}$ $:=$ $\mathop{\bigoplus}\limits_{k \, \in \, \Z} \DP_k$
stands for the $\Z$--graded vector spaces linearly spanned by associate homogeneous distributions with variable degrees $k$ of homogeneity.
\item[$\bullet$]
Note that we have (linear) isomorphisms
\beq\label{IDENTIF}
\frac{\R^{Dn}}{\Delta_n} \, \cong \, \R^N
\quad \text{with} \quad N= D(n-1)
\,,\quad
\frac{\textstyle \R^{Dn} \backslash \Delta_n}{\textstyle\Delta_n}
\, \cong \,
\R^N\backslash \{0\}
\eeq
for every $n=2,3,\dots$.
These identifications can be defined by taking subsequent differences
$(\x_1,\dots,\x_n)$ $\mapsto$ $(\x_1-\x_2,\dots,\x_{n-1}-\x_n)$ 
(as we mentioned in Sect. \ref{NSEC-4.2NN})
but they can be defined also in a more subtle way in order to fulfill further requirements, which we shall consider in Sect. \ref{Unif-sect}.
Then the above maps $\PRIMAP_n$ will be transferred to the primary renormalization maps as introduced in Sect. \ref{Se4.4NEWWW} (Definition \ref{Df-4.4-1}):
\beq\label{TransfP-n}
\begin{array}{ccc}
\DP_{\BLT} \biggl(\frac{\textstyle \R^{Dn} \backslash \Delta_n}{\textstyle\Delta_n} \biggr)
& \mathop{\longrightarrow}\limits^{\PRIMAP_n} 
& \DP_n
\\[-7pt]
\rotatebox{270}{\scalebox{1.2}[1]{$\cong$}}
&
& 
\rotatebox{270}{\scalebox{1.2}[1]{$\cong$}}
\\[3pt]
\DP_{\BLT} \bigl(\R^N \backslash \{0\}\bigr) 
& \mathop{\longrightarrow}\limits^{\PRIMAP_{(N)}} 
& \DP_{\BLT} (\R^N)
\,. 
\end{array}
\eeq
\item[$\bullet$]
The sequence of graded linear maps $\{\SECMAP_n\}_{n \, = \, 2}^{\infty}$
\beq\label{SECRENMAP}
\SECMAP_n :
\AMPSPA_n \, \longrightarrow \, 
\DP_{\BLT} \biggl(\frac{\textstyle \R^{Dn} \backslash \Delta_n}{\textstyle\Delta_n} \biggr)
\,
\eeq
($\Delta_n$ $\cong$ $\R^D$ being the total diagonal in $\R^{Dn}$ $\equiv (\R^D)^{\times \, n}$)
come from the renormalization recursion.
We call $\SECMAP_n$ {\bf secondary renormalization maps}
and their construction is given in Lemma \ref{SECMAP-constr}.
\end{LIST}

In this way we reduce the construction of the sequence of renormalization maps
$\RENMAP_2,\RENMAP_3,\dots$
to a sequence of primary renormalization maps
$\PRIMAP_2,\PRIMAP_3,\dots$.
In the statement below we summarize conditions for $\PRIMAP_n$ that will be  sufficient for proving Theorem \ref{Theorem4.2} in the {\it Euclidean} case
(the Minkowskian case will completed in the 
\ref{Theorem4.2-pr}
together with the proof of the proposition below).

\medskp

\begin{Statement}{Proposition}\label{Df-4.4-1mod}
Assume we are given a system
$\{\PRIMAP_n\}_{n \, = \, 2}^{\infty}$
of linear maps
\(
\PRIMAP_n :
\DP_{\BLT} \biggl(\frac{\textstyle \R^{Dn} \backslash \Delta_n}{\textstyle\Delta_n} \biggr)
\to 
\DP_n
\)
such that the following properties are satisfied
\begin{LIST}{26pt}
\item[$(p_0)$] 
{\rm Extension.} \\
$\PRIMAP_n (\MUDZ) \vrestr{16pt}{}\raisebox{-10pt}{\hspace{-2pt}\small $\frac{\textstyle \R^{Dn} \backslash \Delta_n}{\textstyle\Delta_n}$} = \MUDZ$
for every
$\MUDZ \in \DP_{\BLT} \biggl(\frac{\textstyle \R^{Dn} \backslash \Delta_n}{\textstyle\Delta_n} \biggr)$.
\item[$(p_1)$] 
{\rm Scaling.} \\
$\PRIMAP_n$ is preserving the grading provided by the degree of associate homogeneity.
\item[$(p_2)$] 
{\rm Symmetries.}
\begin{LIST}{30pt}
\item[$(p_{2,1})$]
{\rm Permutation symmetry.} \\
$\PRIMAP_n \circ \, \sigma^*$ $=$ $\sigma^* \circ \PRIMAP_n$,
for every permutation $\sigma=(\sigma_1,\dots,\sigma_n)$
$($similarly to $(r_{2,1})$ of Theorem \ref{Theorem4.2}$)$.
\item[$(p_{2,2})$] 
{\rm Covariance.} \\
$\PRIMAP_n \circ \Lambda^*$ $=$ $\Lambda^* \circ \PRIMAP_n$,
for every Euclidean rotation $\Lambda \in O(D)$
$($as in $(r_{2,2})$ of Theorem \ref{Theorem4.2}$)$.
\end{LIST}
\item[$(p_3)$] 
{\rm Commutativity with multiplication by polynomials.} \\
$\PRIMAP_n (p \hspace{2pt} \MUDZ)$ $=$ $p  \hspace{2pt} \RENMAP_n (\MUDZ)$
for every polynomial $p$ on $\frac{\textstyle \R^{Dn}}{\Delta_n}$
and every $\MUDZ \in \DP_{\BLT} \biggl(\frac{\textstyle \R^{Dn} \backslash \Delta_n}{\textstyle\Delta_n} \biggr)$.%
\end{LIST}
Then under the anzatz $(\ref{RENDECOM})$ we obtain a system of renormalization maps that obeys the Theorem \ref{Theorem4.2}.
We call the sequence $\{\PRIMAP_n\}_{n \, = \, 2}^{\infty}$ a system of {\rm\bf primary renormalization maps}.
\end{Statement}

We note that conditions $(p_0)$, $(p_1)$ and $(p_3)$ of the above proposition exactly match the similarly labeled properties in Theorem \ref{Lemma-C.4-NN},
while the relation between the corresponding symmetry conditions $(p_2)$  is clarified in Appendix \ref{Unif-sect}.

\subsection{Extension (renormalization) of homogeneous two--point amplitudes}\label{Sect-EXT-2pt}

\def\TWPTFN{\tau}
\def\UTWPTFN{\TWPTFN^0}
\def\ETWPTFN{\TWPTFN^{ext}}

The renormalization recursion begins with the two--point Feynman amplitudes. In a massless QFT they are homogeneous distributions $\UTWPTFN(\x_1-\x_2)$ defined outside the diagonal, i.e., for $\x_1\neq\x_2$. We shall consider here the case of the $D$--dimensional Minkowski space $M$ (with a space--like signature $(-,+,\ldots,+)$). 
The Euclidean case is actually simpler, the necessary changes for its incorporation will be indicated on the way. 
Furthermore, we shall restrict our attention to space--time dimensions $D>2$. (For $D=2$ the light cone quadric is factorisable and is thus reduced to the chiral,  i.e., $D=1$ case.)

The vector space of all unrenormalized homogeneous 2--point amplitudes $\UTWPTFN(\x)$, $\x=\x_1-\x_2$, is linearly spanned by basic functions of the form (see also Example~\ref{EXM-1nn}): 
\beq\label{HFUN}
\UTWPTFN_{\MDGR\hspace{1pt};\hspace{1pt}m} (\x) 
\hspace{1pt} (\hspace{1pt} := \, \tau_{\MDGR\hspace{1pt};\hspace{1pt}m\hspace{1pt},\hspace{1pt}\sigma} (\x) \hspace{1pt}) 
\hspace{-2pt} := \, 
(\x^2+i0)^{\MDGR} \, 
h_{m,\sigma}(\x)
\,,\quad
\UTWPTFN_{\MDGR\hspace{1pt};\hspace{1pt}m} \in \DP(M\backslash \{0\})
\eeq
$$
\bigl(\MDGR\in \C 
\,,\quad 
\x^2=-(x^0)^2+\sum_{j=1}^{D-1}(x^j)^2
\bigr)
\,, 
$$
where $\{h_{m\hspace{1pt},\hspace{1pt}\sigma}(\x)\}_{\sigma}$ 
is a basis of harmonic homogeneous polynomials of degree~$m$ (i.e., 
$\Box_{\x}\, h_{m\hspace{1pt},\hspace{1pt}\sigma}(\x)$ $=$ $0$ $=$ $(\x \spr \di_{\x}-m) \, h_{m\hspace{1pt},\hspace{1pt}\sigma}(\x)$, $\Box_{\x} = \di_{\x}^2 = \di_{x^{\mu}}\di_{x_{\mu}}$ being the d'Alembert operator%
; note that we omit the $\sigma$ dependence from $\UTWPTFN_{\MDGR\hspace{1pt};\hspace{1pt}m}$ because of keeping shorter the notations). 
We shall renormalize (i.e., extend) the 2--point function by applying the approach of analytic regularization \cite{Sp} (a ``higher dimensional'' version of Sect.~\ref{An-Sect}).
In particular, we note that for $\MDGR \in \C \backslash \Z$ the distribution $\UTWPTFN_{\MDGR\hspace{1pt};\hspace{1pt}m}$ is homogeneous of degree $2\MDGR+m$ on $M \backslash \{0\}$ and hence, by \cite[Theorem 3.2.3]{H} it has a unique homogeneous extension on $M$ of the same degree,
\beq\label{TWPTFN}
\TWPTFN_{\MDGR\hspace{1pt};\hspace{1pt}m} \in \DP(M)
\,,\quad
(\TWPTFN_{\MDGR\hspace{1pt};\hspace{1pt}m}\vrestr{10pt}{M\backslash \{0\}} \, = \, 
\UTWPTFN_{\MDGR\hspace{1pt};\hspace{1pt}m}
\,,\quad
\x \spr \di_{\x} - 2\MDGR-m) \, \TWPTFN_{\MDGR\hspace{1pt};\hspace{1pt}m} (\x) \, = \, 0 
\,,
\eeq
($\MDGR \in \C \backslash \Z$).

\medskp

\begin{Statement}{Proposition}\label{Prop-N3.1}
The one--parameter family of distributions  $\TWPTFN_{\MDGR} (\x)$ \eqref{TWPTFN} 
for $\MDGR\in\C\backslash \Z$ 
and any fixed $($basic$)$ homogeneous harmonic polynomial $h_m(\x)$ $(:=h_{m\hspace{1pt};\hspace{1pt}\sigma} (\x))$
extends to a meromorphic $\DP(M)$--valued function of $\MDGR$ with simple poles at $\MDGR=-\frac{D}{2}-m-n$, $n=0,1,\ldots$ . Its Laurent--Taylor expansion around the poles is
\beq\label{bas2pt-exp}
\TWPTFN_{-\frac{D}{2} - m - n + \varepsilon\hspace{1pt};\hspace{1pt}m}(\x)=
K_{n,m}\, \Box_{\x}^n \, h_m(\di_{\x}) \, \delta(\x) \, \frac{1}{\varepsilon} 
+ \ETWPTFN_{-\frac{D}{2} - m - n\hspace{1pt};\hspace{1pt}m}(\x)+ O(\varepsilon),
\eeq
where the 
coefficients $K_{n,m}$ are given by 
\beq\label{Eq-Knm}
K_{n.m}=
-i \,
\frac{(-1)^m\,\pi^{\frac{D}{2}}}{2^{2n+m}\,n!\,\Gamma\bigl(\frac{D}{2} +n + m\bigr)} \,.
\eeq
The \textrm{extended} 
distributions
$\ETWPTFN_{-\frac{D}{2} - n - m\hspace{1pt};\hspace{1pt}m}(\x)$ 
$(n,m=0,1,\dots)$ 
are associate homogeneous on the entire Minkowski space $M$ of degree $-2n+m-D$ and 
{\ORder}
$1$, such that 
\beq\label{2ptDilLaw}
(\x\spr\di - 2n - m) \, \ETWPTFN_{-\frac{D}{2} - n - m\hspace{1pt};\hspace{1pt}m}(\x)
=
2\,K_{n,m}\,\Box_{\x}^n\,h_m(\di_{\x}) \,\delta(\x)
\,,
\eeq
and furthermore,
they can be written down explicitly 
in terms of a 
{\rm differential renormalization} $\cite{FJL, HL}$ formulas:
\beqa\label{EEE4.4a}
\ETWPTFN_{-\frac{D}{2}\hspace{1pt};\hspace{1pt}0}(\x)
\podr = \,
\frac{1}{2}\
\di_{x^\mu}\Bigl[\frac{x^\mu\,\ln (\x^2+i0)}{(\x^2+i0)^{\frac{D}{2}}}\Bigr]
\nnb \podr
= \, - \frac{1}{2(D-2)} \, \Box_{\x} \, 
\Bigl[
\frac{\ln(\x^2+i0)}{(\x^2+i0)^{\frac{D}{2}-1}} 
\Bigr]
+\frac{2K_{0,0}}{D-2} \, \delta(\x)
\,,\qquad
\\[5pt] \label{EEE4.4b}
\ETWPTFN_{-\frac{D}{2} - n - m\hspace{1pt};\hspace{1pt}m}(\x)
\podr = \,
\frac{(-1)^m \, \Box_{\x}^n \, h_m(\di_{\x})}{2^{2n+m}\,n!\,\Gamma(\frac{D}{2}+n+m)} \,
\ETWPTFN_{-\frac{D}{2}\hspace{1pt};\hspace{1pt}0}(\x)
\,.
\eeqa
\end{Statement}

\begin{Proof}
The 
distribution $\UTWPTFN_{\MDGR\hspace{1pt};\hspace{1pt}m}$ is originally defined for $\x\neq 0$ and it is homogeneous of degree $2\MDGR+m$ in this domain.
The homogeneity is due to the fact that $\UTWPTFN_{\MDGR\hspace{1pt};\hspace{1pt}m}$ \eqref{HFUN} coincides locally, for $\x\neq0$, with a homogeneous Wightman distribution, i.e. with a boundary value of a homogeneous analytic function and is, hence, itself a homogeneous distribution of degree $2\MDGR+m$ in $\DP(M\backslash 0)$.
By the same argument, all the derivatives in $\MDGR$ of $\UTWPTFN_{\MDGR\hspace{1pt};\hspace{1pt}m}$ are associate homogeneous distributions in $\DP(M\backslash \{0\})$, and thus, $\UTWPTFN_{\MDGR\hspace{1pt};\hspace{1pt}m}$ is an entire function in $\MDGR$ (with values in $\DP(M\backslash \{0\})$).
For noninteger $\MDGR$ or $\RE \, (2\MDGR+m) > -D$ there is a unique associate homogeneous extension on the whole $M$ due to Proposition \ref{PeEXT} and thus, we conclude that $\TWPTFN_{\MDGR\hspace{1pt};\hspace{1pt}m}$ (\ref{TWPTFN}) extends to an analytic $\DP(M)$--valued function for noninteger $\MDGR$ or $\RE \, \MDGR > -\frac{D+m}{2}$.

In particular, for $D>2$ space--time dimensions and 
$\RE \, \MDGR > - \frac{3}{2}$ the distribution 
$\TWPTFN_{\MDGR\hspace{1pt};\hspace{1pt}m}(\x)$ is an analytic $\DP(M)$-valued function.
Let us define
\beq\label{G-func-eq}
G(\x\hspace{1pt};\hspace{1pt}\MDGR\hspace{1pt};\hspace{1pt}m) \, := \,
\frac{\Gamma(-\MDGR) \, \TWPTFN_{\MDGR\hspace{1pt};\hspace{1pt}m}(\x)}{4^{\MDGR} \Gamma\bigl(\MDGR+m+\frac{D}{2}\bigr)}
\, ( \, = \, 
\frac{\Gamma(-\MDGR) \, (\x^2+i0)^{\MDGR} \, h_m(\x)}{4^{\MDGR} \Gamma\bigl(\MDGR+m+\frac{D}{2}\bigr)}
\,)
\,,
\eeq
which is thus an analytic $\DP(M)$--valued function in the strip
$\RE \, \MDGR \in (- \frac{3}{2},0)$.
But $G(\x\hspace{1pt};\hspace{1pt}\MDGR\hspace{1pt};\hspace{1pt}m)$ satisfies the equation
\beq
\quad
\Box_{\x} \, G(\x\hspace{1pt};\hspace{1pt}\MDGR\hspace{1pt};\hspace{1pt}m) \, = \, - G(\x\hspace{1pt};\hspace{1pt}\MDGR-1\hspace{1pt};\hspace{1pt}m)
\,,
\eeq
which implies that it has an analytic continuation for all $\RE \, \MDGR < 0$.
Since
$\TWPTFN_{\MDGR\hspace{1pt};\hspace{1pt}m}(\x)$ $=$
$4^{\MDGR}$ $\Gamma(-\MDGR)^{-1}$ $\Gamma\bigl(\MDGR+m+\frac{D}{2}\bigr)$ $G(\x\hspace{1pt};\hspace{1pt}\MDGR\hspace{1pt};\hspace{1pt}m)$ and $4^{\MDGR} \Gamma(-\MDGR)^{-1}$ is an entire function in $\MDGR$ we conclude that $\TWPTFN_{\MDGR\hspace{1pt};\hspace{1pt}m}$ is a meromorphic $\DP(M)$--valued function in $\MDGR$ with the same poles as those of
$\Gamma\bigl(\MDGR+m+\frac{D}{2}\bigr)$, which in turn are simple and placed at $\MDGR = - m - \frac{D}{2} - n$ for $n =0,1,2,\dots$.

We derive next the Laurent--Taylor expansion (\ref{bas2pt-exp}).
We consider first the special scalar case of $m=n=0$:
\beqa
\podr
\TWPTFN_{-\frac{D}{2}+\varepsilon\hspace{1pt};\hspace{1pt}m}(\x) \, = \, 
\frac{4^{-\frac{D}{2}+\varepsilon} \, \Gamma(\varepsilon) \, G\bigl(\x\hspace{1pt};\hspace{1pt}-\frac{D}{2}+\varepsilon\hspace{1pt};\hspace{1pt}m\bigr)}{\Gamma\bigl(\frac{D}{2}-\varepsilon\bigr)}
\nnbnn \podr
= \, \frac{G\bigl(\x\hspace{1pt};\hspace{1pt}-\frac{D}{2}\hspace{1pt};\hspace{1pt}m\bigr)}{4^{\frac{D}{2}} \Gamma\bigl(\frac{D}{2}\bigr)} \, \frac{1}{\varepsilon} + O(1) \, = \,
- \, \frac{\Box_{\x} \, G\bigl(\x\hspace{1pt};\hspace{1pt}-\frac{D}{2}+1\hspace{1pt};\hspace{1pt}m\bigr)}{4^{\frac{D}{2}} \Gamma\bigl(\frac{D}{2}\bigr)} \, \frac{1}{\varepsilon} + O(1) \,.
\eeqa
Since
$G\bigl(\x\hspace{1pt};\hspace{1pt}-\frac{D}{2}+1\hspace{1pt};\hspace{1pt}0\bigr)$ $=$
$4^{\frac{D}{2}-1} \, \Gamma(\frac{D}{2}-1) \, \TWPTFN_{-\frac{D}{2}+\varepsilon\hspace{1pt};\hspace{1pt}0}(\x)$
and
\beq\label{GEN-GR-FUN}
\Box_{\x} \, \TWPTFN_{-\frac{D}{2}+\varepsilon\hspace{1pt};\hspace{1pt}0}(\x) \, ( \, = \,
\Box_{\x} \, \Bigl[\frac{1}{(\x^2+i0)^{\frac{D}{2}-1}}\Bigr]^{\text{extended on } M} \,) \, = \, 
- \frac{4 \hspace{1pt} i \, \pi^{\frac{D}{2}}}{\Gamma\bigl(\frac{D}{2}-1\bigr)} \, \delta(\x)
\eeq
(cf. Remark \ref{Rm-N3.1} $(b)$)
we obtain Eq. (\ref{bas2pt-exp}) with $K_{0,0}$ $=$ $-i \,\frac{\hspace{1pt}\pi^{\frac{D}{2}}}{\Gamma\bigl(\frac{D}{2}\bigr)}$ (in agreement with (\ref{Eq-Knm})).
For general $m,n=0,1,2,\dots$ we use the equation
\beqa\label{130702-1}
\podr
\Box_{\x}^n \, h_m (\di_{\x}) \, (\x^2+i0)^{\MDGR} 
\bnn \podr
= \, 
2^m
\MDGR (\MDGR-1) \cdots (\MDGR-m+1) \, \Box_{\x}^n \, (\x^2+i0)^{\MDGR-m} \, h_m (\x)
\bnn \podr
= \,
2^{2n+m}
\Biggl(
\mathop{\prod}\limits_{j \, = \, 0}^{n+m-1}
(\MDGR-j)
\Biggr)
\Biggl(
\mathop{\prod}\limits_{k \, = \, 1}^{n}
\Bigl(\MDGR + \frac{D}{2} -k \Bigr)
\Biggr)
(\x^2+i0)^{\MDGR-n-m} \, h_m (\x)
\,,
\eeqa
which is verified first for $\x \neq 0$ by a straightforward computation and then we conclude its validity over the whole $M$ for non integer $\MDGR$ due to the uniqueness of the homogeneous extensions in the case of non integer degree of homogeneity 
(Proposition \ref{PeEXT}). Therefore, Eq. (\ref{130702-1}) is true as an equality of $\DP(M)$--valued meromorphic functions in $\MDGR$.
This allows us to derive (\ref{bas2pt-exp}) and (\ref{Eq-Knm}) for all $m$ and $n$ from the particular case of $m=n=0$.

Equation (\ref{2ptDilLaw}) follows 
from the homogeneity (\ref{TWPTFN})
\beq\label{130207-2nn}
\bigl(\x \spr \di_{\x} + 2n - m + D\bigl) \,
\TWPTFN_{-\frac{D}{2}-n-m+\varepsilon\hspace{1pt};\hspace{1pt}m}(\x)
\, = \, 2\varepsilon \,
\TWPTFN_{-\frac{D}{2}-n-m+\varepsilon\hspace{1pt};\hspace{1pt}m}(\x)
\eeq
valid for small nonzero $\varepsilon$. Indeed, we replace
$\TWPTFN_{-\frac{D}{2}-n-m+\varepsilon\hspace{1pt};\hspace{1pt}m}(\x)$ in both sides of  
(\ref{130207-2nn}) by the expansion (\ref{bas2pt-exp}) and compare term by term in~ $\varepsilon$.

Let us finally prove Eqs. (\ref{EEE4.4a}) and (\ref{EEE4.4b}).
The derivation of the second equality in (\ref{EEE4.4a})
starts with the observation that $(\x^2+i0)^{-\frac{D}{2}+1+\varepsilon} \, \ln(\x^2+i0)$ uniquely extends to an analytic
$\DP(M)$--valued function in $\varepsilon$ in a neighborhood of $\varepsilon = 0$
and for small $\varepsilon>0$ we have:
\beqa
\podr
\frac{1}{4} \, \Box_{\x} \, 
\bigl[
(\x^2+i0)^{-\frac{D}{2}+1+\varepsilon} \, \ln(\x^2+i0) 
\bigr]
^{\text{extended on } M}
\nnbnn \podr
= \,
\frac{1}{4} \, 
\bigl[
\Box_{\x} \, 
(\x^2+i0)^{-\frac{D}{2}+1+\varepsilon} \, \ln(\x^2+i0) 
\bigr]
^{\text{extended on } M}
\bnn \podr
= \,
\Bigl[
\Bigl(2\varepsilon-\frac{D}{2}+1\Bigr) \, 
+ \, \varepsilon\Bigl(\varepsilon-\frac{D}{2}+1\Bigr) \, 
\, \frac{\di}{\di \varepsilon} \,
\Bigr] \,
\TWPTFN_{-\frac{D}{2}+\varepsilon\hspace{1pt};\hspace{1pt}0}(\x)
\bnn \podr
= \,
- \Bigl(\frac{D}{2}-1\Bigr)
\ETWPTFN_{-\frac{D}{2}\hspace{1pt};\hspace{1pt}0}(\x)
+K_{0,0} \, \delta(\x)
+ O(\varepsilon)
\,
\eeqa
(since by (\ref{bas2pt-exp}), 
\(
\TWPTFN_{-\frac{D}{2}+\varepsilon\hspace{1pt};\hspace{1pt}0}(\x)
=
\frac{K_{0,0}}{\varepsilon} \, \delta(\x)
+\ETWPTFN_{-\frac{D}{2}\hspace{1pt};\hspace{1pt}0}(\x)
+ O(\varepsilon)
\)
and the D'Alembert operator commutes with the extension in the case of a non integral degree of homogeneity).
One then derives from this Eq.~(\ref{EEE4.4a}).
\end{Proof}


\begin{remark}\label{Rm-N3.1}
$(a)$
Note that the meromorphic $\DP(M)$--valued function (\ref{G-func-eq}) is not entire although it is analytic for $\RE \, \MDGR < 0$.
We can obtain an entire function in $\MDGR$ if we set
\beqa\label{F-func-eq}
\podr
F(\x\hspace{1pt};\hspace{1pt}\MDGR\hspace{1pt};\hspace{1pt}m) \, := \,
\frac{\TWPTFN_{\MDGR\hspace{1pt};\hspace{1pt}m}(\x)}{4^{\MDGR} \Gamma(\MDGR+1)\Gamma\bigl(\MDGR+m+\frac{D}{2}\bigr)}
\,,
\\[3pt] \nonumber \podr
\Rightarrow \quad
\Box_{\x} \, F(\x\hspace{1pt};\hspace{1pt}\MDGR\hspace{1pt};\hspace{1pt}m) \, = \, F(\x\hspace{1pt};\hspace{1pt}\MDGR-1\hspace{1pt};\hspace{1pt}m)
\,. \
\eeqa
However, it produces unwanted fictitious poles for $\TWPTFN_{\MDGR\hspace{1pt};\hspace{1pt}m}(\x) \, (=(\x^2+i0)^{\MDGR}$ $h_m(\x))$ coming from $\Gamma (\MDGR+1)$ when $\RE \, \MDGR < 0$.

$(b)$
Eq. \eqref{GEN-GR-FUN} is a consequence of the formula for the Fourier transform of the 2--point Feynman amplitude of a general scalar field (of dimension $d$)
\beq\label{eq_4.10}
\frac{\Gamma(d)}{(4\pi)^{\frac{D}{2}}} \, \Bigl(\frac{4}{\x^2+i0}\Bigr)^d
=
-i\Gamma\Bigl(\frac{D}{2}-d\Bigr)\int\frac{e^{i\p\spr\x}}{(\p^2-i0)^{\frac{D}{2}-d}}\frac{\mathrm{d}^D\p}{(2\pi)^D}.
\eeq
(The latter follows from the Schwinger $\alpha$--representation,
\begin{equation*}
\frac{1}{(\p^2-i0)^{\nu}}
\, = \,
\int_0^{\infty}e^{-i\alpha(\p^2-i0)}\,\alpha^{\nu}\,\frac{d\alpha} {\alpha}
\end{equation*}
and the formula for the (conditionally convergent) Fresnel integral
$\int_{-\infty}^{\infty}e^{\pm it^2}$ $dt$ $=$ $\sqrt{\frac{\pi}{2}}(1\pm i)$. Note that for euclidean $\x$ and $\p$ there is no $i$--factor in Schwinger exponent and no $-i$ in the counterpart of \eqref{eq_4.10}.)

$(c)$
Proposition \ref{Prop-N3.1} is valid also in the Euclidean case with a slight correction in Eq. (\ref{Eq-Knm}) by removing the $(-i)$ pre-factor. 
This is because the only place in the proof that needs a special attention in the Euclidean case is Eq. (\ref{GEN-GR-FUN}) 
and it is modified in this case by such a factor.

$(d)$
A physicist 
would place a length $\lambda$ in the expressions like
$\Bigl(\frac{\textstyle \x^2+i0}{\textstyle \lambda^2}\Bigr)^{\MDGR}$ or 
$\ln \Bigl(\frac{\textstyle \x^2+i0}{\textstyle \lambda^2}\Bigr)$
in order to deal with dimensionless expressions.
This is the so called ``renormalization length (or scale)''.
\end{remark}


The simplest examples of divergent 2--point functions are provided by 1 and 2--loop graphs for a massless scalar field in $D=4$ dimensions: 
\begin{center}
\ALTERNATIVE{%
\includegraphics[width=7cm]{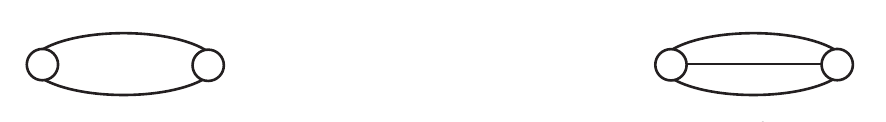}}{%
\includegraphics[width=7cm]{Fig1mod.eps}}
\end{center}
The corresponding Feynman amplitudes are proportional to $(\x^2+i0)^{-d}$ with $d=2$ and 3, respectively.
The first one is logarithmically divergent. Its renormalized expression can be written  according to \eqref{bas2pt-exp} and \eqref{EEE4.4a} as
\beqa\label{130208-3}
\ETWPTFN_{-2\hspace{1pt};\hspace{1pt}0}(\x) \, \equiv \,
\Bigl(\frac{1}{\textstyle (\x^2+i0)^2}\Bigr)^{ext}
\podr = \,
\mathop{\lim}\limits_{\varepsilon \downarrow 0}
\Bigl[
(\x^2+i0)^{-2+\varepsilon}
+
\frac{i\hspace{1pt}\pi^2}{\varepsilon} \, \delta(\x)
\Bigr]
\nnb
\podr=\,
\frac{1}{2} \, \di_{x^\mu}\Big[\frac{x^\mu\ln(\x^2+i0)}{(\x^2+i0)^2}\Big]
\nnb
\podr = \,
-\frac{1}{4} \,
\Box_{\x} \, \frac{\ln(\x^2+i0)}{\x^2+i0}
- i\hspace{1pt}\pi^2 \, \delta(\x)
\qquad
\eeqa
$(D = 4)$,
where the second line corresponds to the differential renormalization:
we present the unrenormalized amplitude as a derivative of a globally defined distribution.
The two loop diagram with an amplitude
$(\x^2+i0)^{-3}$ is similarly treated.
The result is:
\beqa\label{130208-4}
\Bigl(\frac{1}{\textstyle (\x^2+i0)^3}\Bigr)^{ext}
\podr = \,
\mathop{\lim}\limits_{\varepsilon \downarrow 0}
\Bigl[
(\x^2+i0)^{-3+\varepsilon}
-
\frac{\pi^2}{6\varepsilon} \, \delta(\x)
\Bigr]
\nnb
\podr = \,
-\frac{1}{8} \,
\Box_{\x} \, \Bigl(\frac{1}{\textstyle (\x^2+i0)^2}\Bigr)^{ext}
\qquad (D = 4)\,,
\qquad
\eeqa


\begin{remark}\label{Rm-4.2NN}
$(a)$
Note that the presence of a logarithm in Eq. (\ref{130208-3}) already indicates a possible obstruction for the preservation of the homogeneity after the renormalization.
However, this is not a sure sign, as for example, the free massless scalar propagator in $D=4$ has also a similar representation:
$$
\frac{1}{\x^2+i0} \, = \, \frac{1}{4} \, \Box_{\x} \, \ln(\x^2+i0)
\qquad
(D=4)
$$
but nevertheless, it can be extended uniquely as a homogeneous distribution over the whole space--time $M$.

$(b)$ A generalization of the second equality in (\ref{130208-4}) for any space--time dimension and a basic amplitude is that
\beqa
\podr
\Box_{\x} \,
\bigl((\x^2+i0)^{-n-m-\frac{D}{2}} \, h_m(\x)\bigr)^{\text{extended on } M}
\nnb
\podr = \,
\bigl(\Box_{\x} \, (\x^2+i0)^{-n-m-\frac{D}{2}} \, h_m(\x)\bigr)^{\text{extended on } M}
\nnbnn \podr
= \,
4 (n+m+1) \Bigl(n+m+\frac{D}{2}\Bigr)
\, \bigl((\x^2+i0)^{-n-1-m-\frac{D}{2}} \, h_m(\x)\bigr)^{\text{extended on } M}
\,
\eeqa
($n,m=0,1,2,\dots$).
Note however that,
\beqs
\podr
\Box_{\x} \,
\bigl((\x^2+i0)^{-\frac{D}{2}+1}\bigr)^{\text{extended on } M}
\, \equiv \,
\Box_{\x} \,
\ETWPTFN_{-\frac{D}{2}\hspace{1pt};\hspace{1pt}m} (\x) 
\\ \podr
= \,
\bigl(\Box_{\x} \, (\x^2+i0)^{-\frac{D}{2}+1}\bigr)^{\text{extended on } M} 
- \frac{4 \, \pi^{\frac{D}{2}}}{\Gamma\bigl(\frac{D}{2}-1\bigr)} \, \delta(\x)\nnbnn 
\eeqs
(cf. Eq. (\ref{GEN-GR-FUN}))
since before the extension (i.e., for $\x\neq 0$) $(\x^2+i0)^{-\frac{D}{2}+1}$ is harmonic
($\Box_{\x} \, (\x^2+i0)^{-\frac{D}{2}+1} = 0$).
\end{remark}


Let us point out that the analysis of the ultraviolet divergences in the theory of renormalization is usually based on the so called {\it power counting} criterion originally introduced in momentum space (\cite{D}).
In massless QFT on configuration space the power counting criterion is based on a bound of the degree of homogeneity.
The violation of this bound is a necessary and sufficient condition for the absence of {\it unique} homogeneous extensions of the unrenormalized Feynman amplitudes.
In the case of two--point homogeneous Feynman amplitudes $\frac{\textstyle h_m(\x)}{\textstyle (\x^2+i0)^s}$ ($s > 0$) this is reflected by Proposition \ref{PeEXT} and  can be stated as the fact that the difference between the
minus total degree of homogeneity, $2s-m$, and the space time dimension $D$ should be nonnegative.
The latter difference is called 
a ``{\bf degree of divergence}''
and when it is nonnegative we speak about a {\bf superficially divergent diagram}.
Proposition \ref{Prop-N3.1} suggests that there are finer notions of divergence or convergence of Feynman amplitudes.
(We remind the reader that in this paper we consider only ultraviolet divergences and their renormalization.)

\medskp

\begin{Statement}{Corollary}\label{2ptConvergence}
A two--point unrenormalized Feynman amplitude of the form
$\frac{\textstyle h_m(\x)}{\textstyle (\x^2+i0)^s}$ 
$(\x \neq 0)$
has a homogeneous extension iff
\beqa\label{CRITER}
\text{``the degree of harmonicity''} 
\podr 
:= \, m
\nnb 
> \,
\text{``the degree of divergence''} 
\podr
:= \, 2s-m - D 
\,. 
\eeqa
Furthermore, in this case the homogeneous extension is unique if we impose Lorentz covariance.
\end{Statement}

\medskp

\begin{Proof}
The inequality (\ref{CRITER}) is sufficient for the existence of a homogeneous extension of 
$(\x^2+i0)^{-s}$ $h_m(\x)$
according to Eqs. (\ref{bas2pt-exp}) and (\ref{2ptDilLaw}).
The necessity follows by the fact that if 
$(\x^2+i0)^{-n-m+\frac{D}{2}}$ $h_m(\x)$ ($\x\neq 0$)
has a homogeneous extension 
$\bigl((\x^2+i0)^{-n-m+\frac{D}{2}}$ $h_m(\x)\bigr)^{ext_1}$
for some $n,m=0,1,2,\dots$
then the difference between it and the constructed one by Proposition \ref{Prop-N3.1},
$\bigl((\x^2+i0)^{-n-m+\frac{D}{2}}$ $h_m(\x)\bigr)^{ext_1}$
$-$
$\bigl((\x^2+i0)^{-n-m+\frac{D}{2}}$ $h_m(\x)\bigr)^{ext}$,
must be an associate homogeneous of distribution of degree $2n+m$ supported at zero, i.e., a linear combination of $(2n+m)$th derivatives of the delta function.
This would imply that $\bigl((\x^2+i0)^{-n-m+\frac{D}{2}}$ $h_m(\x)\bigr)^{ext}$ is also homogeneous, which contradicts (\ref{2ptDilLaw}).

For the second part we consider again a difference
$Q (\x)$ $:=$
$\bigl((\x^2+i0)^{-s}$ $h_m(\x)\bigr)^{ext_1}$
$-$
$\bigl((\x^2+i0)^{-s}$ $h_m(\x)\bigr)^{ext}$
of two homogeneous extensions.
When the amplitude is {\it not} superficially divergent, the homogeneous extension is unique.
In the opposite case the above difference $Q(\x)$ must be equal to a 
linear combination of derivatives of the delta function,
$P(\di_x) \delta(\x)$, for a homogeneous polynomial $P$ of degree  
$2s - m - D \geqslant 0$.
By Lorentz covariance $Q$, and hence, $P$ are transforming under irreducible representations of the Lorentz group that are isomorphic to the representation on the harmonic homogeneous polynomials of degree $m$.
Therefore,
 $P$ must be divisible by a harmonic polynomial of degree $m$, which is impossible because of the inequality (\ref{CRITER}).
\end{Proof}

\medskp

As an illustration of criterion (\ref{CRITER}) let us consider the propagator of the free electromagnetic field in $D=4$:
\beqa\label{FMN-ex}
\podr
\langle 0| T\bigl(F_{\mu_1\nu_1}(\x_1)F_{\mu_2\nu_2}(\x_2)\bigr)|0\rangle
\, := \,
\frac{h_{\mu_1,\hspace{1pt}\hspace{1pt}\nu_1;\mu_2,\hspace{1pt}\nu_2}(\x)}{4\pi^2 \, (\x^2+i0)^3}
\qquad (\x \, = \, \x_1-\x_2)
\qquad\qquad
\bnn
\podr
h_{\mu_1,\hspace{1pt}\nu_1;\mu_2,\hspace{1pt}\nu_2}
\, := \,
p_{\mu_1,\hspace{1pt}\mu_2}
p_{\nu_1,\hspace{1pt}\nu_2}
-
p_{\mu_1,\hspace{1pt}\nu_2}
p_{\nu_1,\hspace{1pt}\mu_2}
\,,\quad
p_{\mu,\nu}
\, := \,
\eta_{\mu,\nu} \x^2 - 2 x^{\mu} x^{\mu}
\,,
\eeqa
where $h_{\mu_1,\hspace{1pt}\nu_1;\mu_2,\hspace{1pt}\nu_2}(\x)$ is a harmonic homogeneous polynomial of degree~2.
As a Feynman amplitude, (\ref{FMN-ex}) is logarithmically ``divergent'' by power counting
(its degree of homogeneity is $-4$ $=$ $-D$).
However, it possess a homogeneous extension and Lorentz covariance fixes it uniquely.
In fact, in quantum electrodynamics the propagator (\ref{FMN-ex}) is usually expressed as a second derivative of the free massless scalar propagator:
$$
\langle 0| T\bigl(F_{\mu_1\nu_1}(\x_1)F_{\mu_2\nu_2}(\x_2)\bigr)|0\rangle
\, = \, \frac{1}{8} \,
h_{\mu_1,\hspace{1pt}\nu_1;\mu_2,\hspace{1pt}\nu_2}(\di_{\x}) \,
\frac{1}{4\pi^2 \, (\x^2+i0)} \,. 
$$

Motivated by the above statements we call a (ultraviolet) {\bf divergent}
(two--point) Feynman amplitude in the homogeneous case of massless QFT
such an amplitude that does not possess a homogeneous extension as a distribution over the whole Minkowski space.
In view of this one can interpret the ``miraculous'' cancellations of the ultraviolet divergences in some supersymmetric QFT models 
(cf., for example 
\cite{HST}) 
as examples of convergent amplitudes that are superficially divergent (in the ultraviolet regime).


\section{Concluding remarks and outlook}\label{Con-sec}

The Stueckelberg-Bogolubov-Epstein-Glaser approach to perturbative causal QFT has reduced infinite renormalization to a well defined and physically motivated mathematical problem: the extension of distributions in configuration space originally only defined for non-coinciding arguments. In a purely mathematical context (that does not even mention QFT) H\"ormander \cite{H} constructs all such extensions for homogeneous distributions. Applied to the Feynman amplitudes in a massless QFT his results effectively solve the problem of renormalizing primitively divergent amplitudes. To address the general renormalization problem one has to consider the wider class of associate homogeneous distributions \cite{GS}. The present paper translates the Epstein-Glaser recursion of time-ordered products of local quantum fields to a problem of functional analysis that fits the general H\"ormander framework and fills the above mentioned gap by considering extensions of arbitrary associate homogeneous distributions satisfying the causal factorization property. 
The relation of the notion of residue (of Sects. \ref{Se3.6GGG}, \ref{Se4.4NEWWW} and Appendix \ref{Spokes}) in the configuration space approach
should profitably be connected 
with the notion of a Feynman period that is now becoming popular (see e.g. the recent paper \cite{S13} addressed to number theorists and earlier references cited there). 
It should be possible to extend the concepts and results of the present paper to a conformal QFT that includes amplitudes with integrated internal vertices (see e.g. \cite{DDEHPS} and references therein). The leading small distance behavior of massive Feynman amplitudes is also given by associate homogeneous distributions. The systematic treatment of this case in the lines of the present paper requires more work and is worth pursuing.

\section*{Acknowledgments}
 
The authors thank David Broadhurst, Maxim Kontsevich and Dirk Kreimer for discussions.  N.M.N. and I.T. thank CERN and IHES for hospitality during the course of this work.

\bigskip

\appendix
\def\PHANT{${}$\hspace{55pt}${}$}
\renewcommand{\thetheorem}{\Alph{section}.\arabic{theorem}}
\renewcommand{\theproposition}{\Alph{section}.\arabic{proposition}}
\renewcommand{\thelemma}{\Alph{section}.\arabic{lemma}}
\renewcommand{\thecorollary}{\Alph{section}.\arabic{corollary}}
\renewcommand{\theremark}{\Alph{section}.\arabic{remark}}
\renewcommand{\theexample}{\Alph{section}.\arabic{example}}

\renewenvironment{Statement}[1]{%
        \renewcommand{\theStatement}{\Alph{section}.\arabic{Statement}}
        \refstepcounter{Statement}\noindent\textbf{\bf #1\hspace{3.5pt}\Alph{section}.\arabic{Statement}.}${}$\hspace{1pt}${}$\it}{}

\section[{\PHANT}A combinatorial diagonal lemma (generalization of Lemmas \ref{EucDiaLem} and \ref{MinDiaLem})]{A combinatorial diagonal lemma (generalization of Lemmas \ref{EucDiaLem} and \ref{MinDiaLem})}\label{ASe1}
\setcntrs

In both Euclidean and Minkowskian  frameworks 
the ``diagonal lemma'' (Lemmas \ref{EucDiaLem} and \ref{MinDiaLem}) allows to complete each step of the renormalization by the extension of a distribution defined outside the full diagonal.
These lemmas can be formulated more generally in terms of a binary relation $\MREL$ as follows.

Let $\MSET$ be an arbitrary set. A binary relation $\MREL$ on it is a subset
$$
\MREL \subseteq \MSET \times \MSET \,;
\qquad
x\hspace{1pt}\MREL\hspace{1pt}y \quad \text{iff} \quad (x,y) \in \MREL \,.
$$
Here, we take $\MSET=\ESPA$ $(=\R^D)$ or $\MSET=\MSPA$ $(=\R^{D-1,1})$ in the Euclidean or the Minkowski case, respectively.
The relation $\MREL$ is defined in each of these cases as follows:
\begin{LIST}{20pt}
\item[$(E)$]\ $(\x,\y) \in \MREL$\ \ iff\ \ $\x=\y$ \ ($\x,\y \in \ESPA$);
\item[$(M)$]\ $(\x,\y) \in \MREL$\ \ iff\ \ $\x-\y \in \overline{V}^+$ \ ($\x,\y \in \MSPA$). 
\end{LIST}

Let $\MIRL$ be the complementary relation\footnote{so that $(x,y) \in \mIRL \Longleftrightarrow (x,y) \notin \MREL$; alternative notations used in the literature for $(x,y) \in \mIRL$ are $x \mIRL y$ or $x \mNRL\,y$} of $\MREL$:
$$
\MIRL \, \equiv \, (\MSET \times \MSET) \backslash \MREL \,.
$$
Consider an arbitrary splitting of a finite set of labels (indices) $\MISE=\{1,$ $2,$ $\dots,$ $n\}$ into two non-empty, non-intersecting subsets $\MISE_1,\MISE_2$:
\beq\label{APR}
\MISE \, = \, \MISE_1 \dot{\cup} \MISE_2 \,.
\eeq
Define
$$
\RCEL_{(\MISE_1,\MISE_2)} \, := \, 
\bigl\{(x_1,\dots,x_n) \in \MSET^{\times n} \, \bigl| 
(x_{j_1},x_{j_2}) \in \MIRL 
\text{ for all } 
j_1 \in \MISE_1 \text{ and } j_2 \in \MISE_2\bigr\} \,.
$$
Note that in general
$$
\RCEL_{(\MISE_1,\MISE_2)} \, \neq \, \RCEL_{(\MISE_2,\MISE_1)} \,.
$$
Let $\MDIA_n$ be the full (small) diagonal
$$
\MDIA_n \, = \, \bigl\{(x_1,\dots,x_n) \in \MSET^{\times n} \, \bigl| x_1 = \cdots = x_n \bigr\} \,.
$$
The diagonal lemma in both Euclidean and Minkowski spaces states that the domains $\RCEL_{(\MISE_1,\MISE_2)}$ satisfy the {\bf diagonal property}
\beq\label{DP}
\mathop{\text{\footnotesize $\bigcup$}}\limits_{(\MISE_1,\MISE_2)} \, \RCEL_{(\MISE_1,\MISE_2)} \, = \, 
\MCMP \MDIA_n
\, ( \, \equiv \, \MSET^{\times n} \backslash \MDIA_n)
\,,
\eeq
for all $n=2,3,\dots$; here and in what follows, $\mathop{\text{\footnotesize $\bigcup$}}\limits_{(\MISE_1,\MISE_2)}$ stands for the union over all nontrivial enumerated partitions (\ref{APR}).

\medskp

\begin{Statement}{Theorem}\label{GenDiaLem}
The binary relation $\MREL$ satisfies the diagonal property $(\ref{DP})$ 
$($i.e. the Diagonal Lemma holds for $\MREL$$)$
if and only if it is {\rm free of cycles} in the sense that $(x,x) \in \MREL$ and\footnote{or equivalently written,
\(x_1\hspace{1pt}\MREL\hspace{1pt}x_2\hspace{1pt}\MREL\hspace{1pt}x_3\hspace{1pt}\MREL\hspace{1pt} \cdots\hspace{1pt}\MREL\hspace{1pt}x_n\hspace{1pt}\MREL x_1 \ \Longrightarrow \ x_1 = \cdots = x_n\)}
\beqa
\label{Acycl3}
\hspace{-15pt}
(x_1,x_2) \in \MREL \ \, \MAND \ \, 
(x_2,x_3) \in \MREL \ \, \MAND 
\dots
\MAND \
(x_{n-1},x_n) \in \MREL \ \, \MAND \ \,
(x_n,x_1) \in \MREL  \hspace{-3pt}\podr
\nonumber \\
\quad \text{implies} \quad 
x_1 = \cdots = x_n \hspace{-3pt}\podr ,
\qquad
\eeqa
for all $n=2,3,\dots$ and $x,x_1, \dots, x_n \in \MSET$.
\end{Statement}

\medskp

\noindent
{\it Proof.}
For $n=2$ the diagonal property reads
\beqs
&
\RCEL_{(\{1\},\{2\})} \cup \RCEL_{(\{2\},\{1\})} \, = \, \bigl\{(x,y) \in \MSET^{\times 2} \, \bigl| \, x \neq y \bigr\}  \,,
&
\nonumber \\ \nonumber
\text{equivalently,}
\quad
&
\bigl\{(x,y) \in \MSET^{\times 2} \, \bigl| \, (x,y) \in \MIRL \bigr\} 
\cup \bigl\{(x,y) \in \MSET^{\times 2} \, \bigl| \, (y,x) \in \MIRL \bigr\} \, 
&
\nonumber \\ \nonumber
&
= \,
\bigl\{(x,y) \in \MSET^{\times 2} \, \bigl| \, x \neq y \bigr\} \,,
&
\nonumber \\ \nonumber
\text{equivalently,}
\quad
&
\bigl\{(x,y) \in \MSET^{\times 2} \, \bigl| \, (x,y) \in \MREL \bigr\} 
\cap \bigl\{(x,y) \in \MSET^{\times 2} \, \bigl| \, (y,x) \in \MREL \bigr\} \, 
&
\nonumber \\ \nonumber
&
= \,
\bigl\{(x,y) \in \MSET^{\times 2} \, \bigl| \, x = y \bigr\} \,.
&
\eeqs
The last equation is equivalent to relation (\ref{Acycl3}) for $n=2$ together with $(x,x) \in \MREL$.

Next we prove that the implication (\ref{Acycl3}) is necessary for the validity of the diagonal property (\ref{DP}).
To this end note that (\ref{DP}) is equivalent to
\beq\label{DP3}
\mathop{\text{\footnotesize $\bigcap$}}\limits_{(\MISE_1,\MISE_2)} \, \MCMP \RCEL_{(\MISE_1,\MISE_2)} \, = \, 
\MDIA_n
\,,
\eeq
where $\MCMP \RCEL_{(\MISE_1,\MISE_2)}$ is the complement $\MSET^{\times n} \backslash \RCEL_{(\MISE_1,\MISE_2)}$:
\beqa\label{CRCEL}
\MCMP \RCEL_{(\MISE_1,\MISE_2)} \, = \, \bigl\{(x_1,\dots,x_n) \in \MSET^{\times n} \, \bigl| \podr (x_{j_1},x_{j_2}) \in \MREL 
\text{ for {\it some} } \nnb \podr j_1 \in \MISE_1 \text{ and } j_2 \in \MISE_2\bigr\} \,.
\eeqa
Take then an $n$-tuple $(x_1,\dots,x_n) \in \MSET^{\times n}$ that obeys the left hand side of (\ref{Acycl3}), i.e., the conditions
\beqa\label{LHS}
\hspace{-40pt}
\podr
(x_1,x_2) \in \MREL 
\,,\
(x_2,x_3) \in \MREL 
\,,\ \dots\,,\
(x_{n-1},x_n) \in \MREL 
\,,\
(x_n,x_1) \in \MREL \,.
\eeqa
We have to prove that $x_1=\dots=x_n$. Due to (\ref{DP3}) it suffices to show that $(x_1,\dots,x_n) \in \MCMP \RCEL_{(\MISE_1,\MISE_2)}$ for all pairs $(\MISE_1,\MISE_2)$ that split $\MISE$.
Let us take such a pair $(\MISE_1,\MISE_2)$ and then due to (\ref{CRCEL}) and (\ref{LHS}) we need to find a pair $(j_1,j_2)$ of the form  $(k,k+1)$ or $(n,1)$ for some $j_1 \in \MISE_1$ and $j_2 \in \MISE_2$.
If $n \notin \MISE_1$ then take the pair $(k,k+1)$ with $k:= \max \, \MISE_1$. Otherwise, let $\ell$ be the smallest integer such that $\ell,\dots,n \in \MISE_1$. If $1 \notin \MISE_1$ we can take the pair $(n,1)$. Finally, if $1 \in \MISE_1$ we take the pair $(k,k+1)$ with $k=\max \, \{j \in \MISE_1 \,|\, j < \ell\}$.

It remains to show that the conditions $(x,x) \in \MREL$ and (\ref{Acycl3}) are sufficient for the diagonal property (\ref{DP}) to hold.
Denote
\beqs
\MDIA_{j,k} \, := & \bigl\{(x_1,\dots,x_n) \in \MSET^{\times n} & \bigl| \, x_j=x_k\bigr\} \,.
\eeqs

\begin{Statement}{Lemma}\label{ALM1}
The conditions $(x,x) \in \MREL$ for all $x \in \MSET$ and
\beq\label{DP2}
\mathop{\text{\footnotesize $\bigcup$}}\limits_{(\MISE_1,\MISE_2)} \, \RCEL_{(\MISE_1,\MISE_2)} \, \supseteq \,
\MCMP \mathop{\text{\footnotesize $\bigcup$}}\limits_{j,k} \MDIA_{j,k}
\eeq
for all $n=2,3,\dots$ are necessary and sufficient for the validity of the diagonal property~$(\ref{DP})$.
\end{Statement}

\medskp

\noindent
{\it Proof} of Lemma \ref{ALM1}. 
The case $n=2$ of (\ref{DP}) shows that the condition $(x,x) \in \MREL$ is necessary (see also the beginning of Sect.~9). 
Condition~(\ref{DP2}) is also necessary since $\MDIA_n$ $\subseteq$ $\mathop{\bigcup}\limits_{j,k} \MDIA_{j,k}$ and hence, $\MCMP \MDIA_n$ $\supseteq$ $\MCMP \mathop{\bigcup}\limits_{j,k} \MDIA_{j,k}$. 
Assume now that $(x,x) \in \MREL$ and (\ref{DP}) are true for all $n=2,3,\dots$.
Then it follows that
$\mathop{\bigcup}\limits_{(\MISE_1,\MISE_2)} \, \RCEL_{(\MISE_1,\MISE_2)}$ $\subseteq$ 
$\MCMP \MDIA_n$ so it remains to prove the inverse inclusion.
We proceed by induction in $n$. For $n=2$, $\MDIA_{1,2}$ $\equiv$ $\MDIA_2$
so (\ref{DP2}) is exactly what we need.
It remains to prove, by recursion, that the only points within any $\MDIA_{j,k}$, outside $\mathop{\bigcup}\limits_{(\MISE_1,\MISE_2)} \, \RCEL_{(\MISE_1,\MISE_2)}$ are those of the complete diagonal $\MDIA_n$.
We have
$$
\mathop{\text{\footnotesize $\bigcup$}}\limits_{(\MISE_1,\MISE_2)} \, \RCEL_{(\MISE_1,\MISE_2)} \left.\raisebox{12pt}{\hspace{-2pt}}\right|_{\MDIA_{j,k}} 
\, = \,
\MDIA_{j,k} \cap
\Biggl(
\mathop{\text{\footnotesize $\bigcup$}}\limits_{\MJSE_1 \,\dot{\cup}\, \MJSE_2 \, = \, \{1,\dots,\widehat{j},\dots,\widehat{k},\dots,N,(jk)\}} \, \RCEL_{(\MJSE_1,\MJSE_2)}
\Biggr)
\,,
$$
where $(jk)$ stands for one single label and $\widehat{j}$, $\widehat{k}$ means that $j$ and $k$ are omitted.
By the recursion hypothesis the $\RCEL_{(\MJSE_1,\MJSE_2)}$'s cover all of $\MDIA_{i_1,i_2}$
except the complete diagonal
\beqs
&
x_{(j,k)} = x = x_j =x_k 
& \\ & 
= x_1 = \cdots
= \widehat{x}_j=\cdots=\widehat{x}_k=\cdots=x_n
\,. \quad \Box
&
\eeqs

We continue the proof of Theorem \ref{GenDiaLem}.
Note that (\ref{DP2}) is equivalent to
$\mathop{\bigcap}\limits_{(\MISE_1,\MISE_2)} \, \MCMP \RCEL_{(\MISE_1,\MISE_2)}$ $\subseteq$ $\mathop{\bigcup}\limits_{j,k} \MDIA_{j,k}$,
which also means that
$$
\Bigl(\mathop{\text{\footnotesize $\bigcap$}}\limits_{(\MISE_1,\MISE_2)} \, \MCMP \RCEL_{(\MISE_1,\MISE_2)}\Bigr)
\cap \MCMP \mathop{\text{\footnotesize $\bigcup$}}\limits_{j,k} \MDIA_{j,k} \, = \,
\text{\Large $\o$} \,.
$$
On the other hand, $\MCMP \mathop{\text{\footnotesize $\bigcup$}}\limits_{j,k} \MDIA_{j,k} = \mathop{\text{\footnotesize $\bigcap$}}\limits_{j,k} \MCMP \MDIA_{j,k}$ is nothing but the set of all configurations of distinct points $(x_1,\dots,x_n)$.
Thus, we need to show that the conditions $(x,x) \in \MREL$ and (\ref{Acycl3}) imply that 
there are no configurations $(x_1,\dots,x_n)$ of distinct points in the intersection
$\mathop{\bigcap}\limits_{(\MISE_1,\MISE_2)} \, \MCMP \RCEL_{(\MISE_1,\MISE_2)}$.
Let us assume on the contrary that there exists such a configuration $(x_1,\dots,x_n)$. 
Since $(x_1,\dots,x_n)$ $\in$ $\MCMP\RCEL_{(\MISE_1,\MISE_2)}$ (according to (\ref{CRCEL})) for all nontrivial splittings $\MISE_1 \dot{\cup}\MISE_2 = \{1,\dots,n\}$
there exists at least one pair $(x_{i_1},x_{i_2}) \in \MREL$ consisting of different points.
Let us take a subsequence $(x_{i_1},\dots,x_{i_m})$ of distinct points, which satisfies
$$
(x_{i_1},x_{i_2}) \in \MREL 
\,,\ \dots \,,\
(x_{i_{m-1}},x_{i_m}) \in \MREL 
\,
$$
and has a {\it maximal length}.
Take $\MISE_1=\{i_m\}$. 
As $(x_1,\dots,x_n)$ $\in$ $\MCMP\RCEL_{(\MISE_1,\MISE_2)}$
then $(x_{i_m},x_{i'}) \in \MREL$ for some $i' \in \MISE_2 \supseteq \{i_1,\dots,i_{m-1}\}$. 
Since the sequence
$(x_{i_1},\dots,x_{i_m})$ is maximal then $i'$ must be among $i_1,\dots,i_{m-1}$, say $i'=i_{\ell}$.
But then we obtain a cycle
\beqs
\podr
(x_{i_{\ell}},x_{i_{\ell+1}}) \in \MREL 
\,,\ \dots\,,\
(x_{i_{m-1}},x_{i_m}) \in \MREL 
\,,\
(x_{i_m},x_{i_{\ell}}) \in \MREL 
\eeqs
and so, $x_{i_{\ell}}=\cdots=x_{i_m}$, which contradicts the assumption that the points $x_1,\dots,x_n$ are distinct.

This completes the proof of Theorem \ref{GenDiaLem}.$\quad\Box$


\begin{remark}\label{ARM1}
If the binary relation $\MREL$ is a \textit{partial order}, i.e., if $\MREL$ satisfies the conditions
\begin{LIST}{30pt}
\item[$(\Mrel_1)$] $(x,x) \in \MREL\,$;
\item[$(\Mrel_2)$] $(x,y) \in \MREL$ \ $\MAND$ \ $(y,x) \in \MREL$ \ $\Rightarrow$ \ $x=y\,$;
\item[$(\Mrel_3)$] $(x,y) \in \MREL$ \ $\MAND$ \ $(y,z) \in \MREL$ \ $\Rightarrow$ \ $(x,z) \in \MREL\,$
\end{LIST}
$(\forall x,y,z \in X)$, then it is acyclic.
Hence, Theorem~\ref{GenDiaLem} indeed extends the validity of the Diagonal Lemma even for partial order sets since in the proof in Sect.~\ref{TSn1.2-N} one 
restricts 
oneself to orders defined in terms of a convex cone 
(reduced to its tip in the Euclidean context).
\end{remark}

\section[{\PHANT}Partitions and the causal glueing]{Partitions and the causal glueing}\label{TSn1.3-N}
\setcntrs

There is a nice relation between the system of Euclidean like causal domains $\ECEL_{\{\MISE_1,\MISE_2\}}$ and partitions which we proceed to describe.
This however will not play any role in this work and so the reader can skip the present appendix.

\medskp

We start with a brief review of the combinatorics related to partitions.

A partition $\MPRT$ of a set $\MISE$ is a set $\MPRT$ $=$ $\{\MISE_1,\dots,\MISE_k\}$ of disjoint nonempty subsets of $\MISE$ whose union is $\MISE$.
A partition $\MPRT$ can be equivalently described in each of the following two ways:

(i) As an equivalence relation\footnote{%
an {\it equivalence relation} $\sim$ on a set $\MISE$ is a binary relation on $\MISE$ such that: 
it is {\it reflexive}, $j\sim j$ ($\forall j \in \MISE$); 
{\it symmetric}, $j_1\sim j_2$ $\Rightarrow$ $j_2\sim j_1$ ($\forall j_1,j_2 \in \MISE$); 
{\it transitive}, $j_1\sim j_2$ and $j_2\sim j_3$ $\Rightarrow$ $j_1\sim j_3$ ($\forall j_1,j_2,j_3 \in \MISE$). 
Then the set $\MISE$ is uniquely divided into a disjoint subsets, $[j]_{\sim}$ $:=$ $\{i \in J | i \sim j\}$ such that $[j]_{\sim}=[i]_{\sim}$ iff $j\sim i$ and $[j]_{\sim} \cap [i]_{\sim} = \emptyset$ iff $j \nsim i$.
$[j]_{\sim}$ is called an {\it equivalence class} of $j$.
Thus, the relation between the partition $\MPRT$ and its equivalence relation $\SIM{\MPRT}$ is $\MPRT = \{[j]_{\MPRT} | j \in \MISE\}$, where $[j]_{\MPRT} := [j]_{\SIm{\MPRT}}$.} 
$\MEQU_{\MPRT}$ whose collection of equivalence classes coincides with $\MPRT$. 

(ii) As a surjection, $\pi_{\MPRT} : \MISE \to\hspace{-9.5pt}\to \MST{k}$, where $\MST{k}:=\{1,\dots,k\}$ such that $\MISE_{\ell} = \pi_{\MPRT}^{-1}\{\ell\}$,
which is unique up to reordering
(when the set $\MISE$ is linearly ordered then the members $\MISE_{\ell}$ of $\MPRT$ can be enumerated under the convention that $\min I_1 < \cdots  <\min I_k$).

The partitions over a given set $\MISE$ form a poset (partially ordered set) under the order:
\beq\label{teq1}
\MPRT_1 \leqslant \MPRT_2
\quad \Longleftrightarrow \quad
\forall I_1 \in \MPRT_1 \, \exists I_2 \in \MPRT_2 \text{ such that } I_1 \subseteq I_2 \,,
\eeq
i.e., iff every piece in $\MPRT_1$ is a subset of a set in $\MPRT_2$
(one can also say that $\MPRT_1$ is {\it finer} than $\MPRT_2$).
Note 
that if $|\MPRT|$ stands for the cardinality of $\MPRT$ then:
$$
\MPRT_1 \leqslant \MPRT_2 \quad\Longrightarrow\quad |\MPRT_1| \geqslant |\MPRT_2| \,.
$$

Under the above equivalent characterizations of partitions Eq.~(\ref{teq1}) is equivalent to
\beqa\label{teq2}
\MPRT_1 \leqslant \MPRT_2
\quad \Longleftrightarrow \podr\quad
\forall j, j' \in \MISE : \ j \,\SIM{\MPRT_1}\, j' \ \Rightarrow \ j \,\SIM{\MPRT_2}\, j'
\,,
\\ \label{teq3}
\MPRT_1 \leqslant \MPRT_2
\quad \Longleftrightarrow \podr\quad
\pi_{\MPRT_2} \text{ factors trough } \pi_{\MPRT_1} 
\,
\eeqa
(i.e., $\pi_{\MPRT_2} = \psi \circ \pi_{\MPRT_1}$ for some map $\psi$).

Then the poset of all partitions over the set $\MISE$ forms a lattice:
the infimum $\MPRT_1 \wedge \MPRT_2$ is
$$
\MPRT_1 \wedge \MPRT_2 \, := \, \bigl\{\MISE_1 \cap \MISE_2 \,\bigl|\, I_1 \in \MPRT_1\,,\, \MISE_2 \in \MPRT_2\,,\, \MISE_1 \cap \MISE_2 \neq \emptyset\bigr\} \,
$$
and the supremum $\MPRT_1 \vee \MPRT_2$ is more complicated for an explicit description.

\medskp

\noindent
{\it Examples.}
Let $\MISE=\{1,\dots,n\}$. Then the maximal partition $\MPRT$ is $\{\{1,\dots,n\}\}$ and the minimal is $\{\{1\},\dots,\{n\}\}$. In general the maximal partition is characterized by $|\MPRT|=1$ while the minimal partition over $\MISE$ is characterized by $|\MPRT|=|\MISE|$. In particular, $1 \leqslant |\MPRT| \leqslant |\MISE|$. For $n=7$:
\beqs
\{\{1,4,5\},\{2,7\},\{3,6\}\} \wedge \{\{1,2\},\{3,4,7\},\{5,6\}\} \, = \podr
\{\{1\},\dots,\{7\}\}
\,,\\
\{\{1,4,5\},\{2,7\},\{3,6\}\} \vee \{\{1,2\},\{3,4,7\},\{5,6\}\} \, = \podr
\{\{1,\dots,7\}\}
\,.
\eeqs

\medskp

We now proceed with the hierarchy of diagonals in a Cartesian power and its relation to partitions.

For a given set $\MSET$ and index set $\MISE$ we 
consider 
the Cartesian power $\MSET^{\MISE}$.
For every nonempty $\MJSE \subseteq \MISE$ we introduce first the \textit{partial diagonal}
\beq\label{eq-partdia}
\MDIA_{\MJSE} \, := \, \bigl\{(\mxx_j)_{j \in \MISE} \in \MSET^{\MISE} 
\,\bigl|\,
\mxx_i=\mxx_j \, (\forall i,j \in \MJSE)
\bigr\} \,.
\eeq
If $\MJSE$ has one element then $\MDIA_{\MJSE} \equiv \MSET^{\MISE}$.
If $\MJSE=\MISE$ then $\MDIA_{\MISE}$ is the so called {\it total diagonal} (also called ``thin diagonal'').
Then for a partition $\MPRT$ of $\MISE$ we set:
\beq\label{eq-prt-dia}
\MDIA_{\MPRT} \, := \, \mathop{\bigcap}\limits_{\MJSE \in \MPRT}
\MDIA_{\MJSE}.
\eeq
Equivalently:
\beq\label{eq-prt-dia-equ}
(\mxx_j)_{j \in \MISE} \in \MDIA_{\MPRT}
\quad \Longleftrightarrow \quad
\forall i,j \in \MISE :\
i \,\SIM{\MPRT}\, j \ \Rightarrow \ \mxx_i = \mxx_j \,.
\eeq

Then it follows that
\beqa\label{teq4a}
\MPRT_1 \leqslant \MPRT_2 
& \Longleftrightarrow &
\MDIA_{\MPRT_1} \supseteq \MDIA_{\MPRT_2} \,,
\\ \label{teq4b}
\MDIA_{\MPRT_1} \cap \MDIA_{\MPRT_2}
& = &
\MDIA_{\MPRT_1 \vee \MPRT_2} \,.
\eeqa
Equation (\ref{teq4a}) follows from (\ref{eq-prt-dia-equ}).
To prove (\ref{teq4b}) it is convenient to introduce a partition $\MPRT^{\Mxx}\,$, which is canonically defined for every point $\Mxx := (\mxx_j)_{j \in \MISE} \in \MSET^{\MISE}$ by:
\beq\label{teq9}
i \,\SIM{\MPRT^{\Mxx}\,}\, j \quad \Longleftrightarrow \quad \mxx_i = \mxx_j
\eeq
i.e., two indices $i,j$ belong to one piece of $\MPRT^{\Mxx}\,$ iff the corresponding points are equal.
Then it follows that
\beq\label{teq10}
\Mxx \in \MDIA_{\MPRT}
\quad \Longleftrightarrow \quad
\MPRT \leqslant \MPRT^{\Mxx}\, \,.
\eeq
and hence,
\beqs
\Mxx \in \MDIA_{\MPRT_1} \cap \MDIA_{\MPRT_2}
\quad \Longleftrightarrow \podr\quad
\MPRT_1 \leqslant \MPRT^{\Mxx}\, \ \text{ and } \ \MPRT_2 \leqslant \MPRT^{\Mxx}\,
\quad \Longleftrightarrow \quad
\MPRT_1 \vee \MPRT_2 \leqslant \MPRT^{\Mxx}\,
\\
\quad \Longleftrightarrow \podr\quad
\Mxx \in \MDIA_{\MPRT_1 \vee \MPRT_2}
\,,
\eeqs
which proves (\ref{teq4b}).

As a corollary of Eq.~(\ref{teq4b}) it follows that all intersections of the partial diagonals $\MDIA_{\MJSE}$~(\ref{eq-partdia}) are of form $\MDIA_{\MPRT}$ for partitions $\MPRT$ of $\MISE$.
Indeed, $\MDIA_{\MJSE}$ is also of a form $\MDIA_{\MPRT}$ for the partition
$$
\MPRT \, = \,
\MPRT_{\MJSE} \, := \,
\{J\} \cup \bigl\{\{i\} \,\bigl|\, i \in \MISE \backslash \MJSE \bigr\} \,.
$$

Other sets important in the renormalization recursion are the ``causal domains''
defined for every partition $\MPRT$ of $\MISE$ by:
\beq\label{eq-causal-dom}
\ECEL_{\MPRT} 
\, := \,
\, \bigl\{
\Mxx =
(\mxx_j)_{j \in \MISE} \in \MSET^{\MISE} 
\,\bigl|\,
i \,\NSIM{\MPRT}\, j \,\Rightarrow\,
\mxx_i \neq \mxx_j \, (\forall i,j \in \MISE)
\bigr\} \,.
\eeq
It follows then that
\beq\label{teq12}
\Mxx \in \ECEL_{\MPRT}
\quad \Longleftrightarrow \quad
\MPRT^{\Mxx}\, \leqslant \MPRT \,.
\eeq
Hence, similarly to the above arguments we obtain:
\beqa\label{teq13}
\MPRT_1 \leqslant \MPRT_2 
& \Longleftrightarrow &
\ECEL_{\MPRT_1} \subseteq \ECEL_{\MPRT_2} \,,
\\ \label{teq14}
\ECEL_{\MPRT_1} \cap \ECEL_{\MPRT_2}
& = &
\ECEL_{\MPRT_1 \wedge \MPRT_2} \,.
\eeqa
On the other hand,
\beqs
\Mxx \in \ECEL_{\MPRT_1} \cup \ECEL_{\MPRT_2}
\quad \Longleftrightarrow \podr\quad
\MPRT^{\Mxx}\, \leqslant \MPRT_1 \text{ or } \MPRT^{\Mxx}\, \leqslant \MPRT_2
\quad \Longrightarrow \quad
\MPRT^{\Mxx}\, \leqslant \MPRT_1 \vee \MPRT_2
\\
\quad \Longleftrightarrow \podr\quad
\Mxx \in \ECEL_{\MPRT_1 \vee \MPRT_2}
\,,
\eeqs
and hence,
\beq\label{teq15}
\ECEL_{\MPRT_1} \cup \ECEL_{\MPRT_2}
\, \subseteqq \, \ECEL_{\MPRT_1 \vee \MPRT_2}\,.
\eeq
Similarly,
\beq\label{teq16}
\MDIA_{\MPRT_1} \cup \MDIA_{\MPRT_2}
\, \subseteqq \, \MDIA_{\MPRT_1 \wedge \MPRT_2}\,.
\eeq


\noindent
\begin{remark}\label{REM-B-1} $(a)$ 
If $\MSET$ is a vector space then $\MDIA_{\MPRT}$ are vector subspaces of $\MSET^{\MISE}$. The correspondence $\MPRT \to \MDIA_{\MPRT}$ is an injective lattice anti-morphism (i.e., it reverses the order) from the lattice of partitions to the lattice of all linear subspaces. In particular, 
$$
\MDIA_{\MPRT_1 \wedge \MPRT_2} \, = \, 
\MDIA_{\MPRT_1} + \MDIA_{\MPRT_2} \,
$$ 
and thus,
in general
$\MDIA_{\MPRT_1} \cup \MDIA_{\MPRT_2} \subsetneqq \MDIA_{\MPRT_1 \wedge \MPRT_2}$ in Eq.~(\ref{teq16}) above.
The sublattices in the lattice of all linear subspaces of a vector space are called {\it geometric lattices}.

$(b)$ If $\MSET$ is a topological space then $\ECEL_{\MPRT}$ is an open neighbourhood of $\MDIA_{\MPRT}$.


$(c)$ Note that
\beq\label{teq17}
\ECEL_{\MPRT_1} \cap \MDIA_{\MPRT_2} 
\, = \, \emptyset \qquad
\text{iff } \MPRT_2 \nleqslant \MPRT_1 \,.
\eeq
\end{remark}


\begin{Statement}{Lemma}\label{DIAG-LM}
$\bf (The \ Diagonal \ Lemma.)$
$$
\mathop{\bigcup}\limits_{|\MPRT|  \, \geqslant \, 2}
\ECEL_{\MPRT} \, = \,
\mathop{\bigcup}\limits_{|\MPRT| \, = \, 2}
\ECEL_{\MPRT} \, = \, \MSET^{\MISE} \backslash \MDIA_{\MISE} \,.
$$
\end{Statement}

\medskp

The {\it proof} 
of the Diagonal Lemma is straightforward: first the inclusions
\(
\MSET^{\MISE} \backslash \MDIA_{\MISE} 
\supseteq 
\mathop{\textstyle \bigcup}\limits_{|\MPRT|  \, \geqslant \, 2}
\ECEL_{\MPRT} 
\supseteq 
\mathop{\textstyle \bigcup}\limits_{|\MPRT| \, = \, 2}
\ECEL_{\MPRT} \, 
\)
are obvious. To prove $\mathop{\textstyle \bigcup}\limits_{|\MPRT| \, = \, 2} \ECEL_{\MPRT} \supseteq \MSET^{\MISE} \backslash \MDIA_{\MISE}$ we note that if $(\mxx_j)_{j \in \MISE} \notin \MDIA_{\MISE}$ then there exists a pair $\mxx_j \neq \mxx_k$.
Hence, we set $\MPRT := \{\MJSE,\MISE\backslash\MISE\}$ with $\MJSE:=\{j' \in \MISE | \mxx_{j'} = \mxx_j\}$ $\neq\emptyset$ $\neq\MISE$.


\begin{remark}\label{QREM}
A generalization of the diagonal lemma is the following identity:
$$
\mathop{\bigcup}\limits_{\mathop{}\limits^{\MQRT \leqslant \MPRT}_{|\MQRT|  \, \geqslant \, |\MPRT|+1}} \hspace{-4pt}
\ECEL_{\MQRT} \, = \,
\mathop{\bigcup}\limits_{\mathop{}\limits^{\MQRT \leqslant \MPRT}_{|\MQRT|  \, = \, |\MPRT|+1}} \hspace{-4pt}
\ECEL_{\MQRT} \, = \, \ECEL_{\MPRT} \backslash \MDIA_{\MPRT} \,.
$$
\end{remark}

\begin{Statement}{Proposition}\label{UNPrpp}
It follows from the recursively assumed condition $\Mref{teq19}$ that for a general nontrivial partition $\MQRT$ of $\MJSE$ we have:
\beq\label{teq20}
G^{ext}_{\MJSE} 
\hspace{1pt}\Bigl|\raisebox{-7pt}{\hspace{1pt}}_{\ECEL_{\MQRT}} \, = \, \Biggl(\mathop{\prod}\limits_{\MKSE \in \MQRT} G^{ext}_{\MKSE} \Biggr) 
\, G_{\MQRT}
\hspace{1pt}\Bigl|\raisebox{-7pt}{\hspace{1pt}}_{\ECEL_{\MQRT}}
,\quad 
G_{\MQRT} \, = \,
\mathop{\prod}\limits_{\mathop{}\limits^{j_1 \NSIm{\MQRT} j_2}_{j_1 < j_2}}
G_{j_1,j_2} 
\,.
\eeq
\end{Statement}

\medskp

\noindent
{\it Proof.}
We prove the lemma by induction in the cardinality $|\MQRT|$ of $\MQRT$.
For $|\MQRT|=2$ Eq.~(\ref{teq20}) coincides with (\ref{teq19}) for $\MJSE=\MJSE_1 \dot{\cup} \MJSE_2$.
Assume we have established (\ref{teq20}) for all $\MQRT_1$ with $|\MQRT_1|<k=|\MQRT|$. If $\MQRT=\{J_1,\dots,J_k\}$ we apply the induction hypothesis to the partition $\MQRT_1=\{J_1,\dots,J_{k-1}\disjuni J_k\}$:
\beq\label{teq21}
G^{ext}_{\MJSE} 
\hspace{1pt}\Biggl|\raisebox{-12pt}{\,}_{\ECEL_{\MQRT_1}} \, = \, \Biggl(\mathop{\prod}\limits_{\MKSE \in \MQRT_1} G^{ext}_{\MKSE} \Biggr) \mathop{\prod}\limits_{\mathop{}\limits^{j_1 \NSIm{\MQRT_1} j_2}_{j_1 < j_2}}
G_{j_1,j_2} 
\hspace{1pt}\Biggl|\raisebox{-12pt}{\,}_{\ECEL_{\MQRT_1}} \,.
\eeq
Then $\MQRT \leqslant \MQRT_1$ and $\ECEL_{\MQRT} \subseteqq \ECEL_{\MQRT_1}$. 
Hence, restricting further both sides of (\ref{teq21}) from $\ECEL_{\MQRT_1}$ to $\ECEL_{\MQRT}$ and using the equation:
$$
G^{ext}_{\MJSE_{k-1} \disjuni \MJSE_k} 
\hspace{1pt}\Biggl|\raisebox{-12pt}{\,}_{\ECEL_{\{\MJSE_{k-1},\MJSE_k\}}} \, = \, G^{ext}_{\MJSE_{k-1}} \, G^{ext}_{\MJSE_k} \mathop{\prod}\limits_{\mathop{}\limits^{j_1 \in \MJSE_{k-1}}_{j_2 \in \MJSE_k}}
G_{j_1,j_2} 
\hspace{1pt}\Biggl|\raisebox{-12pt}{\,}_{\ECEL_{\{\MJSE_{k-1},\MJSE_k\}}}
$$
we obtain (\ref{teq20}) from (\ref{teq21}).$\quad\Box$.


\section[{\PHANT}Proof of Theorem \ref{Theorem4.2} (Sect. \ref{NSEC-4.2NN})]{Proof of Theorem \ref{Theorem4.2} (Sect. \ref{NSEC-4.2NN})}\label{Theorem4.2-pr}
\setcntrs

\subsection{Constructing the secondary renormalization maps}\label{Sect-ConSecMaps}

\begin{Statement}{Lemma}\label{SECMAP-constr}
Assume that for some $n>1$ we have constructed linear maps $\RENMAP_m$ $(\ref{RENMAP})$ for $m=2,\dots,n-1$ $($$\RENMAP_1:=1$$)$,
which satisfy properties $(r_1)$--$(r_3)$ of Theorem~\ref{Theorem4.2}
$($respectively for the Minkowski or for the Euclidean cases$)$.
Let $G_n \equiv G_{\MISE} \in \AMPSPA_{\MISE} \equiv \AMPSPA_n$,
$\MISE:=\{1,\dots,n\}$ 
be any factorizable unrenormalized amplitude
given by Eq. $(\ref{GISEdef})$
$($cf. the conventions for Eqs. $(\ref{RENMAPCAU-M})$ and $(\ref{RENMAPCAU-E})$$)$.
Then the hypothesis of Theorem \ref{THM-MINK-REC} $($resp., \ref{THM-EUCL-REC} for the Euclidean case$)$ are satisfied with:
\begin{LIST}{26pt}
\item[$\bullet$]
$G^{ext}_{\MJSE}$ $:=$ $\RENMAP_{\MJSE} (G_{\MJSE})$ for every proper subset $\emptyset \subsetneqq \MJSE \subsetneqq \MISE$$;$
\item[$\bullet$]
$W_{j_1,j_2}(\x_{j_2},\x_{j_1})$\hspace{2pt}$:=$\hspace{2pt}%
$\BVMAP_{\hspace{1pt}\mathcal{T}(\{j_1\} \prec \, \{j_2\})} G_{j_1,j_2}(\x_{j_2},\x_{j_1})$,\
for $j_1$\hspace{2pt}$\neq$\hspace{2pt}$j_2$\
in the 
Min-{\linebreak}ko\-w\-s\-ki case
$($then 
$G_{j_2,j_1}(\x_{j_2},$ $\x_{j_1}):=G_{j_1,j_2}(\x_{j_1},$ $\x_{j_2})$\
and\
$W_{j_2,j_1}(\x_{j_2},$ $\x_{j_1})=W_{j_1,j_2}(\x_{j_1},$ $\x_{j_2})$$)$$;$
\item[$\bullet$]
$W_{(\MISE_1,\MISE_2)}$ $:=$ $\BVMAP_{\hspace{1pt}\mathcal{T}(\MISE_1 \prec \, \MISE_2)}$
$G_{\{\MISE_1,\MISE_2\}}$
for every partition 
$\MISE=\MISE_1\disjuni \MISE_2$
$($in the Minkowski case$)$.
\end{LIST}
By Theorem \ref{THM-MINK-REC} $($resp., \ref{THM-EUCL-REC}$)$ and Corollary \ref{CrXX1}
we obtain a distribution belonging to
$\DP_{\BLT} \biggl(\frac{\textstyle \R^{Dn} \backslash \Delta_n}{\textstyle\Delta_n} \biggr)$
--
let us denote it by $\SECMAP_n (G_n)$.
Then the assignment $G_n \mapsto \SECMAP_n (G_n)$
$($defined so far for factorizable $G_n$$)$ 
extends uniquely to a linear map $\SECMAP_n$ $(\ref{SECRENMAP})$.
Furthermore,
the so constructed $\SECMAP_n$ satisfies properties
$(r_1)$--$(r_3)$ of Theorem~\ref{Theorem4.2}
$($note that the latter properties have a straightforward reformulation for $\SECMAP_n$$)$.
\end{Statement}

\medskp

For the proof of the above lemma we shall need the following algebraic statement.

\medskp

\begin{Statement}{Lemma}\label{LEM-C2}
For an arbitrary finite index set $\MJSE$ let us denote by $\REGSPA_{\MJSE}$ the subalgebra of $\AMPSPA_{\MJSE}$, which consists of all polynomials.
Let us also denote by $\AMPSPA_{\{\MISE_1,\MISE_2\}}$ the subalgebra of $\AMPSPA_{\MJSE}$, which is generated by all amplitudes of the form $G_{\{\MISE_1,\MISE_2\}}$ from Eq. $(\ref{GISEdef})$.
Then the assignment
\beq\label{SplitAlg-0}
\bigl(G_{\MISE_1} \otimes G_{\MISE_2}\bigr) \hspace{1pt}
\otimes_{\REGSPA_{\MISE_1} \otimes \hspace{2pt} \REGSPA_{\MISE_2}}
\,
 G_{\{\MISE_1,\MISE_2\}}
\, \mapsto \,
G_{\MISE_1} G_{\MISE_2} G_{\{\MISE_1,\MISE_2\}} \, \equiv \, G_{\MISE}
\,,
\eeq
defined for any factorizable amplitude $G_{\MISE}$ $(\ref{GISEdef})$,
uniquely extends to an algebra isomorphism 
\beq\label{SplitAlg}
\bigl(
\AMPSPA_{\MISE_1}
\otimes 
\AMPSPA_{\MISE_2} 
\bigr)
\hspace{1pt}
\otimes_{\REGSPA_{\MISE_1} \otimes \hspace{2pt} \REGSPA_{\MISE_2}}
\,
\AMPSPA_{\{\MISE_1,\MISE_2\}} 
\ \cong \ 
\AMPSPA_{\MISE}
\,.
\eeq
\end{Statement}

\medskp

\begin{Proof}
Let us introduce the polynomials
$$
\QU_{\MJSE} \, := \,
\mathop{\prod}\limits_{\mathop{}\limits^{j,k \, \in \, J}_{j \, < \, k}}
\bigl(\x_{j}-\x_{k}\bigr)^2 \ \in \ \REGSPA_{\MJSE}
\,, \qquad
\QU_{\{I_1,I_2\}} \, := \,
\mathop{\prod}\limits_{\mathop{}\limits^{j_1 \, \in \, I_1}_{j_2 \, \in \, I_2}}
\bigl(\x_{j_1}-\x_{j_2}\bigr)^2
\ \in \ \REGSPA_{\MISE}
$$
for $J=I,I_1,I_2$, so that
$$
\QU_{\MISE} \, = \, \QU_{\MISE_1} \, \QU_{\MISE_2} \, \QU_{\{I_1,I_2\}} \,,
$$
and
$$
\AMPSPA_{\MJSE} \, = \, \REGSPA_{\MJSE} \bigl[\QU_{\MJSE}^{-1}\bigr]
\,,\qquad
\AMPSPA_{\{I_1,I_2\}} \, = \, \REGSPA_{\MISE} \bigl[\QU_{\{I_1,I_2\}}^{-1}\bigr]
\,,
$$
where $\REGSPA[P^{-1}]$ for an element $P$ of a ring $\REGSPA$ stands for
the localized ring\footnote{%
see e.g.,
Chapter 3 of Atiyah, Macdonald, ``Introduction to commutative algebra'' (1969)}
$\REGSPA$ at $P$ (in case of no zero divisors,
as here, this is the ring of fractions with denominators
that are powers of $P$).

Now, we start with the trivial identity
$$
\REGSPA_{\MISE}
\, \cong \,
\bigl(\REGSPA_{\MISE_1} \otimes \REGSPA_{\MISE_2}\bigr)
\otimes_{\REGSPA_{\MISE_1} \otimes \REGSPA_{\MISE_2}}
\REGSPA_{\MISE} \,
$$
and localize first both sides at $\QU_{\MISE_1}\QU_{\MISE_2}$ to obtain
$$
\REGSPA_{\MISE} \bigl[\QU_{\MISE_1}^{-1}\QU_{\MISE_2}^{-1}\bigr]
\, \cong \,
\Bigl(\AMPSPA_{\MISE_1} \otimes \AMPSPA_{\MISE_2}\Bigr)
\otimes_{\REGSPA_1 \otimes \REGSPA_2}
\REGSPA_{\MISE} \,,
$$
and then localize at $\QU_{\{I_1,I_2\}}$:
$$
\bigl(\REGSPA_{\MISE} \bigl[\QU_{\MISE_1}^{-1}\QU_{\MISE_2}^{-1}\bigr] \bigr)
\bigl[\QU_{\{I_1,I_2\}}^{-1}\bigr]
\, \cong \,
\Bigl(\AMPSPA_{\MISE_1} \otimes \AMPSPA_{\MISE_2}\Bigr)
\otimes_{\REGSPA_1 \otimes \REGSPA_2}
\AMPSPA_{\{I_1,I_2\}} \,
$$
to obtain the isomorphism (\ref{SplitAlg}).
As the factorizable amplitudes $G_{\MISE}$ $(\ref{GISEdef})$ linearly span $\AMPSPA_{\MISE}$ the isomorphism (\ref{SplitAlg}) is the unique linear extension of (\ref{SplitAlg-0}).
\end{Proof}
\medskp

\medskp

\noindent
{\it Proof of Lemma \ref{SECMAP-constr}.}
For any nontrivial partition $\MISE=\MISE_1 \disjuni \MISE_2$ let us consider the assignments
\beqa\label{RENMAPCAU-E-m1}
G_{\MISE} \podr \longmapsto \,
\RENMAP_{\MISE_1} \bigl(G_{\MISE_1}\bigr) \,
\RENMAP_{\MISE_2} \bigl(G_{\MISE_2}\bigr) \,
G_{\{\MISE_1,\MISE_2\}}
\vrestr{12pt}{\ECEL_{\{\MISE_1,\MISE_2\}}} 
\qquad (\text{Euclid})\, ,
\\ \label{RENMAPCAU-M-m1}
G_{\MISE} \podr \longmapsto \,
\RENMAP_{\MISE_1} \bigl(G_{\MISE_1}\bigr) \,
\RENMAP_{\MISE_2} \bigl(G_{\MISE_2}\bigr) \,
\BVMAP_{\hspace{1pt}\mathcal{T}(\MISE_1 \prec \, \MISE_2)} \, G_{\{\MISE_1,\MISE_2\}}
\vrestr{12pt}{\MCEL_{(\MISE_1,\MISE_2)}} 
\quad \ \ (\text{Minkowski})
\,. \nonumber \\[-7pt]
\eeqa
The maps $\RENMAP_{\MISE_k}$ for $k=1,2$ are $\REGSPA_{\MISE_k}$--linear by property $(r_3)$ of Theorem \ref{Theorem4.2} that is assumed for $\RENMAP_m$ for $m < n$.
Hence, assignments (\ref{RENMAPCAU-E-m1}) and (\ref{RENMAPCAU-M-m1}) extend to unique linear maps
\beqa\label{RENMAPCAU-E-m2}
\SECMAP_{\{\MISE_1,\MISE_2\}} \podr : \, \AMPSPA_{\MISE} \, \longrightarrow \, \DP \bigl(\ECEL_{\{\MISE_1,\MISE_2\}}\bigr) 
\qquad (\text{Euclid})\, ,
\\ \label{RENMAPCAU-M-m2}
\SECMAP_{(\MISE_1,\MISE_2)} \podr : \, \AMPSPA_{\MISE} \, \longrightarrow \, \DP \bigl(\MCEL_{(\MISE_1,\MISE_2)}\bigr)
\quad \ \ (\text{Minkowski}) \,,
\eeqa
respectively.
Then, the linear map $\SECMAP_{\MISE}$ is uniquely determined by glueing the values of $\SECMAP_{\{\MISE_1,\MISE_2\}}$
or $\SECMAP_{(\MISE_1,\MISE_2)}$ (respectively).

The proof of properties $(r_1)$--$(r_3)$ of Theorem~\ref{Theorem4.2} stated for $\SECMAP_n$ 
is straightforward.
Property $(r_1)$ is a consequence of Lemma \ref{EucSym}
(cf. Remark \ref{RnXXX1}).
Properties $(r_{2,2})$ and $(r_3)$ follow as each of maps (\ref{RENMAPCAU-E-m2}) or (\ref{RENMAPCAU-M-m2}) (respectively) satisfy them and hence, $\SECMAP_{\MISE}$ will after gluing the values.
Property $(r_{2,1})$ of permutation symmetry follows, from the one hand, by the assumed similar property for $\RENMAP_m$ for $m<n$ and, from the other, by the trivial permutation symmetry of the glueing procedure
(exchanging simultaneously the order of the domains together with the glued distributions on them does not change the result).$\quad\Box$


\begin{remark}\label{REM-C1-NN}
Note that for $n=2$ the decomposition (\ref{RENDECOM}), 
$\RENMAP_2 = \PRIMAP_2 \circ \SECMAP_2$, 
is trivial in the Euclidean case as the map $\SECMAP_2$ is simply the composition of the standard inclusions $\AMPSPA_2 \hookrightarrow \CI(\R^D \backslash \{0\}) \hookrightarrow \DP (\R^D \backslash \{0\})$.
In the Minkowski case this step is again straightforward and $\SECMAP_2 : \AMPSPA_2 \to \DP (\R^D \backslash \{0\})$ coincides locally with some of the boundary value maps $\BVMAP_{\hspace{1pt}\mathcal{T}(\{1\} \prec \, \{2\})}$ or
$\BVMAP_{\hspace{1pt}\mathcal{T}(\{2\} \prec \, \{1\})}$.
Thus, the renormalization map $\RENMAP_2$ essentially coincides with the primary renormalization map $\PRIMAP_2$.
\end{remark}

\subsection{
The choice of the isomorphism $\frac{\R^{Dn}}{\Delta_n} \, \cong \, \R^N$ $(\ref{IDENTIF})$ and
the unification of the permutation and rotation symmetries in the Euclidean case}\label{Unif-sect}

One can combine the permutation and rotation symmetries of the renormalization maps as coming from a unique big rotation symmetry by suitably choosing the isomorphism (\ref{IDENTIF}).

The group $O(N)$ for $N=D(n-1)$ contains two mutually commuting subgroups isomorphic to $O(D)$ and $\Ss_n$, respectively, in the following way. We consider $\R^N\equiv \R^{D(n-1)}$ as a tensor product of two Euclidean spaces, $\R^D \otimes \R^{n-1}$. The first one is $\R^D$ with the {\it standard} ({\it diagonal}) Euclidean metric and on this tensor factor the subgroup $O(D)$ is acting. The second tensor factor is $\R^{n-1}$ on which we shall realize the action of $\Ss_n$ but we take a {\it non-standard} (i.e., {\it non-diagonal}) euclidean metric on $\R^{n-1}$ in order to make the action of $\Ss_n$ orthogonal (Euclidean). Let us explain both: the action of $\Ss_n$ on $\R^{n-1}$ and the Euclidean metric on $\R^{n-1}$. First, we take $\R^n$ (with $n$ not $n-1$) with the standard (diagonal) Euclidean metric and consider the action of $\Ss_n$ on $\R^n$ induced by the permutations of coordinates, \((v^1,\dots,v^{n})$ $\mathop{\mapsto}\limits^{\sigma}$ $(v^{\sigma^{-1}(1)},\dots,v^{\sigma^{-1}(n)})\). Clearly, the latter action of $\Ss_n$ on $\R^n$ is orthogonal (Euclidean). Furthermore, the straight line (row) generated by the vector $(1,\dots,1)$ is invariant with respect to the action of $\Ss_n$, and hence, its orthogonal complement $(1,\dots,1)^{\perp}$ is also $\Ss_n$-invariant. So, $(1,\dots,1)^{\perp}$ is a vector space isomorphic to $\R^{n-1}$ on which the group $\Ss_n$ acts. As an orthogonal complement $(1,\dots,1)^{\perp}$ has Euclidean structure and the action of $\Ss_n$ on $(1,\dots,1)^{\perp}$ is orthogonal (Euclidean). Thus, we construct Euclidean structure on $\R^N \equiv \R^{D(n-1)}$, which differs from the standard (diagonal) one: it is a tensor product of the standard Euclidean space $\R^D$ on which $O(D)$ acts and $\R^{n-1}$ equipped with a non-diagonal Euclidean metric and an action of $\Ss_n$, which is orthogonal for the non-diagonal metric. Since the groups $O(D)$ and $\Ss_n$ act on different tensor factors, their actions commute.

Another way of presenting the above construction is to consider $\R^{Dn} = \R^D \otimes \R^n = (\R^D)^{\times n}$ with the standard Euclidean metric and the mutually commuting actions on it of $O(D)$ and $\Ss_n$: $O(D)$ acts on $\R^D$ and $\Ss_n$ acts on $\R^n$. Then the total diagonal $\Delta_n \subset (\R^D)^{\times n} = \R^{Dn}$ is isomorphic to $\R^D$ and on the other hand, $\Delta_n$ is invariant for both actions: of $O(D)$ and $\Ss_n$. Hence, the orthogonal complement of $\Delta_n$, which is isomorphic to $\R^N\equiv \R^{D(n-1)}$, is also invariant with respect to the both actions.
We can write the obtained form of the isomorphism (\ref{IDENTIF}) as:
\beq\label{IDENTIF2}
\frac{\R^{Dn}}{\Delta_n} \, \cong \, \Delta_n^{\perp} \, \cong \, \R^N \,.
\eeq

\begin{example}\label{Ex-C-1}
Let $D=1, n=3, N=D(n-1)=2$. The total diagonal in $\R^{Dn}\equiv\R^3$ is the straight line spanned by $(1,1,1)$. The orthogonal complement $(1,1,1)^{\perp}$ is isomorphic to $\R^2$ and in it the partial diagonals $\{\x_1=\x_2\}$, $\{\x_1=\x_3\}$ and $\{\x_2=\x_3\}$ are reduced to three straight lines passing trough the origin and crossing each other on an angle of 60 degrees. Thus, the group $\Ss_3$ acts on the plane $\R^2$ as the dihedral group $D_3$.
\end{example}


This picture is equivalent to what in Physics is called ``the system of the center of mass'', or in Mathematics, ``barycentric coordinates'': we consider instead of the basic differences $\x_1-\x_n,\dots,\x_{n-1}-\x_n$ the differences $\x_1-\mathrm{X},\dots,\x_n-\mathrm{X}$, where $\mathrm{X}=(\x_1+\cdots+\x_n)/n$ is the mass center and the differences $\x_1-\mathrm{X},\dots,\x_n-\mathrm{X}$ are subject to one linear relation (their sum is zero).

\medskp

\begin{Statement}{Corollary}\label{Cr-C-1}
Let us chose a sequence of hyper-surfaces $\MHYP_N \subseteq \R^N$ for $N=D(n-1)$ $($$n=2,3,\dots$$)$ that are spheres centered at the origin and set
$\PRIMAP_n$ $\cong$ $\PRIMAP^{\MHYP_N}$ according to the transfer
$(\ref{TransfP-n})$ and the identification $(\ref{IDENTIF2})$.
Then the sequence of maps satisfy the conditions of Proposition~\ref{Df-4.4-1mod}.
\end{Statement}

\subsection{Proof of Proposition \ref{Df-4.4-1mod}
$($completing the proof of Theorem \ref{Theorem4.2} in the Euclidean case$)$}\label{Unif-sect-1}

We start with the remark that recursive condition
(\ref{RENMAPCAU-E}) (as well as, (\ref{RENMAPCAU-M}) in the Minkowski case)
are already ensured for $\RENMAP_n$ according to decomposition $\RENMAP_n = \PRIMAP_n \circ \SECMAP_n$ (\ref{RENDECOM}) and the construction of $\SECMAP_n$ by Lemma \ref{SECMAP-constr}.
Since both maps, $\SECMAP_n$ and $\PRIMAP_n$, are graded according to Lemma \ref{SECMAP-constr} and $(p_1)$, respectively, it follows that $\RENMAP_n$ satisfies $(r_1)$ of Theorem \ref{Theorem4.2}.
Properties $(r_{2,1})$ and $(r_{2,2})$ are similarly implied by the corresponding equivariance properties of $\SECMAP_n$ combined with property $(p_2)$ of $\PRIMAP_n$.
Property $(r_3)$ follows by $(p_3)$ and the corresponding property of $\SECMAP_n$.

\subsection{Restoration of Lorentz covariance}

Theorem \ref{Theorem4.2} was proven by Proposition \ref{Df-4.4-1mod} and Corollary \ref{Cr-C-1} in the Euclidean case.
However, the argument used in the proof works also in the Minkowski case apart from covariance property $(r_{2,2})$.
Indeed, in Lemma \ref{SECMAP-constr} we can drop condition $(r_{2,2})$ for both, $\RENMAP_m$ ($n=2,\dots,n-1$) and $\SECMAP_n$.
Thus, we can construct a system of renormalization maps $\RENMAP_n$ that satisfy the Minkowski version of Theorem \ref{Theorem4.2} possibly without property $(r_{2,2})$.
We shall show in this subsection how we can additionally fulfill the latter property by a standard cohomological argument (\cite{S82/93} \cite{DF}).

We proceed by induction in the number of points $n=2,3,\dots$.
For $n=2$ we have already explicitly constructed $\RENMAP_2$ in the Minkowski case
(cf. Remark \ref{RMR4.3} $(b)$).
Assume that for some $n>2$ we have constructed linear maps $\RENMAP_m$ $(\ref{RENMAP})$ for $m=2,\dots,n-1$ $($$\RENMAP_1:=1$$)$,
which satisfy properties $(r_1)$--$(r_3)$ of Theorem~\ref{Theorem4.2} in the Minkowski case.
Let $\SECMAP_n$ be the linear map constructed by Lemma \ref{SECMAP-constr}.
Applying to it some map $\PRIMAP_n$ $\cong$ $\PRIMAP_{(N)}$ ($N=D(n-1)$, cf. Eq. (\ref{TransfP-n}))
we get by setting $\RENMAP_n = \PRIMAP_N \circ \SECMAP_n$ a renormalization map that satisfies all the properties of Theorem \ref{Theorem4.2} except $(r_{2,2})$.
We would like to modify $\RENMAP_n' := \RENMAP_n + \mathcal{Q}_n$
in order to fulfill all properties $(r_1)$--$(r_3)$.
Here, $\mathcal{Q}_n$ is a linear map $\AMPSPA_n \to \DP_{\BLT} [\ZERN]$
and $\DP_{\BLT} [\ZERN]$ stands for the subspace of $\DP_{\BLT} (\R^N) \equiv \DP_n$ of the distributions supported at zero.
In order not to destroy the condition $(r_3)$ we have to choose $\mathcal{Q}_n$ so that it commutes with multiplication by polynomials on $\R^N$.
According to Theorem~\ref{Th-4.6-IMP} $\mathcal{Q}_n$ must be constructed by a linear functional $q_n$ on $\AMPSPA_n$ via the ansatz:
$$
\mathcal{Q}_n (G) (\xx) \, = \,
\mathop{\sum}\limits_{\rr \, \in \, \{0,1,\dots\}^{\times N}} \, 
q_n 
\Bigl(
\frac{(-\xx)^{\hspace{1pt}\rr}}{\rr\hspace{1pt}!} \, G 
\Bigr)
\,
\delta^{(\rr)} (\xx) 
\,.
$$
In order not to destroy the scaling property $(r_1)$ we need to impose that $q_n$ 
has a degree $-N$ with respect to the degree of homogeneity.
What remains is only to make $q_n$ Lorentz invariant.
To this end let us first  decompose $\AMPSPA_n$ into an (infinite) direct sum 
$$
\AMPSPA_n \, = \,
\mathop{\bigoplus}\limits_{\Xi} \AMPSPA_{n;\Xi}
$$
of {\it finite dimensional} irreducible representations $\AMPSPA_{n;\Xi}$ of the Lorentz Lie algebra and let us consider the restrictions
\beq\label{AMPSPA-deco}
q_n\vrestr{12pt}{\AMPSPA_{n;\Xi}} \in \AMPSPA_{n;\Xi}' \,.
\eeq

Now, if $X$ is the vector field on $\R^{Dn}$ induced by some infinitesimal Lorentz transformation of $\R^D$ let us consider the commutator $[X,\RENMAP_n]$. 
By the induction hypothesis will take values that are distributions supported at the origin, i.e.,
$[X,\RENMAP_n]:\AMPSPA_n \to \DP_{\BLT} [\ZERN]$.
Furthermore,
by $(r_3)$
$[X,\RENMAP_n]$ commutes with the multiplication by polynomials on $\R^N$ and hence, applying Theorem~\ref{Th-4.6-IMP} we obtain a decomposition:
$$
[X,\RENMAP_n] (G) (\xx) \, = \,
\mathop{\sum}\limits_{\rr \, \in \, \{0,1,\dots\}^{\times N}} \, 
\Omega_n (X)
\Bigl(
\frac{(-\xx)^{\hspace{1pt}\rr}}{\rr\hspace{1pt}!} \, G 
\Bigr)
\,
\delta^{(\rr)} (\xx)
$$
for some $\Omega_n (X) \in \AMPSPA_n'$ linearly depending on $X$.
Then, let us compute:
\beqa\label{COCYCL-COND}
[X,\RENMAP'_n] \podr = \, 
[X,\RENMAP_n] + [X,\mathcal{Q}_n]
\,,
\bnn {}
[X,\mathcal{Q}_n] (G) (\xx) \podr = \,
\mathop{\sum}\limits_{\rr \, \in \, \{0,1,\dots\}^{\times N}} \, 
(- q_n \circ X)
\Bigl(
\frac{(-\xx)^{\hspace{1pt}\rr}}{\rr\hspace{1pt}!} \, G 
\Bigr)
\,
\delta^{(\rr)} (\xx) 
\,.
\eeqa
Note, $q_n \mapsto X^{\#}(q_n):=- X^*(q_n) \equiv - q_n \circ X$ 
is the dual action of the Lorentz Lie algebra on $\AMPSPA_n$ and furthermore,
$X \mapsto \Omega_n (X)$ is a $1$--cocycle for this action.
Thus, 
the condition $[X,\RENMAP'_n]=0$ according to (\ref{COCYCL-COND}) is equivalent to the exactness of the $1$--cocycle $\Omega_n (X)$:
\beq\label{1cocycl-cond}
\Omega_n (X) 
\, = \,
d q_n (X)
\, \equiv \,
X^{\#} (q_n)
\,.
\eeq
It is crucial now that 
$\AMPSPA_n$ decomposes into a direct sum of finite--dimensional irreducible representations (\ref{AMPSPA-deco}).
Then we can solve Eq. (\ref{1cocycl-cond}) by restricting it on each irreducible summand:
$$
\Omega_n \vrestr{10pt}{\AMPSPA_{n;\Xi}}
\, = \,
d q_n \vrestr{10pt}{\AMPSPA_{n;\Xi}} \,.
$$
By the semisimplicity of the Lorentz Lie algebra it follows that the $1$-cocycle $\Omega_n \vrestr{12pt}{\AMPSPA_{n;\Xi}}$ is trivial for every $\AMPSPA_{n;\Xi}$, and hence, $\Omega_n$.
In other words, we can construct the desired linear functional $q_n$.
Then, $\RENMAP_n' := \RENMAP_n + \mathcal{Q}_n$ will satisfy property $(r_{2,2})$ without
destroying the validity of the other properties.


\section[{\PHANT}Examples of residues of Feynman amplitudes]{Examples of residues of Feynman amplitudes}\label{Spokes-0}

\subsection{An example of a primitively divergent Feynman amplitude: the wheel with $n$ spokes}\label{Spokes}

A primitively divergent Feynman amplitude is such an amplitude that admits a homogeneous extension outside the total diagonal.
We shall consider here an example of such an amplitude in $D=4$ dimensional Euclidean space--time,
which corresponds to an $n$--loop graph with $n+1$ vertices:
\begin{equation}
\label{eq3.4}
G_n = \biggl(\prod_{j \, = \, 1}^n (\x_0-\x_j)^2 (\x_j-\x_{j+1})^2\biggr)^{-1}
\,,\qquad \text{with} \quad \x_{n+1} \equiv \x_1 \,.
\end{equation}
It can be parametrized by the spherical coordinates of the $n$ independent 4-vectors:
\beqa\label{To-eq3.5}
&
\xx \, = \, (\x_0 - \x_1,\dots,\x_0-\x_n) \, = \,
(x^1,\dots,x^N) \qquad (N = 4n) \,,
&
\\[5pt] \label{To-eq3.5-1}
&
\x_0 - \x_j = r_j \, \omega_j 
\,, \quad 
r_j \geqslant 0 
\,, \quad
\omega_j \in \Sr^3 
\quad
(\omega_j^2 = 1)
\,, \quad 
j=1, 2, \dots, n.
& \nnb[-3pt]
\eeqa
An important special case is given by the complete 4-point graph on 
following picture
(presenting the tetrahedron graph in the $(\varphi^4)_4$-theory):
\beq\label{Fig_2}
\text{${}$\hfill
$
\ALTERNATIVE{%
\includegraphics[width=3cm]{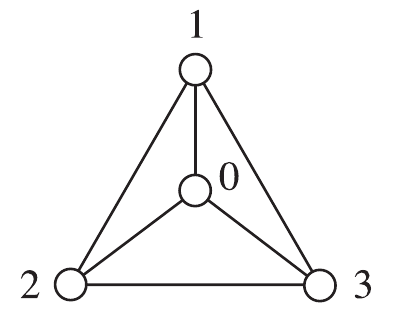}}{%
\includegraphics[width=3cm]{Fig2.pdf}}
$
\qquad
\raisebox{1cm}{\(
G_3 = \frac{\textstyle 1}{\textstyle \x_{0,1}^2 \, \x_{0,2}^2 \, \x_{0,3}^2 \, \x_{1,2}^2 \, \x_{2,3}^2 \, \x_{1,3}^2}
\qquad
(\x_{j,k} := \x_j-\x_k)\)}
\hfill\raisebox{-5pt}{${}$}}
\eeq
According to Eq. (\ref{AnSubtr-RES}) the extension of $G_n$ is done by
\beq\label{AnSubtr-2-G}
G_n^{\MHYP} (\xx) 
\, = \, 
\mathop{\lim}\limits_{\varepsilon \, \to \, 0}
\,\Bigl\{
\Bigl[\MRAD_{\MHYP} (\xx)^{\varepsilon}\, G (\xx) \Bigr]
\raisebox{12pt}{\hspace{0pt}}^{\text{extended on } \R^N}
-
\frac{1}{\varepsilon} \,
\RES 
(G_n)
\,
\delta (\xx)
\Bigr\} \,,
\eeq
where
\beq\label{W-RES-Gn}
\RES 
(
G_n
)
\, = \,
\mathop{\text{\Large $\textstyle \int$}}\limits_{\hspace{-5pt}\MHYP}
G_n (\xx)
\, \Bigl(
\mathop{\sum}
\limits_{j \, = \, 1}^{N} (-1)^{j-1} \, x^j \, d x^1 \wedge \cdots \wedge \widehat{dx^j} \wedge \cdots \wedge dx^N
\Bigr) \,
\
\eeq
as $G_n (\xx)$ is locally integrable for $\xx \neq 0$ and homogeneous of degree $-N$.
Here, $\MHYP$ is a hyper-surface encircling the origin (as in Sect. \ref{SECC-3}) and although the extension $G_n^{\MHYP}$ depends on $\MHYP$ the residue term $\RES(G_n)  \, \delta (\xx)$ does not depend on it as we explained in Sect. \ref{Se3.6GGG}
and for this reason we dismiss $\MHYP$ from $\RES(G_n)$ in (\ref{AnSubtr-2-G}).
We shall use the letter fact in order to compute this residue term by changing the hyper-surface $\MHYP$ with another hyper-surface $\MHYP'$ that is possibly either non-smooth (although continuous) or non-compact.
Note that in the first term of the right hand side of Eq. (\ref{AnSubtr-2-G}) we will not replace the norm $\MRAD_{\MHYP} (\xx)$ by the ``norm'' coresponding to $\MHYP'$ as in this case our results from Sect. \ref{SECC-3} do not ensure that the polar part at $\varepsilon = 0$ will be only of the form $\frac{1}{\varepsilon} \, \RES(G_n) \, \delta(\xx)$.

We consider first the residue in the above ``physical'' case of $n+1=4$ points using a (continuous but not smooth) hyper-surface defined by the norm:
$$
R (\xx) \, := \, \max(r_1, r_2, r_3) \,,
\qquad
\MHYP' \, := \, \bigl\{\xx \in \R^{4 \times 3} \bigl| R(\xx)=1\bigr\} \,.
$$
The case of an arbitrary $n$ we shall treat applying a result by Broadhurst 
\cite{B93} using a seminorm instead.
We shall compute $\RES(G_3)$ by first integrating $G_3$ over the angles $\omega_j$ using the 
identification of the propagators $\frac{1\textstyle }{\textstyle \x_{j,k}^2}$ with the generating functions for the {\it Gegenbauer polynomials}. 
Having in mind applications to a scalar field theory in $D$  dimensions we shall write down the corresponding more general formulas. 
The propagator $(\x_{1,2}^2)^{-\lambda}$ of  a free massless scalar field in 
$D = 2\lambda + 2$ dimensional space-time is expanded as follows in (hyperspherical) Gegenbauer polynomials:
$$
(\x_{j,k}^2)^{-\lambda} = (r_j^2 + r_k^2 -2 r_j r_k \, \omega_j \spr \omega_k)^{-\lambda}= \frac{1}{R_{j,k}^{2\lambda}} \sum_{n=0}^{\infty} \left( \frac{r_{j,k}}{R_{j,k}} \right)^n C_n^\lambda (\omega_j \spr \omega_k) \, ,
$$
\begin{equation}
\label{eq3.7}
R_{j,k} := \max (r_j , r_k) \, , \quad r_{j,k} := \min (r_j , r_k) \, , \quad i \ne j \, , \ i,j = 1,2,3.
\end{equation}
We shall also use the integral formula
\begin{equation}
\label{eq3.8}
\int_{\Sr^{2\lambda+1}} d \omega \, C_m^\lambda (\omega_1 \spr \omega) \, C_n^\lambda (\omega_2 \spr \omega) = \frac{\lambda |\Sr^{2\lambda +1}|}{n+\lambda} \, \delta_{mn} \,
 C_n^\lambda (\omega_1 \spr \omega_2) \, ,
\end{equation}
where $|\Sr^{2\lambda+1}| = \frac{2 \,\pi^{\lambda+1}} {\Gamma(\lambda+1)}$ is the volume of the unit hypersphere in $D = 2\lambda + 2$ dimensions.

\smallskip

Clearly, the expansion (\ref{eq3.7}) requires an ordering of the lengths $r_j$. In general, one should consider separately $n!$ sectors, obtained from one of them, say
\begin{equation}
\label{eq3.9}
r_n \geqslant r_{n-1} \geqslant \cdots \geqslant r_1 \, (\geqslant 0)
\end{equation}
by permutations of the indices. For $n=3$ it is, in fact, sufficient to consider just the sector (\ref{eq3.9}) and multiply the result for the residue by six. (Because of the symmetry of 
the tetrahedron graph (\ref{Fig_2}) this is obvious for $n=3$; unfortunately, there is no full permutation symmetry for $n \, (\geqslant 4)$, so the general case will require a more complicated 
argument.) The result of the angular integration in the sector (\ref{eq3.9}) involves a polylogarithmic function:
\begin{eqnarray}
\label{eq3.10}
\podr 
\RES (G_n) 
:= \, 
n \,
\int_{\max\{r_1,\dots,r_{n-1}\} \, \leqslant \, r_n \, = \, 1} 
r_1^3 dr_1 \cdots r_{n-1}^3 dr_{n-1}
\nnb
\podr \times \, \int_{\Sr^3} \mathrm{Vol}_{\Sr^3}(\omega_1) \cdots \int_{\Sr^3} \mathrm{Vol}_{\Sr^3} (\omega_n) \, G_n \vrestr{12pt}{\x_{0,j} \, = \, r_j \, \omega_j \text{ for } j=1,\dots,n} \,  \nonumber \\
\podr = \, 
n \, \int_{1 \, = \, r_n \, \geqslant \, r_{n-1} \, \geqslant \, \cdots \, \geqslant \, r_1} dr_1 \cdots dr_{n-1} \
\frac{(2 \pi^2)^n}{r_1 \cdots r_n} \ Li_{n-2}\Bigl(\frac{r_n^2}{r_1^2}\Bigr)
+ \cdots
,\qquad\qquad \\ \nonumber
& & Li_{n-2}(\xi) = \sum_{m=1}^\infty \frac{1}{m^{n-2}}\, \xi^{m} \quad \, (\xi = \frac{r_1^2}{r_n^2})
\,.
\end{eqnarray}
To derive the last equation we have applied once more (\ref{eq3.8}) and used
$$
\left(C_m^1 (\omega_1^2) \, = \, \right) \, C_m^1 (1) = m+1.
$$
Then for $n=3$ the residue (\ref{W-RES-Gn}) is given by
\begin{equation}
\label{eq3.12}
\RES \, G_3^0 =3! \, (2\pi^2)^3 \int_0^1\frac{dr_2}{r_2}  \int_0^{r_2}\frac{dr_1}{r_1} \ln \, \frac{1}{1-r_1^2} = 6 \cdot 2(\pi^2)^3 Li_3(1) = 12 \pi^6 \zeta(3).
\end{equation}

For arbitrary $n\geqslant 3$ we shall compute $\RES (G_n)$ with $R:=r_1$ viewed as a {\it seminorm} (without assuming the inequalities (\ref{eq3.9})). 
This is justified by considering a limit in (\ref{W-RES-Gn}) for a hyper-surface $\MHYP_{\alpha}$ corresponding to the norm 
$$
\MRAD_{\alpha} \, := \, r_1^2 + \alpha \bigl(r_2^2+\cdots+r_n^2\bigr) \,,
$$
when $\alpha \to 0$.
It is straightforward to obtain in this way that 
\begin{equation}
\label{eq3.13}
\RES
\, G_n^0 = \vert \Sr^3 \vert \, r_1^4 \int d^4x_2 \cdots \int d^4x_n \, G_n^0(\x_0,\x_1, \x_2, \dots, \x_n)
\end{equation}
(the result being independent of $\x_0$).
In order to evaluate the resulting integral we use, following \cite{B93}
(for a more detailed argument see \cite[Example 3.31]{S13}), 
the relations 
\beqa
\label{eq3.14}
P_L(\x_0, \x_{L+1}) \podr := \, \prod_{k=1}^L \int\frac{d^4x_k}{\pi^2 \x_k^2}\prod_{m=0}^L\frac{1}{\x_{m,m+1}^2} 
\nnb
\podr = \, \int\frac{d^4x_k}{\pi^2 \x_k^2} P_{k-1}(\x_0, \x_k) P_{L-k}(\x_k, \x_{L+1})
\,,
\eeqa
which yields
\begin{eqnarray}
\label{eq3.15}
\podr
\x^2 P_L(\x, \y) = \sum_{n=1}^\infty C_{n, L}(r) r^{n-1} \frac{\sin n\theta}{\sin\theta}
\,,\quad 
\text{for}  \quad  r^2 = \frac{\y^2}{\x^2},  \, \, \cos\theta := \frac{\x \spr \y}{r\x^2} \,, \nonumber \\
\podr
C_{n, L}(r) = \frac{1}{n^{2L}}\sum_{k=0}^L \binom{2l-k}{L} \frac{(\ln \frac{1}{r^{2n}} )^k}{k!}, \, \, \, 
\end{eqnarray}
Inserting the result in (\ref{eq3.13}) we find (for $L=n-1, r = 1$):
\begin{equation}
\label{eq3.16}
\RES (G_n^0) \, = \,
 2\pi^{2n} \x^2 P_{n-1}(\x,\x) =2\pi^{2n} \binom{2n-2}{n-1} \zeta(2n-3).
\end{equation}
In particular, for the tetrahedron graph, $n = 3$, we reproduce the known result (\ref{eq3.12}) -- see, for instance, \cite{G-B}.  

\smallskip

The integration technique based on the properties of Gegenbauer polynomials has been introduced in the study of $\x$-space Feynman integrals in \cite{CKT}. 
The appearance of $\zeta$-values in similar computations has been detected in early work of Rosner \cite{R} and Usyukina \cite{U}. It was related to the non-trivial 
topology of graphs by Broadhurst and Kreimer \cite{BrK}, \cite{Kr}.

\subsection{Residues of recursively renormalized Feynman amplitudes}\label{Res_Rec}

One of the simplest examples of a Feynman amplitude that needs a recursive renormalization corresponds to the following diagram in the $(\varphi^4)_4$--theory: 
\beq\label{Fig_3}
\text{${}$\hfill
$
\ALTERNATIVE{%
\includegraphics[width=3cm]{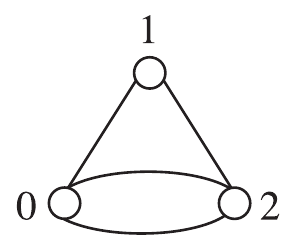}}{%
\includegraphics[width=3cm]{Fig3.pdf}}
$
\qquad
\raisebox{1cm}{\(
G_{\{0,1,2\}} = \frac{\textstyle 1}{\textstyle \x_{0,1}^2 \, (\x_{0,2}^2)^2 \, \x_{1,2}^2}
\,.
\)}
\hfill\raisebox{-5pt}{${}$}}
\eeq
Let $G_{\{0,1,2\}}^0$ be the extended (renormalized) amplitude outside the total diagonal $\x_{0,1}=0=\x_{0,2}$ under the construction of Theorem \ref{THM-EUCL-REC} with $G^{ext}_{\{0,1\}}$ $=$ $\bigl(\x_{0,1}^2\bigr)^{-1}$, $G^{ext}_{\{1,2\}}$ $=$ $\bigl(\x_{1,2}^2\bigr)^{-1}$ and $G^{ext}_{\{0,2\}}$ $=$ $\bigl(\bigl(\x_{0,2}^2\bigr)^{-2}\bigr)^{\text{extended on $\R^4$}}$.
The extension of $G^{ext}_{\{0,2\}}$ on $\R^4$ ($\ni \x_{0,2}$) can be performed for instance according to Proposition \ref{Prop-N3.1}, and $G^{ext}_{\{0,1\}}$ and $G^{ext}_{\{1,2\}}$ are uniquely defined as homogeneous distributions on $\R^4$.
As $G^{ext}_{\{0,2\}}$ is associate homogeneous of order 1 then according to Lemma \ref{EucSym} the distribution $G_{\{0,1,2\}}^0$ will be associate homogeneous of order $1$ (and degree $-8$) on $(\R^4 \times \R^4) \backslash \{0\} \ni (\x_{0,1},\x_{0,2})$.
We will set in this example 
$\xx$ $:=$ $(\x_{0,1}$, $\x_{0,2})$ as independent variables on $\R^8$ and fix the norm 
$$
\MRAD_{\ell_1,\ell_2}(\xx) \, = \, \biggl(\frac{\x_{0,1}^2}{\ell_1^2}+\frac{\x_{0,2}^2}{\ell_2^2}\biggr)^{\frac{1}{2}}
\,,\quad
\MHYP \ ( \, = \,\MHYP_{\ell_1,\ell_2} \, ) \, = \, \bigl\{\xx \,\bigl|\, \MRAD_{\ell_1,\ell_2}(\xx) =1 \bigr\}
\,.
$$
Then according to Eqs. (\ref{AnSubtr-RES}) and (\ref{Res-def1}) we have
\beqa\label{G0123-RES}
\podr
G_{\{0,1,2\}}^{ext} (\xx) \, = \, 
\mathop{\lim}\limits_{\varepsilon \, \to \, 0}
\,\Bigl(
\Bigl[\MRAD_{\ell_1,\ell_2} (\xx)^{\varepsilon}\,G_{\{0,1,2\}}^0 (\xx) \Bigr]
\raisebox{12pt}{\hspace{0pt}}^{\text{extended on } \R^8}
\nn \podr
- \,
\frac{1}{\varepsilon} \ 
\RES_{\MHYP}\bigl(G_{\{0,1,2\}}^0\bigr) \, \delta (\xx)
+ \,
\frac{1}{\varepsilon^2} \
\RES_{\MHYP}\bigl((\MEUL+8) \, G_{\{0,1,2\}}^0\bigr) \, \delta (\xx)
\Bigr)
\,,
\qquad\quad
\eeqa
where $\RES_{\MHYP}\bigl(G_{\{0,1,2\}}^0\bigr)$ and 
$\RES_{\MHYP}\bigl((\MEUL+8) \, G_{\{0,1,2\}}^0\bigr)$
are two real numbers from which the second one is independent on the choice of the norm $\MRAD_{\ell_1,\ell_2}$ and furthermore if $\RES_{\MHYP}\bigl((\MEUL+8) \, G_{\{0,1,2\}}^0\bigr)$ vanish then the order of $G_{\{0,1,2\}}^{ext}$ will remain~$1$ (as an associate homogeneous distribution). 
We shall see however that the latter is not the case.

Let us consider first the highest pole part in (\ref{G0123-RES}) (which is renormalization independent).
It is determined by the successor $(\MEUL+8) \, G_{\{0,1,2\}}^0$ of $G_{\{0,1,2\}}^0$.
Let us start with the general observation that outside the large diagonal $G_{\{0,1,2\}}^0$ every massless Feynman amplitude is homogeneous (and even smooth in the Euclidean case).
Therefore, all the successors of such amplitudes, and in particular, $(\MEUL+8) \, G_{\{0,1,2\}}^0$, must be supported at the large diagonal.
In the case of $G_{\{0,1,2\}}^0$ we have:
$$
(\MEUL+8) \, G_{\{0,1,2\}}^0 \, = \, 2\pi^2 \, (\x_{0,1}^2)^{-2} \, \delta (\x_{0,2})
\qquad
(\x_{0,1} \neq 0) \,,
$$
according tot Eq. (\ref{2ptDilLaw}) (for $D=4$).
We observe that residue $\RES_{\MHYP} \bigl((\x_{0,1}^2)^{-2} \, \delta (\x_{0,2})\bigr)$ is the same as the residue of $(\x_{0,1}^2)^{-2}$ that is again extracted from Eq. (\ref{2ptDilLaw}), and is equal to $2\pi^2$.
So, we obtain
$$
\RES_{\MHYP}\bigl((\MEUL+8) \, G_{\{0,1,2\}}^0\bigr) \, = \, 4\pi^4 \, \neq \, 0
$$ 
and thus, $G_{\{0,1,2\}}^{ext}$ has always order $2$ as an associate homogeneous distribution.

Computing $\RES_{\MHYP}\bigl(G_{\{0,1,2\}}^0\bigr)$ is a less trivial task as it uses ``more information'' from the amplitude $G_{\{0,1,2\}}^0$.
(In addition, this number will depend on $\MHYP$.)
We will give the calculation in sketch only.
First we obtain a suitable coordinate form for the integrations:
setting $\x_{0,1} = r_1 \u_1$ and $\x_{0,2} = r_2 \u_2$,
where $\u_1,\u_2 \in \Sr^3$ we have
\beqs
\MVOL \, = \podr r_1^3 \, r_2^3 \, dr_1 \wedge dr_2 \wedge \MVOL_{\Sr^3}(\u_1) \wedge \MVOL_{\Sr^3}(\u_2) 
\,,\quad
\\
\MEUL \, = \podr \x_{0,1} \spr \di_{\x_{0,1}} + \x_{0,2} \spr \di_{\x_{0,2}}
\, = \, r_1 \, \di_{r_1} + r_2 \, \di_{r_2}
\,,\quad
\\
\MVOL_{\MHYP} \, = \podr \iota_{\MEUL} \MVOL \,\vrestr{10pt}{\MHYP} \, = \,
r_1^3 \, r_2^3 \, (r_1 \, dr_2 - r_2 \, dr_1) \wedge \MVOL_{\Sr^3}(\u_1) \wedge \MVOL_{\Sr^3}(\u_2) 
\,\vrestr{10pt}{\MHYP}
\,.
\eeqs
In these coordinates the domain of integration is
$$
\frac{r_1^2}{\ell_1^2}+\frac{r_2^2}{\ell_2^2} =1 \,,\quad r_1,r_2 \geqslant 0 \,,\quad \u_1,\u_2 \in \Sr^3 \,.
$$
Then integrating the part in $\RES_{\MHYP}\bigl(G_{\{0,1,2\}}^0\bigr)$ that is over $\u_1$ and $\u_2$ for $r_1 \neq r_2$ we obtain
$$
\mathop{\int}\limits_{\hspace{-5pt}\Sr^3 \times \Sr^3} \MVOL_{\Sr^3}(\u_1) \wedge \MVOL_{\Sr^3}(\u_2) \,
\bigl((r_1 \u_1-r_2\u_2)^2\bigr)^{-1}
\, = \,
4 \pi^4 \max(r_1,r_2)^{-2} \,
$$
and together with the remaining part from the integration in $\RES_{\MHYP}\bigl(G_{\{0,1,2\}}^0\bigr)$ we get formally
$$
\RES_{\MHYP}\bigl(G_{\{0,1,2\}}^0\bigr)
\, = \,
4 \pi^4 \,
\mathop{\int}\limits_{(r_1,r_2) \, \in \, \sigma}
r_1 \, r_2^{-1} \, \max(r_1,r_2)^{-2} \,
(r_1 \, dr_2 - r_2 \, dr_1) \,,
$$
where $\sigma=\sigma_{\ell_1,\ell_2}$ is the arc
$\Bigl\{\frac{\textstyle r_1^2}{\textstyle \ell_1^2}+\frac{\textstyle r_2^2}{\textstyle \ell_2^2} = 1,\,  r_1,r_2 \, \geqslant \, 0\Bigr\}$.
The resulting integral is divergent at $r_2=0$, which is due to the necessity of renormalization of the subdivergent amplitude $(\x_{0,2}^2)^{-2} \equiv r_2^{-4}$.
At this point we should specify the renormalization (extension) procedure used before the recursive construction of $G_{\{0,1,2\}}^0$.
If we use the scheme of analytic renormalization then 
$$
G_{\{0,1,2\}}^0 \, = \, 
\mathop{\lim}\limits_{\nu \, \downarrow \, 0}
\biggl(
\bigl(\x_{0,1}^2\bigr)^{-1}
\bigl(\x_{1,2}^2\bigr)^{-1}
\bigl(\x_{0,2}^2\bigr)^{-2} \, \biggl(\frac{\x_{0,2}^2}{\ell_0^2}\biggr)^{\frac{\nu}{2}} 
- 
\frac{1}{\nu} \, F
\biggr)
\,,
$$
where the extension of $(\x_{0,2}^2)^{-2}$ is parametrized by a new scale parameter $\ell_0$
and the distribution coefficient $F$ to $1/\nu$ is proportional to
$(\x_{0,1}^2)^{-2} \, \delta (\x_{0,2})$.
Hence,
we need to calculate the regularized integral
with $\nu > 0$ and then expand in $\nu$ and subtract the polar part:
\beqs
\podr
\mathop{\int}\limits_{(r_1,r_2) \, \in \, \sigma}
r_1 \, r_2^{-1} \, \biggl(\frac{r_2}{\ell_0}\biggr)^{\nu} \, \max(r_1,r_2)^{-2} \,
(r_1 \, dr_2 - r_2 \, dr_1)
\, = \,
\frac{1}{\nu} + 
\frac{1}{2}+\ln \frac{\ell_1}{\ell_0}
+ O(\nu)
\,.
\eeqs
In this way we obtain:
\quad
\(
\RES_{\MHYP}\bigl(G_{\{0,1,2\}}^0\bigr) = 4 \pi^4 \, 
\Bigr(\frac{\textstyle 1}{\textstyle 2}+\ln \frac{\textstyle \ell_1}{\textstyle \ell_0} \Bigl) \,.
\)


\label{BIBLL}
\addtocontents{toc}{\protect\vspace{-11pt}}
\addtocontents{toc}{\contentsline {section}{{\bf\small }{\bf\small References}}{\small\rm \pageref{BIBLL}}}

\end{document}